\documentclass[apj,numberedappendix,appendixfloats]{emulateapj}
\usepackage{apjfonts}

\newcommand{\etal}{et~al.}
\newcommand{\MgIIdblt}{{\rm Mg}\kern 0.1em{\sc ii}~$\lambda\lambda 2796, 2803$}
\newcommand{\MgII}{\hbox{{\rm Mg}\kern 0.1em{\sc ii}}}
\newcommand{\SiII}{\hbox{{\rm Si}\kern 0.1em{\sc ii}}}
\newcommand{\CII}{\hbox{{\rm C}\kern 0.1em{\sc ii}}}
\newcommand{\FeII}{\hbox{{\rm Fe}\kern 0.1em{\sc ii}}}
\newcommand{\CaII}{\hbox{{\rm Ca}\kern 0.1em{\sc ii}}}
\newcommand{\HI}{\hbox{{\rm H}\kern 0.1em{\sc i}}}
\newcommand{\kms}{\hbox{km~s$^{-1}$}}
\newcommand{\cmsq}{\hbox{cm$^{-2}$}}

\newcommand{\magiicat}{\hbox{{\rm MAG}{\sc ii}CAT}}

\shorttitle{\sc {CGM Virial Mass Self-Similarity}}
\shortauthors{\sc Churchill et~al.}
\slugcomment{Submitted to the Astrophysical Journal}

\begin{document}


\title{~{\magiicat} III. Interpreting Self-Similarity of the
  Circumgalactic Medium \\ with Virial Mass using {\MgII} Absorption}



\author{\sc
Christopher W. Churchill\altaffilmark{1},
Sebastian Trujillo-Gomez\altaffilmark{1},
Nikole M. Nielsen\altaffilmark{1},
and
Glenn G. Kacprzak\altaffilmark{2,3}
}
\altaffiltext{1}{New Mexico State University, Las Cruces, NM 88003}
\altaffiltext{2}{Swinburne University of Technology, Victoria 3122, Australia}
\altaffiltext{3}{Australian Research Council Super Science Fellow}

\begin{abstract}
In Churchill {\etal}, we used halo abundance matching applied to 182
galaxies in the {\MgII} Absorption-Galaxy Catalog ({\magiicat},
Nielsen {\etal}) and showed that the mean {\MgII}~$\lambda 2796$
equivalent width follows a tight inverse-square power law, $W_r(2796)
\propto (D/R_{\rm vir})^{-2}$, with projected location relative to the
galaxy virial radius and that the {\MgII} absorption covering fraction
is effectively invariant with galaxy virial mass, $M_{\rm\, h}$, over
the range $10.7 \leq \log M_{\rm\, h}/M_{\odot} \leq 13.9$.  In this
work, we explore multivariate relationships between $W_r(2796)$,
virial mass, impact parameter, virial radius, and the theoretical
cooling radius that further elucidate self-similarity in the cool/warm
($T=10^{4-4.5}$ K) circumgalactic medium (CGM) with virial mass.  We
show that virial mass determines the extent and strength of the
{\MgII} absorbing gas such that the mean $W_r(2796)$ increases with
virial mass at fixed distance while decreasing with galactocentric
distance for fixed virial mass.  The majority of the absorbing gas
resides within $D \simeq 0.3 R_{\rm vir}$, independent of both virial
mass and minimum absorption threshold; inside this region, and perhaps
also in the region $0.3 < D/R_{\rm vir} \leq 1$, the mean $W_r(2796)$
is independent of virial mass.  Contrary to absorber-galaxy
cross-correlation studies, we show there is no anti-correlation
between $W_r(2796)$ and virial mass.  We discuss how simulations and
theory constrained by observations support self-similarity of the
cool/warm CGM via the physics governing star formation, gas-phase
metal enrichment, recycling efficiency of galactic scale winds,
filament and merger accretion, and overdensity of local environment as
a function of virial mass.
\end{abstract}

\keywords{galaxies: halos --- quasars: absorption lines}


\section{Introduction}
\label{sec:intro}

Early models of galaxy formation were based on relatively simple
scenarios in which baryonic gas collapsed in a monolithic structure
due to gravitational instability \citep{els62} modulated by transient
infalling post-collapse protogalactic fragments that were chemically
evolving \citep{searle78}.  Following the adoption of the dark matter
paradigm, this scenario matured into a model in which gas cooled as it
accreted into dark matter halos, condensed and relaxed in the halo
center, and formed stars \citep[e.g.,][]{white78, silk81,
  blumenthal86, white91, mo96, maller04}.  As observational details 
emerged and theoretical ideas evolved over the last decades, we
collectively developed a more complex picture of galaxy evolution in
which stars and gas are intimately linked in complex cycles involving
galactic scale outflowing stellar driven winds, filamentary accretion,
major and minor galaxy mergers, and the development of a hot coronal
gas medium, all in the context of dark matter halo evolution
\citep[e.g.,][]{keres05, keres09, dekel09, ceverino09, oppenheimer10,
  schaye10, danovich12, vandevoort+schaye11, ceverino13}.

Due to the primary role of gas in the global evolution of galaxies,
the connection between gas processes and galaxy stellar masses,
colors, luminosities, and morphologies, have been explored with
increasing sophistication using semi-analytic models
\citep[e.g.,][]{bullock01b, somerville01, hernquist03, croton06,
  henriques10}, simulations of isolated galaxies
\citep[e.g.,][]{birnboim03, dekel06, birnboim11}, and hydrodynamic
cosmological simulations that incorporate the context of local
overdensity and environment \citep[e.g.,][]{keres05, keres09,
  ceverino09, oppenheimer10, vandevoort+schaye11, ceverino13}.  The
studies indicate that the gas bound within and/or inflowing,
outflowing, or recycling through galaxy dark matter halos governs the
large scale physics driving galaxy evolution and therefore controls
the global distribution of observed galaxy properties.

We now fully accept the reality of an extended gaseous medium
surrounding galaxies that regulates the rhythms of star-formation in
the gaseous interstellar medium (ISM) and the accretion of gaseous
structures from the surrounding intergalactic medium (IGM).  This
complex, multi-phase, highly dynamic ``circumgalactic medium'' (CGM)
is where chemically-enriched galactic scale outflowing stellar winds
interact and mix with infalling gas-rich satellites and intergalactic
filaments.  The CGM is the reservoir that buffers the ISM from the
IGM and controls the efficiency at which baryonic gas is converted
into stars.

The mass of the dark matter halo dictates the depth and concentration
of the gravitational potential \citep{nfw95, klypin01}, and correlates
with local overdensity and environment \citep[e.g.,][]{mo-white96,
  klypin11}.  Thus, the physics of the CGM is intimately connected to
its dark matter halo mass, which dictates the hot coronal gas
temperature, density profile, and pressure gradient.  This physics
also governs the cloud infall, compression, cooling, formation, and
disruption timescales \citep[e.g.,][]{mo96, maller04, dekel06}.
Furthermore, the dark matter plus baryonic matter halo profile
provides the radial profile of the escape velocity
\citep[see][]{steidel10}.  As such, halo mass governs the overall
balance and efficiency of gas and metallicity transport via infall,
outflow, and recycling.  It is highly probable that the CGM forged the
observed shape of the stellar-mass to halo-mass relation of
galaxies \citep[cf.,][]{behroozi13}.

Based on isolated galaxy and cosmological simulations, a strong
dependence of CGM properties on dark matter virial mass, $M_{\rm\,
  h}$, has been found \citep[e.g.,][]{birnboim03, keres05, keres09,
  dekel06, stewart11, vandevoort11}.  The simulations indicate that
$\log M_{\rm\, h}/M_{\odot} \sim 12$ is a critical mass, above which
``cold-mode'' accreting gas is expected to be suppressed since the
cooling time and/or compression time of the gas is longer than the gas
dynamical time.  As such, accreting cool/warm clouds are not expected
to survive as the gas shock heats near the virial radius, resulting in
``hot-mode'' accreting gas that remains in the hot gaseous corona.  In
halos of $\log M_{\rm\, h}/M_{\odot} \leq 12$, the accreting gas can
cool on a shorter time scale than the dynamical time, so that
cool/warm accreting clouds are expected to survive and accrete into
the ISM and fuel star formation.  ``Cold-mode'' or ``hot-mode''
accretion in a given galaxy halo would first and foremost govern the
mass and chemical enrichment of gas infalling from the IGM, and
secondly, through its interaction with stellar driven winds, govern
the recycling of cool/warm clouds through the CGM and back into the
ISM.

As such, the chemical composition, temperatures, densities, geometric
distributions, ionization conditions, and kinematics of the various
gaseous structures in the CGM are expected to reflect the dark matter
halo mass, and thus provide a detailed snapshot of the complex recent
history of a galaxy and its future evolution.  Charting these CGM
properties across a range of galaxies (i.e., dark matter halo masses)
over cosmic time promises highly detailed insight into the physics
underlying galaxy evolution and places important constraints on galaxy
evolution theory.

Currently, the best approach to measuring CGM gas properties out to
large galactocentric distances is to analyze absorption lines in the
spectra of background luminous objects whose lines of sight
serendipitously pass near intervening galaxies.  One approach is to
use stacking techniques of large numbers of sightlines to gain insight
through statistically significant global behaviors
\citep[e.g.,][]{zibetti07, steidel10, bordoloi11, rudie12, zhu13,
  bordoloi13}, but for which the detailed complexity of the CGM and
its relationship to galaxies and dark matter halos is smoothed over.
Alternatively, samples of high-quality spectra can be studied on a
CGM-to-galaxy basis, which provide insights into the complexity of the
CGM environment in relation to galaxy properties
\citep[e.g.,][]{sdp94, lanzetta95, chen01a, chen01b, kacprzak08,
  chen10a, kcems, rao11, nikki-cat1, nikki-cat2, stocke13, werk13},
but yield smaller numbers for which statistically significant insight
is mitigated.

Recent studies of far ultraviolet metal-line transitions using the
Cosmic Origins Spectrograph on the {\it Hubble Space Telescope} to
study the $z<0.3$ CGM in detail have revealed a metal-enriched
environment comprising $\sim 50$\% of the baryonic gas mass in dark
matter halos \citep{tumlinson11}.  CGM gas exhibits a wide range of
density, metallicity, and localized ionizing conditions
\citep[e.g.,][]{stocke13, werk13}.  The kinematics indicate that the
majority of the gas is gravitationally bound, recycling material
\citep{tumlinson11}.  However, \citet{stocke13} report that some
clouds seen in absorption may be escaping the galaxy.  They also find
that almost all cool/warm CGM clouds reside within the inner 50\% of
the virial radius, and that there are no trends in the cool/warm CGM
cloud properties with galactocentric distance, relative velocity, or
galaxy luminosity once they scale the cloud locations with respect to
virial radius.

At $z>0.3$, the CGM is mostly studied with ground-based facilities
using the near ultraviolet {\MgIIdblt} transitions.  With an ionization
potential slightly above that of {\HI}, {\MgII} probes the cool/warm
component of the CGM.  Here, we define cool/warm gas to have a
temperature range of $T = 10^{4-5}$~K, though this gas is often dubbed
``cold'' gas.  The strength of {\MgII} as a tracer of the CGM is that
it arises in low-ionization gas over five decades of {\HI} column
density, $10^{16.5} \leq N(\HI) \leq 10^{21.5}$~{\cmsq} \citep{weakI,
  archiveI, rao00, weakII}, and is detected out to projected distances
of $\sim 150$~kpc \citep[see][for a review]{cwc-china}.  Furthermore,
{\MgII} has been directly observed or indirectly inferred to probe a
wide range of CGM structures, such as galactic scale winds
\citep[e.g.,][]{tremonti07, martin09, weiner09, rubin10, martin12},
infalling material \citep[e.g.,][]{ggk-sims, ribaudo11, rubin11,
  thom11, ggk-q1317}, co-rotating material \citep{steidel02, kcems},
superbubble structures \citep{cwc-garching05, bond01, ellison03}, and
the complex disk/extra-planer/CGM interface \citep{ggk-smallD}.

In an effort to facilitate further studies of the {\MgII} absorbing
CGM, \citet[][Paper~I]{nikki-cat1} compiled the {\MgII}
Absorber-Galaxy Catalog ({\magiicat})\footnote{
  http://astronomy.nmsu.edu/cwc/Group/magiicat/.}.  The general
characteristics of the {\MgII} absorbing CGM, including systematic
luminosity, color, and redshift dependencies of the {\MgII} absorption
covering fractions as a function of absorption threshold, are
presented in \citet[][Paper~II]{nikki-cat2}.  \citet{ggk-orientations}
used the {\magiicat} sample to show that the covering fraction has a
dependency on galaxy orientation.

In \citet{cwc-massesI}, we used halo abundance matching to obtain the
virial masses for the galaxies in {\magiicat} and studied how the
{\MgII}~$\lambda 2796$ equivalent width, $W_r(2796)$, behaves with
galaxy virial mass, impact parameter, $D$, and virial radius, $R_{\rm
  vir}$.  We presented four main results: [1] A substantial component
of the scatter in the $W_r(2796)$--$D$ anti-correlation is explained
by a systematic segregation of virial mass on the $W_r(2796)$--$D$
plane; higher virial mass absorbing galaxies are found at higher $D$
and larger $W_r(2796)$ compared to lower virial mass absorbing
galaxies.  [2] The data are well described by the relation $W_r(2796)
\propto (D/R_{\rm vir})^{-2}$ with significantly reduced scatter and a
vanishing of virial mass segregation on the $W_r(2796)$--$D/R_{\rm
  vir}$ plane.  [3] The covering fraction at a given impact parameter
is higher for higher mass halos, especially at $D<50$~kpc, than for
low mass halos, but the covering fraction at a given $D/R_{\rm vir}$
is independent of virial mass.  [4] As a function of both $D/R_{\rm
  vir}$ and $W_r(2796)$ absorption threshold, the covering fraction is
effectively independent of virial mass and does not show a precipitous
drop for $\log M_{\rm\,h}/M_{\odot} \geq 12$ as predicted by the
scenario of a suppressed ``cold-mode'' accretion in higher mass halos.
The data indicate that the absorption strength and covering fraction
of cold CGM gas is primarily governed by how far out in the virial
radius the gas resides, and that this behavior holds over a virial
mass range of $10.7 \leq \log M_{\rm\,h}/M_{\odot} \leq 13.8$.  These
results were interpreted to suggest a self-similar behavior of the
cool/warm CGM with virial mass.

In this paper, we further explore the connection between virial mass
and the {\MgII} absorbing CGM and elucidate the interrelationships
between absorption strength, virial mass, impact parameter, virial
radius, and the theoretical cooling radius.  In
\S~\ref{sec:sample+ham} we briefly overview the characteristics of the
  {\magiicat} galaxy sample and describe the application of halo
  abundance matching to estimate galaxy virial masses.  Additional
  details are provided in Appendix~\ref{app:ham}.  We characterize and
  quantify several interrelationships between the measured quantities
  in \S~\ref{sec:results}.  In \S~\ref{sec:discuss}, we discuss the
  multivariate relations in the data, and compare, contrast, and
  interpret our results with respect to other works.  As we will show,
  the data strongly support a self-similar cool/warm CGM with virial
  mass.  In \S~\ref{sec:conclude}, we summarize our findings and
  conclude with a discussion in which we draw upon observations and
  theory to address the question ``what drives the self-similarity of
  the CGM?''  Throughout this work, we adopt a flat $\Lambda$CDM
  cosmological model with $h=0.70$, $\Omega_{\rm M}=0.3$, and
  $\Omega_{\Lambda}=0.7$.  When discussing the gas phase metallicity,
  we employ the term $Z_{\rm gas}$ to designate $Z/Z_{\odot}$.


\section{The Sample and Virial Masses}
\label{sec:sample+ham}

\subsection{The Galaxy-Absorption Sample}
\label{sec:sample}

Our sample comprises the 182 ``isolated'' galaxies in the ``{\MgII}
Absorber-Galaxy Catalog'' \citep[{\magiicat},][Paper~I]{nikki-cat1}.
Each galaxy has a published spectroscopic redshift, with the sample
spanning the range $0.07 \leq z \leq 1.12$.  The galaxy-quasar impact
parameters range from $5.4 \leq D \leq 194$~kpc.  The {\sc ab}
absolute $B$- and $K$- band magnitudes cover the ranges $-16.1 \geq
M_B \geq -23.1$ and $-17.0 \geq M_K \geq -25.3$, with rest-frame $B-K$
colors $0.04 \leq B-K \leq 4.09$.  The range of detected rest-frame
{\MgII} $\lambda 2796$ equivalent widths is $0.03 \leq W_r(2796) \leq
2.90$~{\AA} with one system at $W_r(2796) = 4.42$~{\AA}.  Upper limits
($3~\sigma$) on $W_r(2796)$ were measured for 59 of the 182 systems
over the range $W_r(2796) \leq 0.003$~{\AA} to $W_r(2796) \leq
0.3$~{\AA}.  Apart from the details of how the virial masses of the
galaxies have been determined, which we present in this work, the
particulars of the galaxy-absorber sample and standardization of
photometric and absorption properties have been presented in Paper~I
\citep{nikki-cat1}.

In Table~\ref{tab:obsprops}, we present the data employed for this
work.  Columns (1) through (4) list the quasar field name (B1950
designation or identification of a quasar as having been discovered in
the Sloan Digital Sky Survey, SDSS), the quasar J2000 designation, the
galaxy redshift, $z_{\rm gal}$, and the impact parameter, $D$. Column
(13) lists the {\MgII} $\lambda 2796$ rest-frame equivalent width,
$W_r(2796)$.  These data are taken from Paper~I \citep{nikki-cat1}.
The remaining columns, which are newly published data, are: (5) the
galaxy $r$-band absolute {\sc ab} magnitude, $M_r$, (6) the virial
mass, $M_{\rm\, h}$, (7) the maximum circular velocity, $V_{c}^{\rm
  max}$, (8) the virial radius, $R_{\rm vir},$ (9) the ratio
$\eta_{\rm v} = D/R_{\rm vir}$, (10) the theoretical cooling radius,
$R_{\rm c}$, (11) the ratio $\eta_{\rm c} = D/R_{\rm c}$, and (12) the
ratio $R_{\rm c}/R_{\rm vir}$.

The $M_r$ were computed using the methods applied to obtain $M_B$ and
$M_K$ as described in Paper~I \citep{nikki-cat1}; The resulting range
is $-22.2 \leq M_r \leq -16.4$.  Calculation of the virial radius,
$R_{\rm vir}$, was presented in \citet{cwc-massesI}.  Calculation of
the theoretical cooling radius is discussed in
\S~\ref{sec:rcooldiscuss}.

\begin{deluxetable*}{llccccccccccc}
\tablecolumns{13}
\tablewidth{0pt}
\setlength{\tabcolsep}{0.03in}
\tablecaption{Galaxy Properties \label{tab:obsprops}}
\tablehead{
 \colhead                            {(1)} &
 \colhead                            {(2)} &
 \colhead                            {(3)} &
 \colhead                            {(4)} &
 \colhead                            {(5)} &
 \colhead                            {(6)} &
 \colhead                            {(7)\tablenotemark{a}} &
 \colhead                            {(8)\tablenotemark{a}} &
 \colhead                            {(9)\tablenotemark{a}} &
 \colhead                           {(10)\tablenotemark{a,b}} &
 \colhead                           {(11)\tablenotemark{a}} &
 \colhead                           {(12)\tablenotemark{a}} &
 \colhead                          {(13)} \\
 \colhead                          {Field} &
 \colhead                         {J-Name} &
 \colhead                  {$z_{\rm gal}$} &
 \colhead                            {$D$} &
 \colhead                          {$M_r$} &
 \colhead   {$\log M_{\rm\, h}/M_{\odot}$} &
 \colhead                {$V_c^{\rm max}$} &
 \colhead                  {$R_{\rm vir}$} &
 \colhead                 {$\eta_{\rm v}$} &
 \colhead                    {$R_{\rm c}$} &
 \colhead                 {$\eta_{\rm c}$} &
 \colhead        {$R_{\rm c}/R_{\rm vir}$} &
 \colhead                   {$W_r(2796)$} \\
 \colhead                              { } &
 \colhead                              { } &
 \colhead                              { } &
 \colhead                          {[kpc]} &
 \colhead                     {({\sc ab})} &
 \colhead                              { } &
 \colhead                       {[{\kms}]} &
 \colhead                          {[kpc]} &
 \colhead                              { } &
 \colhead                          {[kpc]} &
 \colhead                              { } &
 \colhead                              { } &
 \colhead                        {[{\AA}]} }
\startdata
 $ 0002\!-\!422 $ & J$000448.11\!-\!415728.8  $ & $ 0.8400 $ & $  53.8$ & $ -21.7$ & $ 12.1_{-0.1}^{+0.2} $ & $  262_{-26}^{+35} $ & $  218_{-24}^{+32} $ & $ 0.25_{+0.03}^{-0.03} $ & $   50_{+ 3}^{ -4} $ & $ 1.07_{-0.06}^{+0.09} $ & $ 0.23_{+0.03}^{-0.04} $ & $ 4.422\pm 0.002$ \\[3pt]
 $ 0002\!+\!051 $ & J$000520.21\!+\!052411.80 $ & $ 0.2980 $ & $  59.2$ & $ -20.9$ & $ 12.0_{-0.2}^{+0.3} $ & $  211_{-26}^{+45} $ & $  191_{-26}^{+45} $ & $ 0.31_{+0.06}^{-0.05} $ & $  103_{+ 5}^{ -7} $ & $ 0.57_{-0.02}^{+0.04} $ & $ 0.54_{+0.08}^{-0.13} $ & $ 0.244\pm 0.003$ \\[3pt]
 $ 0002\!+\!051 $ & J$000520.21\!+\!052411.80 $ & $ 0.5920 $ & $  36.0$ & $ -22.0$ & $ 12.3_{-0.2}^{+0.2} $ & $  291_{-29}^{+38} $ & $  257_{-28}^{+37} $ & $ 0.14_{+0.02}^{-0.02} $ & $   59_{+ 4}^{ -4} $ & $ 0.61_{-0.04}^{+0.05} $ & $ 0.23_{+0.03}^{-0.04} $ & $ 0.102\pm 0.002$ \\[3pt]
 $ 0002\!+\!051 $ & J$000520.21\!+\!052411.80 $ & $ 0.8518 $ & $  25.9$ & $ -21.2$ & $ 11.8_{-0.2}^{+0.2} $ & $  220_{-24}^{+40} $ & $  179_{-22}^{+36} $ & $ 0.14_{+0.02}^{-0.02} $ & $   60_{+ 3}^{ -5} $ & $ 0.43_{-0.02}^{+0.04} $ & $ 0.33_{+0.05}^{-0.07} $ & $ 1.089\pm 0.008$ \\[3pt]
  SDSS            & J$003340.21\!-\!005525.53 $ & $ 0.2124 $ & $  21.7$ & $ -21.3$ & $ 12.2_{-0.2}^{+0.2} $ & $  232_{-27}^{+41} $ & $  214_{-27}^{+42} $ & $ 0.10_{+0.02}^{-0.01} $ & $  107_{+ 4}^{ -6} $ & $ 0.20_{-0.01}^{+0.01} $ & $ 0.50_{+0.07}^{-0.10} $ & $ 1.050\pm 0.030$ \\[3pt]
  SDSS            & J$003407.34\!-\!085452.07 $ & $ 0.3617 $ & $  33.1$ & $ -20.1$ & $ 11.7_{-0.2}^{+0.4} $ & $  176_{-24}^{+55} $ & $  154_{-23}^{+54} $ & $ 0.21_{+0.06}^{-0.04} $ & $  106_{+ 5}^{ -9} $ & $ 0.31_{-0.01}^{+0.03} $ & $ 0.69_{+0.12}^{-0.24} $ & $ 0.480\pm 0.050$ \\[3pt]
  SDSS            & J$003413.04\!-\!010026.86 $ & $ 0.2564 $ & $  30.4$ & $ -20.7$ & $ 11.9_{-0.2}^{+0.3} $ & $  195_{-25}^{+47} $ & $  176_{-25}^{+47} $ & $ 0.17_{+0.04}^{-0.03} $ & $  112_{+ 5}^{ -7} $ & $ 0.27_{-0.01}^{+0.02} $ & $ 0.63_{+0.10}^{-0.17} $ & $ 0.610\pm 0.060$ \\[3pt]
 $ 0058\!+\!019 $ & J$010054.15\!+\!021136.52 $ & $ 0.6128 $ & $  29.5$ & $ -19.8$ & $ 11.4_{-0.2}^{+0.4} $ & $  151_{-20}^{+51} $ & $  125_{-18}^{+47} $ & $ 0.24_{+0.06}^{-0.04} $ & $   92_{+ 4}^{ -8} $ & $ 0.32_{-0.01}^{+0.03} $ & $ 0.74_{+0.12}^{-0.28} $ & $ 1.684\pm 0.004$ \\[3pt]
 $ 0058\!+\!019 $ & J$010054.15\!+\!021136.52 $ & $ 0.6800 $ & $  45.6$ & $ -21.2$ & $ 11.9_{-0.2}^{+0.2} $ & $  225_{-25}^{+42} $ & $  190_{-24}^{+40} $ & $ 0.24_{+0.04}^{-0.03} $ & $   69_{+ 4}^{ -5} $ & $ 0.66_{-0.03}^{+0.05} $ & $ 0.36_{+0.05}^{-0.08} $ & $<\! 0.003$ \\[3pt]
  SDSS            & J$010135.84\!-\!005009.08 $ & $ 0.2615 $ & $  50.9$ & $ -21.4$ & $ 12.2_{-0.2}^{+0.2} $ & $  242_{-28}^{+40} $ & $  223_{-28}^{+40} $ & $ 0.23_{+0.03}^{-0.03} $ & $   99_{+ 4}^{ -5} $ & $ 0.51_{-0.02}^{+0.03} $ & $ 0.44_{+0.06}^{-0.08} $ & $<\! 0.110$ \\[3pt]
  SDSS            & J$010156.32\!-\!084401.74 $ & $ 0.1588 $ & $  28.4$ & $ -19.2$ & $ 11.3_{-0.2}^{+0.6} $ & $  121_{-17}^{+64} $ & $  106_{-16}^{+63} $ & $ 0.27_{+0.10}^{-0.05} $ & $  146_{+ 6}^{-15} $ & $ 0.20_{-0.01}^{+0.02} $ & $ 1.38_{+0.25}^{-0.82} $ & $ 0.360\pm 0.030$ \\[3pt]
  SDSS            & J$010352.47\!+\!003739.79 $ & $ 0.3515 $ & $  48.3$ & $ -20.1$ & $ 11.7_{-0.2}^{+0.4} $ & $  178_{-24}^{+54} $ & $  157_{-23}^{+53} $ & $ 0.31_{+0.08}^{-0.05} $ & $  107_{+ 5}^{ -9} $ & $ 0.45_{-0.02}^{+0.04} $ & $ 0.68_{+0.12}^{-0.23} $ & $ 0.380\pm 0.030$ \\[3pt]
 $ 0102\!-\!190 $ & J$010516.82\!-\!184641.9  $ & $ 1.0250 $ & $  40.0$ & $ -22.3$ & $ 12.1_{-0.1}^{+0.1} $ & $  284_{-25}^{+31} $ & $  230_{-22}^{+27} $ & $ 0.17_{+0.02}^{-0.02} $ & $   36_{+ 3}^{ -3} $ & $ 1.12_{-0.08}^{+0.11} $ & $ 0.16_{+0.02}^{-0.02} $ & $ 0.670\pm 0.050$ \\[3pt]
 $ 0109\!+\!200 $ & J$011210.18\!+\!202021.79 $ & $ 0.5340 $ & $  44.7$ & $ -20.4$ & $ 11.6_{-0.2}^{+0.4} $ & $  173_{-23}^{+53} $ & $  147_{-21}^{+50} $ & $ 0.30_{+0.08}^{-0.05} $ & $   92_{+ 4}^{ -8} $ & $ 0.49_{-0.02}^{+0.05} $ & $ 0.63_{+0.11}^{-0.22} $ & $ 2.260\pm 0.050$ \\[-5pt]
\enddata
\tablenotetext{a}{Uncertainties are based upon uncertainties in the virial masses (column 6).  For some quantities a larger (smaller) virial mass results in smaller (larger) value such that the uncertainties anti-correlate.}
\tablenotetext{b}{Because the slope of $R_{\rm c}$ changes sign as function of virial mass, where the slope is positive the uncertainties correlate and where the slope is negative they anti-correlate (see Figure~\ref{fig:rcool}).
In the narrow virial mass ranges where the slope of $R_{\rm c}$ changes sign, it is possible that both the upward and downward uncertainties
in virial mass can result in an upward (or downward) uncertainty in $R_{\rm c}$.}
\tablenotetext{ }{Table~\ref{tab:obsprops} is published in its entirety
  in the electronic edition of ApJ. A portion is shown here for
  guidance regarding its form and content.}
\end{deluxetable*}

\subsection{Determining Galaxy Virial Masses}
\label{sec:ham}

Here, we elaborate on the method employed to determine the galaxy
virial masses that were originally studied in \citet{cwc-massesI}.
For each galaxy in the sample, the virial mass (dark + baryonic
matter), $M_{\rm\,h}$, was obtained by halo abundance matching.  The
virial mass is the total mass enclosed within the virial radius.  The
virial radius is defined as the radius enclosing an average density
$\Delta _c(z) \rho_c$, where $\Delta_c(z)$ \citep[see Eq.~A15
  of][]{eke96} is a cosmology and redshift dependent multiplier under
the assumption of virialization of a collapsed spherical top-hat
perturbation, and $\rho_c$ is the critical density.

Halo abundance matching assigns galaxies to dark matter halos in a
simulation based on number density with no free parameters. The method
has been thoroughly explored and applied to various astronomical
problems \citep{kravtsov04, tasitsiomi04, vale04, conroy06, conroy09,
  guo10, behroozi10, firmani10, trujillo-gomez11, rodriguez-puebla12,
  behroozi13, moster13, reddick13}.  In practice, the technique has
been extremely successful in reproducing many galaxy statistics, such
as the two-point correlation function as a function of redshift
\citep{conroy06}, luminosity \citep{trujillo-gomez11}, stellar mass
\citep{reddick13}, and color \citep[][accounting for halo formation
  times]{hearin13a}, as well as the luminosity-velocity relation,
baryonic Tully-Fisher relation, and galaxy velocity function
\citep{trujillo-gomez11}.  Halo abundance matching also yields galaxy
stellar-to-halo mass relations that agree with direct estimates from
lensing and satellite kinematics within the uncertainties of the
observations \citep[see][]{dutton10}

In essence, halo abundance matching links a given property (i.e.,
stellar mass, luminosity, etc.) of galaxies to a given halo property
(circular velocity, virial mass, etc.) in a monotonic fashion.  
For this work, the dark matter halo catalogs are taken from the
Bolshoi $N$-body cosmological simulation \citep{klypin11}.  

For the halo property, we adopt the maximum circular velocity,
\begin{equation}
V_{\rm c}^{\rm max} = \left[ \frac{GM_{\rm\, h}(\!<\!r)}{r} \right] ^{1/2} \,
\Bigg|_{\rm max} \, ,
\end{equation}
which properly accounts for the depth of the galactic potential and is
unambiguously defined for both central halos and sub-halos
\citep[halos within the virial radius of larger
  halos,][]{trujillo-gomez11}.  At a given redshift, the halo catalog
comprises individual halos for which both $V_c^{\rm max}$ and
$M_{\rm\, h}$ are tabulated.

For the galaxy property, we adopt the $r$-band luminosity, $M_r$.  For
the number density of galaxies with a given $M_r$, we adopt the
COMBO-17 $r$-band luminosity function (LF) of \citet{wolf03}, which
covers the redshift of our galaxy sample in a band that successfully
reproduces the clustering of galaxies at both low and high redshifts
\citep{trujillo-gomez11, gerke12}.

For the galaxy sample, we solve for the $V_c^{\rm max}$ for a galaxy
with $M_r$ such that the fractional area under the observed galaxy LF
corresponds to an equal fractional area under the curve of the
distribution of maximum circular velocities of halos,
\begin{equation}
\frac{1}{N_{M_r}(z\,)} \int _{-\infty}^{M_r} 
\!\!\!\!\!\!\! n(M_{r},z\,) \, dM_r = 
\frac{1}{N_{V_c^{\rm max}}(z\,)} \int _{V_c^{\rm max}}^{\infty} 
\!\!\!\!\!\!\!\! n(V_c^{\rm max}\!,z\,) \, dV_c^{\rm max}   \, ,
\label{eq:ham}
\end{equation} 
where the denominators are the total number density in the respective
distributions.  The LF is preserved by construction.  The only
assumption in the method is that there is only one galaxy inhabiting
each dark matter halo.

The redshift of a given galaxy determines the redshift of both the
Bolshoi halo catalog and the LF for which Eq.~\ref{eq:ham} was
applied.  \citet{wolf03} published the $r$-band LFs for five
redshifts, $z=0.3$, 0.5, 0.7, 0.9, and 1.1.  We abundance match a
given galaxy $M_r$ to $V_c^{\rm max}$ in a $\Delta z = 0.2$ redshift
bin bracketing the galaxy redshift, where the bin centers correspond
to the five COMBO-17 redshifts. For $z<0.2$, we opted to not use the
``local'' $r$-band LFs from SDSS \citep{blanton01} or 2dFGRS
\citep{madgwick02M} due to inconsistencies with the COMBO-17 LF, which
may be due to different sensitivities of the surveys at the bright end
\citep[see][]{wolf03}.  To maintain self-consistency, we adopt the
COMBO-17 LF in the bin $0.2 < z < 0.4$ under the assumption that the
LF does not evolve significantly below $z=0.3$.

There is intrinsic scatter in $V_c^{\rm max}$ for a given $M_{\rm\,
h}$ due to variation in formation times of halos of the same mass
\citep[see][]{trujillo-gomez11}.  Once the halo abundance matching is
solved (a $V_c^{\rm max}$ for each dark matter halo in the Bolshoi
catalog at the appropriate redshift is assigned to an $M_r$ for a
galaxy in the sample), we account for the scatter in $M_{\rm\, h}$
with $V_c^{\rm max}$ by computing the average $M_{\rm\, h}$ of
all the halos that fall in a fixed luminosity bin, $\Delta M_{\rm r}$,
centered on the measured value for that galaxy.  We adopted $\Delta
M_{\rm r} = 0.1$ (for details see Appendix~\ref{app:ham}).  Since halo
abundance matching is a statistical method, each derived $M_{\rm\, h}$
should be interpreted as the average mass of a halo which hosts a
galaxy of a given $M_r$.  

\begin{figure*}[thb]
\epsscale{1.15}
\plotone{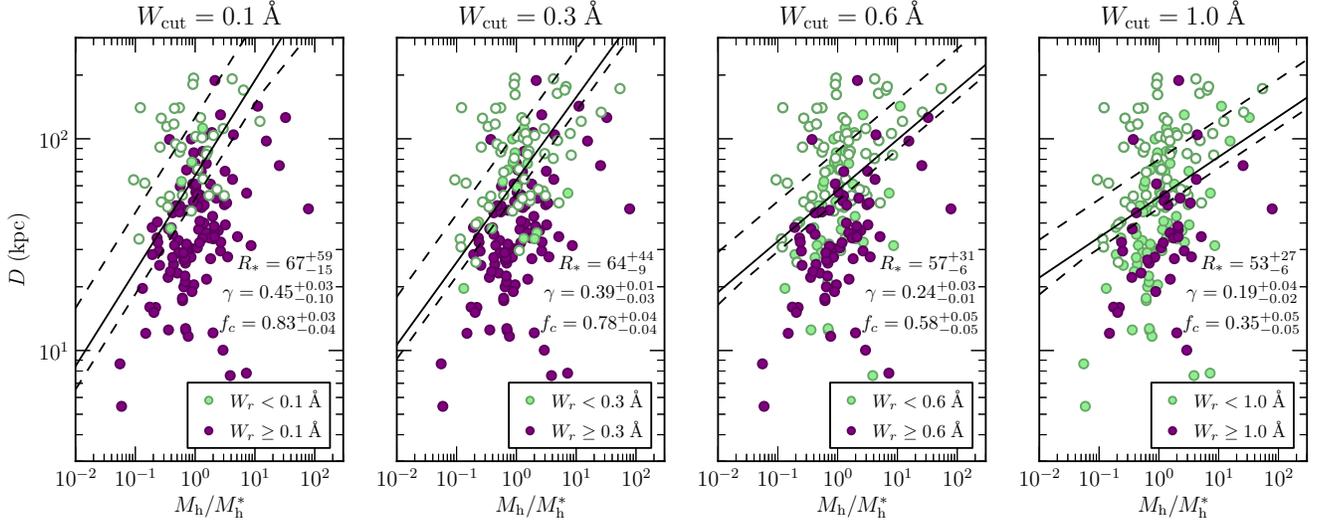}
\caption{The virial mass dependence of the {\MgII} CGM ``absorption
  radius'', $R(M_{\rm\, h})$, for the four $W_r(2796)$ absorption
  thresholds, $W_{\rm cut}=0.1$, 0.3, 0.6, and 1.0~{\AA}. The virial
  mass scale is normalized to $\log M_{\rm\, h}^{\ast} = \log M_{\rm\,
    h}/M_{\odot} = 12$.  Purple points are systems for which
  $W_r(2796) \geq W_{\rm cut}$ and green points are those for which
  $W_r(2796) < W_{\rm cut}$; an open point denotes that the
  measurement of $W_r(2796)$ is an upper limit.  The solid line is the
  maximum likelihood fit and the dashed curves provide the $1~\sigma$
  uncertainty envelope in the fit.  The absorption radius of an
  $M_{\rm\, h}^{\ast}$ galaxy decreases with increasing $W_{\rm cut}$
  (though the boundary remains equally ``fuzzy'') and the virial mass
  dependence, $\gamma$, decreases with increasing  $W_{\rm cut}$.}
\label{fig:DMhenv}
\end{figure*}

\begin{figure*}[htb]
\epsscale{1.15}
\plotone{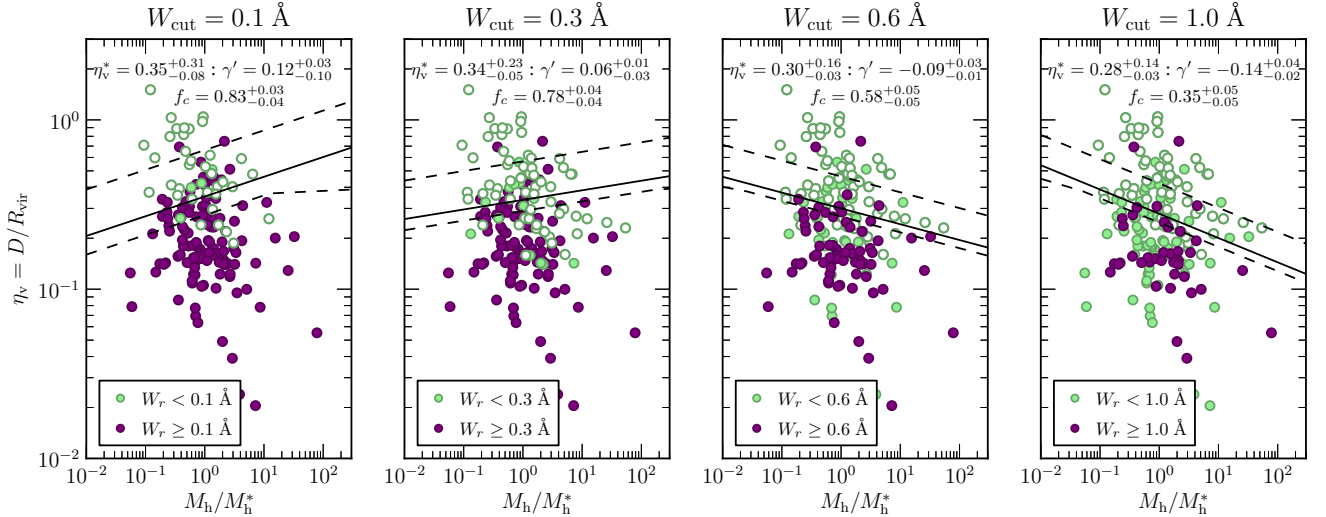}
\caption{The virial mass dependence of the mass-normalized {\MgII} CGM
  absorption envelope, $\eta_{\rm v}(M_{\rm\, h})$, given by
  Eq.~\ref{eq:etaMh}, for the four $W_r(2796)$ thresholds, $W_{\rm
    cut}=0.1$, 0.3, 0.6, and 1.0~{\AA}.  The virial mass scale is
  normalized to $\log M_{\rm\, h}^{\ast} = \log M_h/M_{\odot} = 12$.
  The data points and curves are as described for
  Figure~\ref{fig:DMhenv}. The mass-normalized absorption envelope of
  an $M_{\rm\, h}^{\ast}$ galaxy is $\eta^{\ast}_{\rm v} \simeq 0.3$ for all
  $W_{\rm cut}$. The virial mass dependence is weak, with some
  indication of reversing from slightly positive dependence to
  slightly negative dependence as $W_{\rm cut}$ is increased.}
\label{fig:DMRhenv}
\end{figure*}

In columns (6)--(8) of Table~\ref{tab:obsprops}, we present the galaxy
properties derived from halo abundance matching.  The resulting virial
masses have the range $10.7 \leq \log M_{\rm\,h}/M_{\odot} \leq 13.9$.
Including both systematics and scatter, the uncertainties are $\delta
\log M_{\rm\,h} \simeq 0.1$ at $\log M_{\rm\,h}/M_{\odot}=10$
increasing quasi-linearly to $\delta \log M_{\rm\,h} \simeq 0.35$ at
$\log M_{\rm\,h}/M_{\odot}=13$.  However, for each galaxy, we adopt
the $1~\sigma$ standard deviation in the scatter of the average
$M_{\rm\, h}$ in the luminosity bin as the uncertainty in $M_{\rm\,
  h}$.

We obtained the virial radius, $R_{\rm vir}$, for each galaxy using
the relation with $M_{\rm\, h}$ given by \citet{bryan98}.  The
resulting virial radii have the range $70 \leq R_{\rm vir} \leq 800$ proper
kpc.  The uncertainties in $R_{\rm vir}$ were obtained from the
uncertainties in the virial masses using standard error propagation.
The typical uncertainty is $\delta R_{\rm vir}/R_{\rm vir} \simeq
0.1$.

In Appendix~\ref{app:ham}, we quantify and discuss the systematic and
statistical uncertainties in $M_{\rm\, h}$ associated with our
methodology and quantify the effects of observational uncertainties.


\section{Results}
\label{sec:results}

In this section we report (1) the virial mass scaling of the {\MgII}
absorption radius, (2) the virial mass dependence of the
mass-normalized {\MgII} absorption radius, (3) the relationship
between $W_r(2796)$, virial mass, and impact parameter, (4) the
relationship between $W_r(2796)$, virial mass, and virial radius, (5)
the relationship between $W_r(2796)$, virial mass, and the theoretical
cooling radius, and (6) the covering fraction as a function of
$W_r(2796)$ threshold and fractional distance of the absorption with
respect to the theoretical cooling radius.

\subsection{Virial Mass Scaling of the ``Absorption Radius''}
\label{sec:halorad}

For {\MgII} absorption, many works have measured the luminosity
dependence of the ``absorption radius'' assuming the Holmberg scaling
$R(L) = R_{\ast} (L/L^{\ast})^{\beta}$, where $R_{\ast}$ is the
absorption radius of an $L^{\ast}$ galaxy and $\beta$ parameterizes
the luminosity scaling \citep[e.g.,][and references
  therein]{nikki-cat2}.  The ``absorption radius'' is interpreted as
an average physical extent out to which absorption is detected above a
given absorption threshold; it represents an idealistic projected
radius within which CGM gas is detected and outside of which CGM gas
is not detected.

In Paper~II \citep{nikki-cat2}, the two parameters $R_{\ast}$ and
$\beta$ were obtained for various absorption thresholds, $W_{\rm
cut}$, by maximizing the number of systems with $W_r(2796) \geq W_{\rm
cut}$ residing at $D \leq R(L)$ and maximizing the number of systems
with $W_r(2796)<W_{\rm cut}$ residing at $D > R(L)$.  The covering
fraction, $f_c$, of the absorption within $R(L)$ for each threshold is
also directly computed in their analysis, where the uncertainties are
determined using binomial statistics \citep[see][]{gehrels86}.


Following the methods applied by \citet{nikki-cat2}, we investigated
whether there is a virial mass dependence of the {\MgII} CGM absorption
radius, $R(M_{\rm\, h})$, for the four $W_r(2796)$ absorption
thresholds, $W_{\rm cut}=0.1$, 0.3, 0.6, and 1.0~{\AA}.  In place of
the galaxy luminosity relative to $L^{\ast}$, we define $M_{\rm\,
  h}^{\ast} = 10^{12}~$M$_{\odot}$ (the median mass of the sample) and
write
\begin{equation}
R(M_{\rm\, h}) = R_{\ast} \left( \frac{M_{\rm\, h}}{M_{\rm\,
    h}^{\ast}} \right) ^{\gamma} \, .
\label{eq:RMh}
\end{equation}

\begin{figure*}[thb]
\epsscale{1.0} 
\plotone{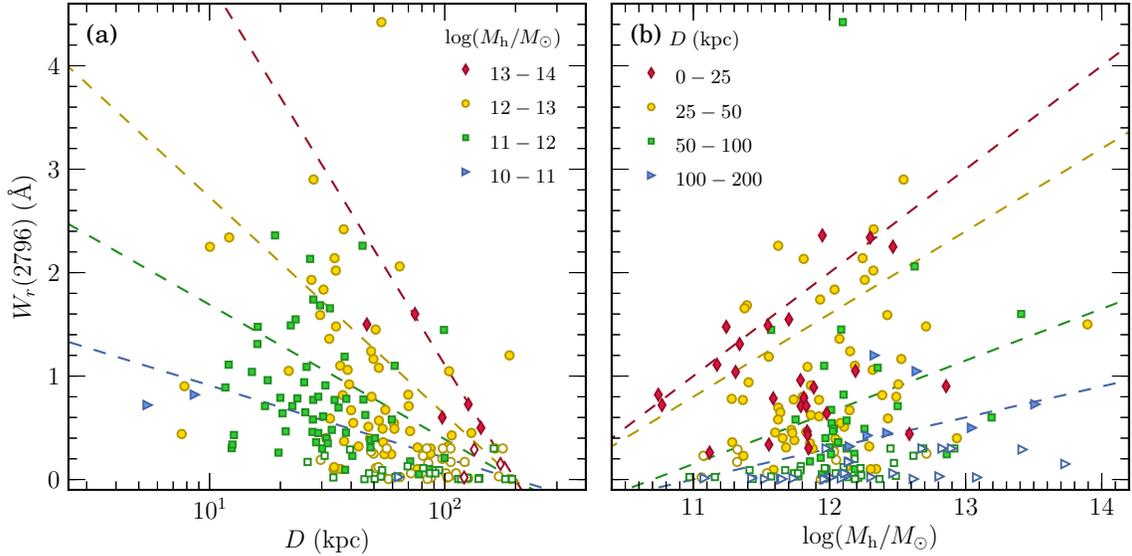}
\caption{(a) The $W_r(2796)$--$D$ plane with data points colored by
  virial mass range.  Higher mass halos, $\log M_{\rm\, h}/M_{\odot}
  \geq 12 $, are yellow and red, and lower mass halos, $\log M_{\rm\,
    h}/M_{\odot} < 12 $, are green and blue.  The data within each
  range of virial mass shows the anti-correlation between $W_r(2796)$
  and $D$, but each mass range has a different upper envelope, as
  represented by the dashed lines based upon the minimization fit.
  (b) The $W_r(2796)$--$M_{\rm\, h}$ plane with data points colored by
  impact parameter range.  Lower impact parameter data, $D < 50$ kpc,
  are yellow and red, and higher impact parameter data, $D > 50$ kpc,
  are green and blue.  The data within each range of impact parameter
  show a proportionality with virial mass in fixed impact parameter
  ranges, but with an increasing and steepening envelope as impact
  parameter becomes lower as shown by the dashed lines representing
  the minimization fit.}
\label{fig:EW2D}
\end{figure*}

In Figure~\ref{fig:DMhenv}, we plot $D$ versus $M_{\rm\, h}/M_{\rm\,
  h}^{\ast}$ for the sample. In each panel, purple points have
$W_r(2796) \geq W_{\rm cut}$ and green points have $W_r(2796) < W_{\rm
  cut}$.  Thus, purple points indicate absorption detections and green
points indicate absorption ``misses'' for the given $W_{\rm cut}$.
Open points indicate that the measurement of $W_r(2796)$ is an upper
limit to the detection sensitivity of the quasar spectrum. The solid
line is the maximum likelihood fit to Eq.~\ref{eq:RMh} and shows the
virial mass dependence of the {\MgII} CGM ``absorption radius'',
$R(M_{\rm\, h})$.  The dashed curves provide the $1~\sigma$ envelope
to the best fit parameters.  Note that there are fewer data points
included in the analysis for the $W_{\rm cut}=0.1$~{\AA} subsample.
This is because we exclude non-detections with $W_r(2796)$ limits
greater than $W_{\rm cut}$.  The fitting results suggest that the
absorption radius of an $M_{\rm\, h}^{\ast}$ galaxy, $R_{\ast}$,
decreases with increasing $W_{\rm cut}$, from $\simeq 70$~kpc for
$W_{\rm cut} = 0.1$~{\AA} to $\simeq 50$~kpc for $W_{\rm cut} =
1.0$~{\AA}.  However, the uncertainties in $R_{\ast}$ are larger,
which reflects the degree the absorption radius is actually a
``fuzzy'' boundary.  As the equivalent width threshold, $W_{\rm cut}$,
is raised, we find that the virial mass dependence systematically
decreases from $\gamma \simeq 0.45$ for $W_{\rm cut}=0.1$~{\AA} to
$\gamma \simeq 0.20$ for $W_{\rm cut}=1.0$~{\AA}.


The resulting $R_{\ast}$ and $\gamma$ values can be applied to quantify
the absorption radius relative to the virial radius. We will refer to
this quantity as the ``mass-normalized absorption envelope'', and
denote the quantity as $\eta_{\rm v} (M_{\rm\, h})$.  Defining
$\eta_{\rm v} = D/R_{\rm vir}$ and $\eta^{\ast}_{\rm v} = R_{\ast}/R^{\,
  \ast}_{\rm vir}$, where $R^{\, \ast}_{\rm vir}$ is the virial radius
for an $M_{\rm\, h}^{\ast} =10^{12}~$M$_{\odot}$ halo taken at the
median redshift of the sample, and invoking $R_{\rm vir} \propto
M_{\rm\, h}^{1/3}$ \citep{bryan98}, we obtain the relation
\begin{equation}
\eta_{\rm v} (M_{\rm\, h}) = \eta^{\ast}_{\rm v} 
\left( \frac{M_{\rm\, h}}{M_{\rm\, h}^{\ast}} \right) ^{\gamma'} \, ,
\label{eq:etaMh}
\end{equation}
where $\gamma' = \gamma - 1/3$.  

In Figure~\ref{fig:DMRhenv}, we plot $\eta_{\rm v} (M_{\rm\, h})$
versus $M_{\rm\, h}/M_{\rm\, h}^{\ast}$ for the $W_r(2796)$ absorption
thresholds, $W_{\rm cut}=0.1$, 0.3, 0.6, and 1.0~{\AA}.  The mean
mass-normalized absorption envelope for $M_{\rm\, h}^{\ast}$ galaxies
is $\eta^{\ast}_{\rm v} \simeq 0.3$ and is remarkably consistent
within uncertainties as being independent of the absorption threshold.
However, the mean covering fraction decreases by a factor of two as
$W_{\rm cut}$ is increased from 0.1 to 1.0~{\AA}.  The virial mass
dependence is quite weak, ranging from $\gamma' \simeq +0.1$ to
$\simeq -0.14$ as $W_{\rm cut}$ is increased.  Overall, by scaling the
absorption radius parameters, we find that the parameters describing
the mass-weighted absorption envelope, $\eta^{\ast}_{\rm v}$ and
$\gamma'$, indicate a very weak dependence on virial mass and that
this holds for all absorption thresholds.

\subsection{Absorption Strength, Virial Mass, and Impact Parameter}
\label{sec:W-M-D}


In view of the fitted relation $\log W_r(2796)\propto -2 \log
(D/R_{\rm vir})$ obtained by \citet{cwc-massesI}, and given that $\log
R_{\rm vir} \propto (1/3) \log M_{\rm\,h}$, one could infer that $\log
W_r(2796) \propto -2 \log D +  (2/3) \log M_{\rm\,h}$. That is, $\log
W_r(2796) \propto -2 \log D$ for a narrow range of $M_{\rm\,h}$ and
$\log W_r(2796) \propto (2/3) \log M_{\rm\,h}$ for a narrow range of
$D$.  This behavior is consistent with the virial mass segregation on
the $W_r(2796)$--$D$ plane presented by \citet{cwc-massesI}, in which
stronger absorption is preferentially associated with higher mass
halos and found at larger impact parameter.

To further investigate the relationships between $W_r(2796)$, virial
mass, and impact parameter, we explored the $W_r(2796)$--$D$ plane for
differential virial mass behavior, and the $W_r(2796)$--$M_{\rm\, h}$
plane for differential impact parameter behavior.  In
Figure~\ref{fig:EW2D}a, we present the $W_r(2796)$--$D$ plane in which
we colored the data points according virial mass using the mass
decades $\log M_{\rm\, h}/M_{\odot}=10$--11, 11--12, 12--13, and
13--14.  Consistent with many works \citep[see][and references
  therein]{nikki-cat1}, $W_r(2796)$ tends to decrease with increasing
impact parameter.  There is a clear visual trend for higher mass halos
to host larger $W_r(2796)$.  This is especially apparent in that the
``upper envelope'' of $W_r(2796)$ is dominated by the higher mass
galaxies, i.e., $\log M_{\rm\, h}/M_{\odot} >12$.  Furthermore, it
appears that the slope of each upper envelope increases with
increasing virial mass.

To quantify this differential virial mass behavior in the upper
absorption envelope, we used a maximum likelihood
approach to solving the relation $W_r(2796) = \alpha_1 \log D +
\alpha_2$, for each of the four virial mass ranges presented in
Figure~\ref{fig:EW2D}a.  We minimized the function
\begin{equation}
{\cal L}(\alpha_1,\alpha_2) = 
{\rm min} \left| \, N_{\rm abv}/N_{\rm tot} - {\rm erfc}(1) \, \right| \, ,
\label{eq:maxL}
\end{equation}
where $N_{\rm abv}/N_{\rm tot}$ is the ratio of systems above the
envelope to the total number of systems in the mass range.  The
complimentary error function is employed to account for the scatter in
the data and the different number of data points in each virial mass
range.  We allow 15.7\% ($1~\sigma$) of the data points in each mass
range to reside above the envelope when ${\cal L}(\alpha_1,\alpha_2)$
is a minimum. Thus, the resulting envelope encloses 84.3\% of the
data.  The envelopes have all been normalized to $W_r(2796)=0$~{\AA}
at $D=200$ kpc.

The resulting fitted parameters are listed in Table~\ref{tab:WDenv},
as are the values of the likelihood function.  The envelopes for each
of the respective virial mass ranges are plotted on
Figure~\ref{fig:EW2D}a.  The exercise quantifies the degree to which
the upper absorption envelope on the $W_r(2796)$--$D$ plane is virial
mass dependent.  At a given impact parameter, larger $W_r(2796)$ tends
to arise in higher mass halos.

\begin{deluxetable}{cccc}
\tablecolumns{4}
\tablewidth{0pt}
\setlength{\tabcolsep}{0.06in}
\tablecaption{Envelope $W_r(2796) = \alpha_1 \log D + \alpha_2$ \label{tab:WDenv}}
\tablehead{
  \colhead{(1)}             &
  \colhead{(2)}             &
  \colhead{(3)}             &
  \colhead{(4)}             \\
  \colhead{$\log M_{\rm\, h}/M_{\odot}$}       &
  \colhead{$\alpha_1$}      &
  \colhead{$\alpha_2$}      &
  \colhead{${\cal L}(\alpha_1,\alpha_2)$}     }
\startdata
(10--11] & $-0.7^{+0}_{-0.2}$       & $1.6^{+0.2}_{-0.5}$ & $1.57 \times 10^{-1}$ \\[3pt]
(11--12] & $-1.3^{+0.1}_{-0.2}$ & $3.0^{+0.3}_{-0.6}$ & $2.54 \times 10^{-3}$ \\[3pt]
(12--13] & $-2.1^{+0.1}_{-0.3}$ & $4.8^{+0.1}_{-0.2}$ & $3.62 \times 10^{-3}$ \\[3pt]
(13--14) & $-3.7^{+0.2}_{-0.1}$ & $8.5^{+0.2}_{-0.4}$ & $3.23 \times 10^{-2}$ \\[-5pt]
\enddata
\end{deluxetable}

\begin{deluxetable}{cccc}
\tablecolumns{4}
\tablewidth{0pt}
\setlength{\tabcolsep}{0.06in}
\tablecaption{Envelope $W_r(2796) = \alpha_1 \log (M_{\rm\, h}/M_{\odot}) + \alpha_2$ \label{tab:WMenv}}
\tablehead{
  \colhead{(1)}             &
  \colhead{(2)}             &
  \colhead{(3)}             &
  \colhead{(4)}             \\
  \colhead{$D$, kpc}       &
  \colhead{$\alpha_1$}      &
  \colhead{$\alpha_2$}      &
  \colhead{${\cal L}(\alpha_1,\alpha_2)$}     }
\startdata
(0--25]   & $1.0^{+0.1}_{-0.1}$ & $-10.0^{+1.5}_{-1.9}$ & $3.45 \times 10^{-3}$ \\[3pt]
(25--50]  & $0.8^{+0.1}_{-0.1}$ & $-8.0^{+2.2}_{-2.1}$  & $4.47 \times 10^{-3}$ \\[3pt]
(50--100] & $0.5^{+0.2}_{-0.1}$ & $-5.3^{+1.1}_{-2.1}$  & $6.80 \times 10^{-2}$ \\[3pt]
(100-200) & $0.3^{+0.2}_{-0.1}$ & $-3.3^{+1.1}_{-2.2}$  & $1.04 \times 10^{-3}$ \\[-5pt]
\enddata
\end{deluxetable}

In Figure~\ref{fig:EW2D}b, we show the $W_r(2796)$--$M_{\rm\, h}$
plane where we have color coded the data points binned by impact
parameter using the bins $0 < D \leq 25$ kpc, $25 < D \leq 50$ kpc,
$50 < D \leq 100$ kpc, and $100 < D \leq 200$ kpc.  Again, for a fixed
impact parameter range, there is a clear general trend for increasing
$W_r(2796)$ with increasing virial mass.  Though a range of
$W_r(2796)$ are present at a given virial mass in each impact
parameter range, the upper envelope of the absorption for an impact
parameter range clearly increases with virial mass.  Parameterizing the
envelope as $W_r(2796) = \alpha_1 \log M_{\rm\, h}/M_{\odot} +
\alpha_2$ for each impact parameter range, we applied
Eq.~\ref{eq:maxL} to quantify this behavior.

The resulting fitted parameters and values of the likelihood function
are listed in Table~\ref{tab:WMenv}.  The results indicate that, in a
finite range of impact parameter, stronger $W_r(2796)$ absorption is
preferentially found around higher mass halos and that this trend is
more pronounced for smaller impact parameters (i.e., $D<50$~kpc).
Though the data exhibit substantial scatter, the trends highlighted in
Figures~\ref{fig:EW2D}a and \ref{fig:EW2D}b are consistent with the
notion of a correlation between virial mass and {\MgII} equivalent width
at a fixed impact parameter.  In order to further investigate the
relationships between $W_r(2796)$, virial mass, and impact parameter, we
examined the behavior of the means in these quantities.

\begin{figure}[thb]
\epsscale{1.15} 
\plotone{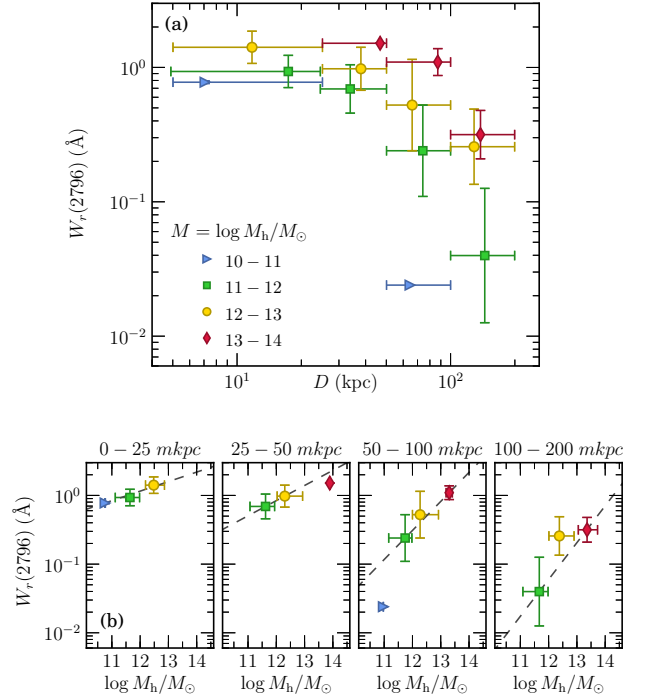}
\caption{(a) The $W_r(2796)$--$D$ plane illustrating the mean
  $W_r(2796)$ in a given impact parameter and virial mass range.  The
  impact parameter bins are $D=0$--25, 25-50, 50-100, and 100-200 kpc.
  The data points are colored by virial mass bin, with $\log M_{\rm\,
    h}/M_{\odot}=10$--11 (blue), 11--12 (green), 12--13 (yellow), and
  13--14 (red).  The data points are plotted at the mean $D$ for the
  galaxies in each virial mass range.  The horizontal error bars give
  the width of the impact parameter bin and the vertical error bars
  give the $1~\sigma$ variance in the mean {\MgII}~$\lambda 2796$
  equivalent width.  (b) For each impact parameter bin, the mean
  $W_r(2796)$ is plotted as a function of virial mass, $\log M_{\rm\,
    h}/M_{\odot}$. The horizontal error bars provide the actual virial
  mass range within the mass bins.  We find that, in each impact
  parameter bin, the mean $W_r(2796)$ increases as virial mass
  increases.  Note that not all mass ranges are represented in all
  impact parameter bins.  The dashed lines are the maximum likelihood
  fits presented in Table~\ref{tab:fits}.}
\label{fig:EWvsD}
\end{figure}

In Figure~\ref{fig:EWvsD}a, we present the $W_r(2796)$--$D$ plane in
which we plot the mean $W_r(2796)$ in a fixed impact parameter range
for the virial mass decades $\log M_{\rm\, h}/M_{\odot}=10$--11, 11--12,
12--13, and 13--14.  The impact parameter ranges are $0 < D \leq 25$
kpc, $25 < D \leq 50$ kpc, $50 < D \leq 100$ kpc, and $100 < D \leq
200$ kpc.  Vertical bars provide the $1~\sigma$ variances in the mean
$W_r(2796)$ and horizontal bars indicate the width of the
impact parameter bin.  Note that in the virial mass decade $\log
M_{\rm\, h}/M_{\odot}=10$--11 (blue points), there are only three
galaxies in the sample.

The general trend seen in Figure~\ref{fig:EWvsD}a is that, in each
impact parameter bin, the mean $W_r(2796)$ increases with virial mass.
We further illustrate the trend in the four panels of
Figure~\ref{fig:EWvsD}b, where we plot the mean $W_r(2796)$ as a
function of virial mass in fixed impact parameter bins.  The points are
plotted at the mean virial mass and mean $W_r(2796)$.  The data clearly
show a trend of increasing mean $W_r(2796)$ as a function of $M_{\rm\,
  h}$ in each impact parameter bin.

Since the appearance of binned data can be sensitive to the choice of
binning, and since we clearly do not have equal numbers of galaxies in
each virial mass decade, we divided the sample into virial mass tertiles
and virial mass quartiles.  We obtain the same qualitative results
presented in Figure~\ref{fig:EWvsD}.  

To determine whether the correlations between $W_r(2796)$ and
$M_{\rm\, h}$ in each impact parameter range are statistically
significant, we performed a non-parametric rank correlation test on
the unbinned data represented in each panel of
Figure~\ref{fig:EWvsD}b.  Since a substantial number of our
$W_r(2796)$ values are upper limits, we employed the Brown, Hollander,
\& Korwar {\sc bhk}-$\tau$ test \citep{bhk74}, which allows for upper
limits in either the dependent or the independent variable \citep[also
  see][]{feigelson85, isobe86, wang00}.  The tests do not
significantly rule out the null hypothesis of no correlation between
$W_r(2796)$ and $M_{\rm\, h}$ to better than $3~\sigma$.  However,
they suggest a strong trend to better than $2.5~\sigma$ in each impact
parameter range.

In view of the prediction that $\log W_r(2796)$ is proportional to $
(2/3) \log M_{\rm\, h} -2 \log D $, we might expect the data presented
in Figure~\ref{fig:EWvsD}b would obey $\log W_r(2796) = (2/3)\log
M_{\rm\, h} + C$ in finite impact parameter bins.  Assuming the
relation $\log W_r(2796) = \alpha_1 \log M_{\rm\, h}/M_{\odot} +
\alpha_2$, we applied a maximum likelihood linear fit to the unbinned
data presented in each panel of Figure~\ref{fig:EWvsD}b to obtain an
estimate of the slope between $W_r(2796)$ and $M_{\rm\, h}$ in fixed
impact parameter bins.  Accounting for upper limits for some of our
$W_r(2796)$ measurements, we employed the Expectation-Maximization
algorithm {\sc emalgo} \citep{wolynetz79}.

\begin{deluxetable}{cccrrrr}
\tablecolumns{7}
\tablewidth{0pt}
\setlength{\tabcolsep}{0.06in}
\tablecaption{Fit to $\log W_r(2796) = \alpha_1 \log (M_{\rm\, h}/M_{\odot}) + \alpha_2$ \label{tab:fits}}
\tablehead{
  \colhead{(1)}             &
  \colhead{(2)}             &
  \colhead{(3)}             &
  \colhead{(4)}             &
  \colhead{(5)}             &
  \colhead{(6)}             &
  \colhead{(7)}             \\
  \colhead{$D$ range}       &
  \colhead{$\alpha_1$}      &
  \colhead{$\alpha_2$}      &
  \colhead{$N_{10,11}$}  &
  \colhead{$N_{11,12}$}  &
  \colhead{$N_{12,13}$}  &
  \colhead{$N_{13,14}$}  \\
  \colhead{[kpc]}           &
  \colhead{}                &
  \colhead{}                &
  \colhead{}     &
  \colhead{}     &
  \colhead{}     &
  \colhead{}     }
\startdata
0--25    & $0.14\pm0.06$ & $-1.6\pm0.7$ &        2 & 19 &  5 & $\cdots$ \\
25--50   & $0.22\pm0.04$ & $-2.7\pm0.5$ & $\cdots$ & 37 & 30 &        1 \\
50--100  & $0.42\pm0.35$ & $-5.6\pm4.3$ &        1 & 19 & 33 &        2 \\
100--200 & $0.55\pm0.25$ & $-7.7\pm3.2$ & $\cdots$ &  9 & 19 &        5 \\[-8pt]
\enddata
\end{deluxetable}

The resulting fitted parameters are presented in Table~\ref{tab:fits}.
Columns (4)--(7) provide the number of galaxy-absorber pairs in each
mass decade.  The fits are overplotted on the binned data presented in
Figure~\ref{fig:EWvsD}b.  For all impact parameter bins, we find that
the slope, $\alpha_1$, is always less than the predicted $2/3$.
However, the slopes for the $D>50$ kpc bins are consistent with $2/3$
within uncertainties.  The best fit slope increases as impact
parameter is increased, though the large uncertainties for $D>50$ kpc
reflect the increased scatter in $W_r(2796)$ due to the decreasing
covering fraction as impact parameter increases \citep{nikki-cat2}.
Within uncertainties, the zero-point of the fit, $\alpha_2$, decreases
with increasing impact parameters consistent with the $D^{-2}$
scaling.


\subsection{Absorption Strength, Virial Mass, and Virial Radius}
\label{sec:W-M-DoR}

In view of the results of \citet{cwc-massesI} in which the
{\MgII}~$\lambda 2796$ equivalent width is tightly anti-correlated with
$\eta_{\rm v} = D/R_{\rm vir}$, the projected impact parameter in
units of the virial radius, we further investigate the relationship
between $W_r(2796)$, virial mass, and $\eta_{\rm v}$.

\begin{figure}[thb]
\epsscale{1.15} 
\plotone{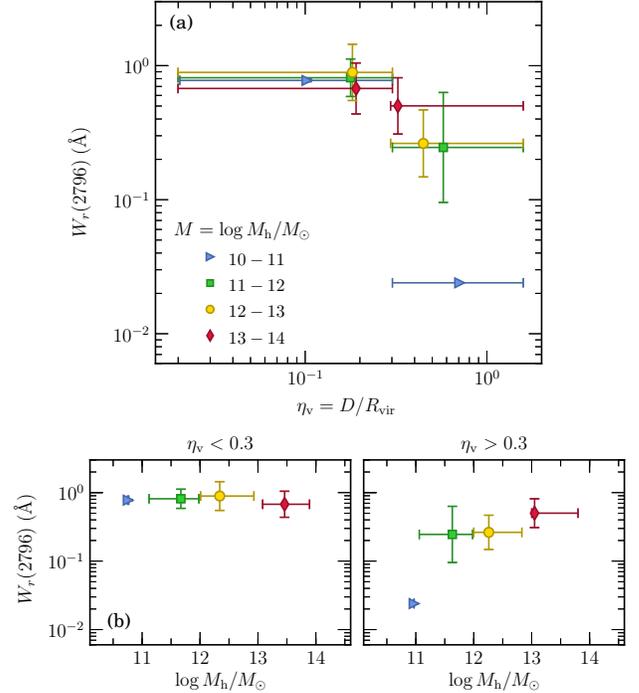}
\caption{(a) The $W_r(2796)$--$\eta_{\rm v}$ plane illustrating the
  mean $W_r(2796)$ for $\eta_{\rm v} \leq 0.3$ and $\eta_{\rm v} >
  0.3$.  The data points and ``error bars'' are the same as described
  for Figure~\ref{fig:EWvsD}a.  (b) For each finite $\eta_{\rm v}$
  range, the mean $W_r(2796)$ is plotted as a function of virial mass,
  $\log M_{\rm\, h}/M_{\odot}$.  The data points and ``error bars''
  are the same as described for Figure~\ref{fig:EWvsD}b.  We find that
  for $\eta_{\rm v} \leq 0.3$, where the majority of {\MgII}
  absorbing gas resides, the mean $W_r(2796)$ is independent of virial
  mass.}
\label{fig:EWvsDR}
\end{figure}

Of interest is that on the $W_r(2796)$--$D$ plane, there is a
virial mass segregation \citep[at the $4~\sigma$ significance
  level;][]{cwc-massesI}, in which higher virial mass galaxies are
seen to have larger $W_r(2796)$ and $D$ than galaxies with lower
virial masses.  This behavior induces a substantial scatter on the
$W_r(2796)$--$D$ plane, and shows that the scatter is systematic with
virial mass \citep[this systematic scatter is also seen at the
  $4~\sigma$ significance level with $B$- and
  $K$-luminosity,][]{nikki-cat2}.  However, when $D$ is normalized to
$R_{\rm vir}$, the virial mass segregation vanishes on the
$W_r(2796)$--$\eta_{\rm v}$ plane and the scatter is reduced to a very
high significance level relative to the scatter on the
$W_r(2796)$--$D$ plane.

Overall, this behavior might suggest that the role of virial mass is
manifest in the virial radius, such that $W_r(2796)$ should show
little to no trend with virial mass when examined as a function of
$\eta_{\rm v}$.  In Figure~\ref{fig:EWvsDR}a, we present the
$W_r(2796)$--$\eta_{\rm v}$ plane in which we plot the mean
$W_r(2796)$ for $\eta_{\rm v} \leq 0.3$ and $\eta_{\rm v} > 0.3$.  The
cut $\eta_{\rm v} = 0.3$ is motivated by the above result (see
Figure~\ref{fig:DMRhenv}) in which $\eta^{\ast}_{\rm v} \simeq 0.3$
and that virial mass scaling of $\eta_{\rm v} (M_{\rm\, h})$ is very
weak.  Data points are colored as in Figure~\ref{fig:EWvsD}.  The mean
$W_r(2796)$ is clearly independent of virial mass for $\eta_{\rm v}
\leq 0.3$.  The data do not present as clear a picture between mean
absorption strength and virial mass for $\eta_{\rm v} > 0.3$; however,
for $\log M_{\rm\, h}/M_{\odot} > 11$, the $W_r(2796)$ are consistent
with being independent of virial mass within the $1~\sigma$ variances
of their distributions (note that there is only a single data point
for $\log M_{\rm\, h}/M_{\odot} < 11$).

In Figure~\ref{fig:EWvsDR}b, we plot the mean $W_r(2796)$ directly
as a function of virial mass.  For both $\eta_{\rm v} \leq 0.3$ and
$\eta_{\rm v} > 0.3$, {\sc bhk}-$\tau$ non-parametric rank correlation
tests on the unbinned data represented in each panel of
Figure~\ref{fig:EWvsDR}b are consistent with the null-hypothesis of no
correlation between $W_r(2796)$ and $M_{\rm\, h}$.  This is a
remarkable behavior that strongly suggests a self-similarity between
the cool/warm CGM over a wide range of virial mass.  Within the inner
third of the virial radius, the strength of the absorption is invariant
with virial mass.  The degree of invariance we find outside the inner
third of the virial radius is also remarkable.  Whatever the physical
source governing the column density and kinematic distribution of
{\MgII} absorbing gas (chemical enrichment, stellar feedback, infall
accretion, cooling and heating, and/or destruction and creation
mechanisms), the net result is one in which a uniform behavior in the
average properties of the gas is constant as a function of $\eta_{\rm
  v}$ for all virial masses.

\subsection{Absorption Strength, Virial Mass, and Cooling Radius}
\label{sec:rcooldiscuss}

In the first modern models of galaxy formation in the dark matter
paradigm, theorists proposed that the cooling of gas in the halo is a
key mechanism governing galaxy mass \citep[cf.,][]{white78, white91}
and the extent and mass of cool/warm CGM gas \cite[cf.,][]{mo96}.  In
such models, it was stipulated that gas falling into dark matter halos
shock heats and sets up an initial hot phase ($T \geq 10^6$~K) at the
virial temperature with gas density decreasing with increasing radius.
The models were developed based on the notion of a theoretical cooling
radius, $R_{\rm c}$, inside of which gas cools, falls into the galaxy,
and feeds star formation, and outside of which the gas does not have
time to cool and remains in the hot phase.  Later works show the
cooling time scale in lower mass halos is shorter than the infall
dynamical and/or compression time scale, and cold-mode accretion feeds
the central galaxy \citep[e.g.,][]{birnboim03, keres05, dekel06,
  keres09, stewart11, vandevoort11}.

\begin{figure*}[bth]
\epsscale{0.95} 
\plotone{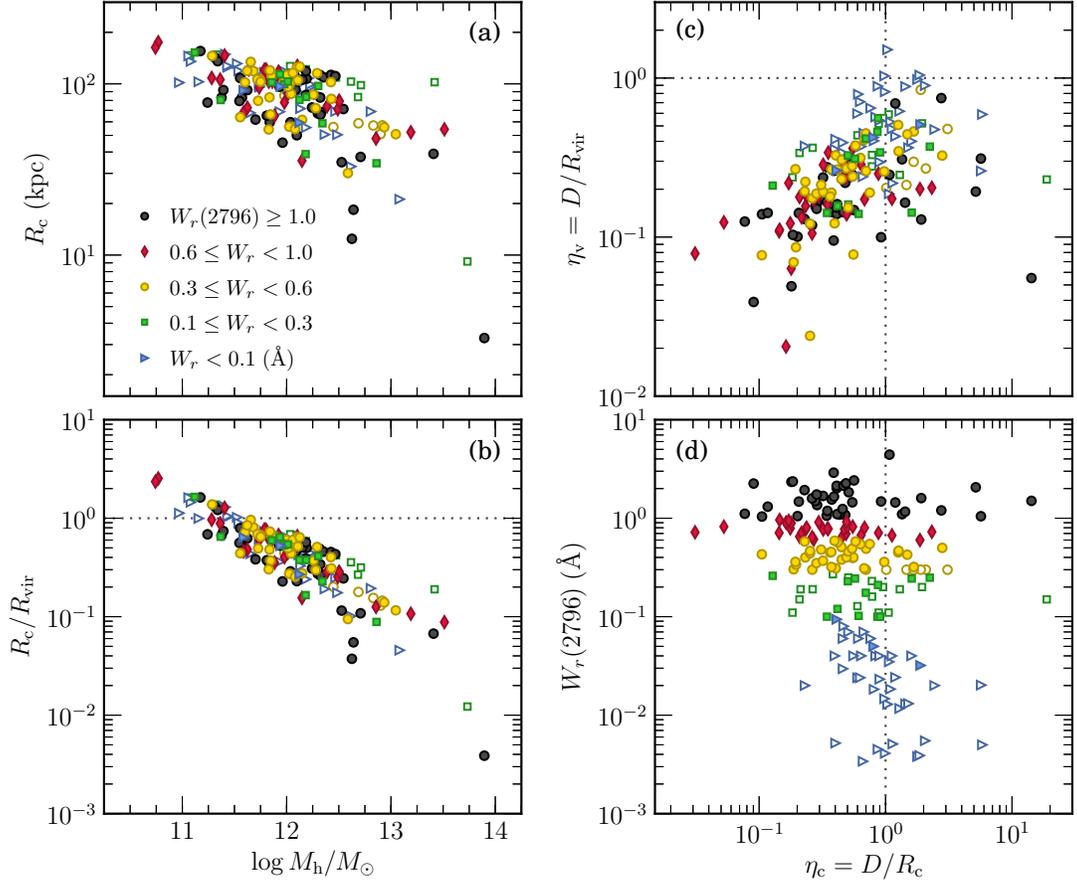}
\caption{(a) The theoretical cooling radius, $R_{\rm c}$ versus virial
  mass for the galaxies in the sample.  The data are colored in the
  bins $W_r(2796) \leq 0.1$~{\AA} (blue), $0.1 < W_r(2796) \leq
  0.3$~{\AA} (green), $0.3 < W_r(2796) \leq 0.6$~{\AA} (yellow), $0.6
  < W_r(2796) \leq 1.0$~{\AA} (red), and $W_r(2796) >0.1$~{\AA}
  (black).  Filled points are detections and open points are upper
  limits on $W_r(2796)$.  (b) The ratio $R_{\rm c}/R_{\rm vir}$ versus
  virial mass.  (c) The ratio $\eta_{\rm v}=D/R_{\rm vir}$ versus the
  ratio $\eta_{\rm c}=D/R_{\rm c}$. (d) $W_r(2796)$ versus $\eta_{\rm
    c}$.  All panels illustrate $Z_{\rm gas} = 0.1$.}
\label{fig:etacool}
\end{figure*}

In the model of \citet{mo96}, the cool/warm gas ($10^{4} \leq T \leq
10^{5}$~K) traced by {\HI} and {\MgII} absorption is predicted to have
a high covering fraction inside the cooling radius.  \citet{maller04}
expanded the \citet{mo96} model to account for a multi-phase gas
medium in which they allow for a hot gas core to persist at $r<R_{\rm
  c}$, while invoking thermal and dynamical instabilities to provide
for the fragmentation and condensation of some of the hot gas into
cool clouds.  This multi-phase model predicts a non-unity covering
fraction of cool/warm absorbing gas inside the cooling radius, which
is more in line with {\MgII} absorption observations
\citep{kacprzak08, chen10a, nikki-cat2}.  The model also predicts that
cool/warm gas which originated via condensation from the initial hot
halo gas will reside exclusively inside $R_{\rm c}$.


To investigate where the {\MgII} absorbing CGM resides in relation to
the theoretical cooling radius and to determine the covering fraction
both inside and outside the cooling radius, we estimated $R_{\rm c}$
using the model of \citet{maller04}.  Other analytical dark matter
halo models that predict {\MgII} absorption have been developed
\citep[e.g.,][]{tinker08, chelouche08, chelouche10}, but the model of
\citet{maller04} is best suited for our study because it is based upon
physical principles that provide a clear formalism for computing the
theoretical cooling radius as a function of virial mass.

Formally, the cooling radius is defined at the radial distance, $r$,
from the center of the halo at which the initial gas density,
$\rho_{\rm gas}(r)$, equals the characteristic density at which gas can
cool, $\rho_{\rm c}$, known as the ``cooling density''.  As such,
cooling of the gas can occur for $r\leq R_{\rm c}$ when $\rho_{\rm
  gas}(r) \geq \rho_{\rm c}$. The theoretical cooling radius is
defined when
\begin{equation}
\rho_{\rm gas}(R_{\rm c}) = \rho_{\rm c}    
\end{equation}
is satisfied.  Following \citet{maller04}, we applied their Eq.~9 for
$\rho_{\rm gas}(r)$ and Eq.~12 for $\rho_{\rm c}$ to obtain $R_{\rm
  c}$ for each galaxy in our sample.  The required input quantities
are the virial mass, virial radius, redshift, formation time of the dark
matter halo, and metallicity of the hot gas halo.  In
Appendix~\ref{app:rcool}, we describe our computation of $R_{\rm c}$
and provide a brief discussion of how the value of $R_{\rm c}$
responds to the input quantities (see Figure~\ref{fig:rcool}).  For
our work, the most uncertain quantities are the halo formation time
and the metallicity of the hot halo gas, $Z_{\rm gas}$.

For fixed redshift, the formation time, $\tau_{\rm f}$, is shorter for
higher mass halos. For fixed virial mass, $\tau_{\rm f}$ decreases with
increasing redshift.  Since the cooling density scales as $\rho_{\rm
  c} \propto \tau^{-1}_{\rm f}$, a shorter formation time yields a
smaller cooling radius.  We describe our estimation of the formation
time in Appendix~\ref{app:rcool} and present $\tau_{\rm f}$ as a
function of redshift and virial mass in Figure~\ref{fig:rcool}a.

\begin{figure*}[thb]
\epsscale{1.18} 
\plotone{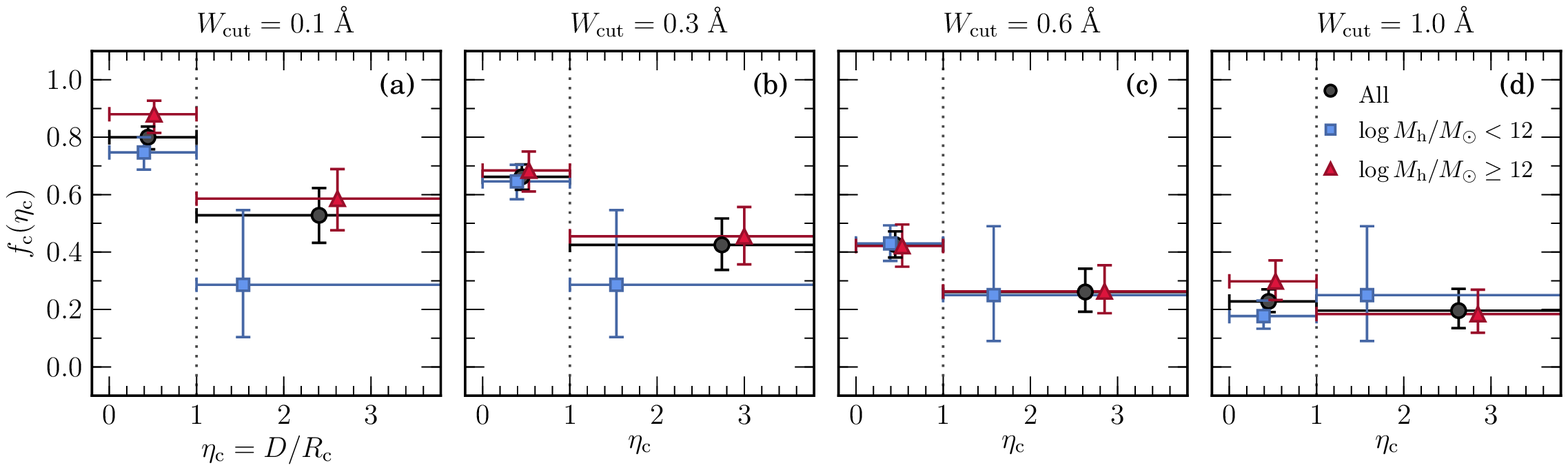}
\caption{The {\MgII} absorbing gas covering fraction,
  $f_c(\eta_{\rm c})$, as a function of fractional projected distance
  in units of the theoretical cooling radius for $Z_{\rm gas}=0.1$ and
  $W_r(2796)$ absorption threshold, $W_{\rm cut}$.  The data are
  binned by $\eta_{\rm c} \leq 1.0$ (inside the cooling radius) and
  $\eta_{\rm c} > 1.0$ (outside the cooling radius).  (a) $W_{\rm cut}
  = 0.1$~{\AA}. (b) $W_{\rm cut} = 0.3$~{\AA}. (c) $W_{\rm cut} =
  0.6$~{\AA}. (d) $W_{\rm cut} = 1.0$~{\AA}.  Red points are for
  galaxies with $ \log M_{\rm\, h}/M_{\odot} \geq 12$, blue points are
  for $ \log M_{\rm\, h}/M_{\odot} < 12$, where $ \log M_{\rm\,
    h}/M_{\odot} = 12$ is the median of the sample, and black points
  are for the full observed range of virial mass.  The non-negligible
  $f_c(\eta_{\rm c})$ outside the theoretical cooling radius implies a
  substantial population of cool/warm CGM clouds that are not formed
  via fragmentation and condensation out of the hot coronal gas
  component of the CGM.}
\label{fig:fDbar}
\end{figure*}

The cooling radius is inversely proportional to the cooling function,
$\Lambda (T,Z_{\rm gas})$.  A fixed volume of solar metallicity gas
can cool at a rate 3--10 times more rapidly than zero metallicity gas,
depending upon the temperature regime.  Since the cooling density
follows $\rho_{\rm c} \propto \Lambda ^{-1} (T,Z_{\rm gas})$, the
value of $R_{\rm c}$ can be as much as a factor of $\simeq 1.5$ larger
in a halo of the same mass and redshift but with $\simeq 1$ dex higher
metallicity (see Figure~\ref{fig:rcool}b).  However, this applies only
in the lower mass halos; at higher mass, and therefore higher initial
gas temperatures, the cooling rate is metallicity independent (where
bremsstrahlung cooling dominates).

It is important to keep in mind that, in the framework of the
\citet{maller04} model, the gas metallicity corresponds to the hot
phase of the halo, for which evidence is mounting that the mixing
between stellar feedback and accretion of the intergalactic medium
converges on a mean metallicity of $Z_{\rm gas} \sim 0.1$
\citep[cf.,][]{crain13}.  For our presentation of $R_{\rm c}$, we adopt
$Z_{\rm gas} \sim 0.1$, as motivate in Appendix~\ref{app:rcool}.  

The computed theoretical cooling radii for the galaxies in our sample
are listed in column (10) of Table~\ref{tab:obsprops} and plotted in
Figure~\ref{fig:etacool}a as a function of virial mass.  Points are
colored by {\MgII} $\lambda 2796 $ equivalent width bins.  Filled
points are detections and open points are upper limits.  In
Figure~\ref{fig:etacool}b, we plot the ratio $R_{\rm c}/R_{\rm vir}$
for the galaxies [also see column (12) of Table~\ref{tab:obsprops}].
Most of the galaxies in the sample have $0.1 \leq R_{\rm c}/R_{\rm
  vir} \leq 1.0$, in that the cooling radius lies inside the virial
radius.  For $\log M_{\rm\, h}/M_{\odot} < 11.3$, the cooling radius
resides outside the virial radius.  The scatter in these two diagrams
is due solely to the different galaxy redshifts at a given virial
mass.  Higher redshift galaxies have shorter formation times, and a
shorter formation time yields a larger cooling density, and therefore
a smaller cooling radius.

In Figure~\ref{fig:etacool}c, we plot the fractional projected
distance within the theoretical cooling radius at which absorption is
probed, $\eta_{\rm c}= D/R_{\rm c}$, versus the fractional projected
distance within the virial radius, $\eta_{\rm v} = D/R_{\rm vir}$, at
which absorption is probed.  These quantities are listed in columns
(9) and (11) of Table~\ref{tab:obsprops}. Whereas virtually all the
{\MgII} absorption is found within the virial radius (noting that our
sample probes $\eta_{\rm vir} \leq 1$ in all but three cases), we find
that {\MgII} absorption is detected well outside the theoretical
cooling radius.  In Figure~\ref{fig:etacool}d, we present $W_r(2796)$
versus $\eta_{\rm c}$.  Of interest is that the strongest absorbers
are detected over a wide range of $\eta_{\rm c}$, including well
outside the theoretical cooling radius.

In Figure~\ref{fig:fDbar} and Table~\ref{tab:fcov}, we present the
covering fraction, $f_c(\eta_{\rm c})$, for different $W_r(2796)$
absorption thresholds, $W_{\rm cut}$, for $\eta_{\rm c} \leq 1.0$
(inside the cooling radius) and $\eta_{\rm c} > 1.0$ (outside the
cooling radius) for the fiducial model with $Z_{\rm gas} = 0.1$.  For
this exercise, we examined the full range of virial masses (black
points), and the two subsamples defined by $ \log M_{\rm\,
  h}/M_{\odot} \geq 12$, and by $ \log M_{\rm\, h}/M_{\odot} < 12$,
where $ \log M_{\rm\, h}/M_{\odot} = 12$ is the median of the sample.
We also computed $f_c(\eta_{\rm c})$ for the ranges $0 \leq \eta_{\rm
  c} \leq 0.5$ and $0.5 < \eta_{\rm c} \leq 1$.

\begin{deluxetable}{ccccc}
\tablecolumns{5}
\tablewidth{0pt}
\setlength{\tabcolsep}{0.06in}
\tablecaption{Covering Fraction with Cooling Radius\tablenotemark{a}\label{tab:fcov}}
\tablehead{
  \colhead{(1)}                &
  \colhead{(2)}                &
  \colhead{(3)}                &
  \colhead{(4)}                &
  \colhead{(5)}                \\
  \colhead{$W_{\rm cut}$}      &
  \colhead{$\eta_{\rm c}$ Range}       &
  \colhead{$f_c(\eta_{\rm c})$}   &
  \colhead{$f_c(\eta_{\rm c})$}   &
  \colhead{$f_c(\eta_{\rm c})$}   \\
  \colhead{[{\AA}]}        &
  \colhead{$D/R_{\rm c}$}           &
  \colhead{(All)}    &
  \colhead{($M_{\rm\, h}< 10^{12}$)}    &
  \colhead{($M_{\rm\, h}\geq 10^{12}$)}    }
\startdata
 $0.1$ &   $\leq 0.5$ & $ 0.91_{- 0.04}^{+ 0.03}$  & $ 0.89_{- 0.06}^{+ 0.04}$  & $ 0.96_{- 0.09}^{+ 0.03}$  \\[3pt]
       &   0.5--1.0   & $ 0.59_{- 0.09}^{+ 0.08}$  & $ 0.32_{- 0.12}^{+ 0.14}$  & $ 0.80_{- 0.11}^{+ 0.08}$  \\[3pt]
       &   $\leq1.0$  & $ 0.80_{- 0.04}^{+ 0.04}$  & $ 0.74_{- 0.06}^{+ 0.05}$  & $ 0.88_{- 0.07}^{+ 0.05}$  \\[3pt]
       &   $>1.0$     & $ 0.53_{- 0.10}^{+ 0.10}$  & $ 0.29_{- 0.18}^{+ 0.26}$  & $ 0.59_{- 0.11}^{+ 0.10}$  \\[8pt]
 $0.3$ &   $\leq 0.5$ & $ 0.83_{- 0.05}^{+ 0.04}$  & $ 0.83_{- 0.06}^{+ 0.05}$  & $ 0.81_{- 0.11}^{+ 0.08}$  \\[3pt]
       &   0.5--1.0   & $ 0.38_{- 0.07}^{+ 0.08}$  & $ 0.10_{- 0.06}^{+ 0.12}$  & $ 0.57_{- 0.11}^{+ 0.10}$  \\[3pt]
       &   $\leq1.0$  & $ 0.66_{- 0.05}^{+ 0.04}$  & $ 0.65_{- 0.06}^{+ 0.06}$  & $ 0.69_{- 0.07}^{+ 0.07}$  \\[3pt]
       &   $>1.0$     & $ 0.42_{- 0.09}^{+ 0.09}$  & $ 0.29_{- 0.18}^{+ 0.26}$  & $ 0.46_{- 0.10}^{+ 0.10}$  \\[8pt]
 $0.6$ &   $\leq 0.5$ & $ 0.58_{- 0.06}^{+ 0.06}$  & $ 0.58_{- 0.07}^{+ 0.07}$  & $ 0.59_{- 0.12}^{+ 0.11}$  \\[3pt]
       &   0.5--1.0   & $ 0.16_{- 0.05}^{+ 0.07}$  & $ 0.00_{- 0.00}^{+ 0.09}$  & $ 0.27_{- 0.09}^{+ 0.10}$  \\[3pt]
       &   $\leq1.0$  & $ 0.43_{- 0.05}^{+ 0.05}$  & $ 0.43_{- 0.06}^{+ 0.06}$  & $ 0.42_{- 0.07}^{+ 0.08}$  \\[3pt]
       &   $>1.0$     & $ 0.26_{- 0.07}^{+ 0.08}$  & $ 0.25_{- 0.16}^{+ 0.24}$  & $ 0.26_{- 0.08}^{+ 0.09}$  \\[8pt]
 $1.0$ &   $\leq 0.5$ & $ 0.31_{- 0.05}^{+ 0.06}$  & $ 0.24_{- 0.06}^{+ 0.07}$  & $ 0.48_{- 0.11}^{+ 0.11}$  \\[3pt]
       &   0.5--1.0   & $ 0.08_{- 0.04}^{+ 0.06}$  & $ 0.00_{- 0.00}^{+ 0.09}$  & $ 0.13_{- 0.06}^{+ 0.09}$  \\[3pt]
       &   $\leq1.0$  & $ 0.23_{- 0.04}^{+ 0.04}$  & $ 0.18_{- 0.04}^{+ 0.05}$  & $ 0.30_{- 0.07}^{+ 0.07}$  \\[3pt]
       &   $>1.0$     & $ 0.20_{- 0.06}^{+ 0.08}$  & $ 0.25_{- 0.16}^{+ 0.24}$  & $ 0.18_{- 0.07}^{+ 0.09}$  \\[-5pt]
\enddata
\tablenotetext{a}{Values apply for $Z_{\rm gas}=0.1$.}
\end{deluxetable}

As documented in Table~\ref{tab:fcov}, for $\eta_{\rm c} \leq 0.5$,
the covering fraction decreases from $\simeq 0.9$ to $\simeq 0.3$ as
$W_{\rm cut}$ is increased from $0.1$ to $1.0$~{\AA} and shows little
to no dependence on virial mass for all $W_{\rm cut}$, except for a
suggestion that higher mass halos have larger $f_c(\eta_{\rm c})$ for
$W_{\rm cut} = 1.0$~{\AA} (though the values are consistent within
uncertainties).  For $0.5 < \eta_{\rm c} \leq 1.0$, the covering
fraction also decreases as $W_{\rm cut}$ is increased, but in this
regime $f_c(\eta_{\rm c})$ exhibits virial mass dependence such that
higher mass halos have substantially higher covering fraction than
lower mass halos.  In fact, for $W_{\rm cut} > 0.6$~{\AA},
$f_c(\eta_{\rm c})$ is consistent with zero in lower virial mass
galaxies.

Given that $R_{\rm c}$ is strongly anti-correlated with $M_{\rm\, h}$,
a fixed $D$ would probe further out into the theoretical cooling
radius for higher mass halos.  Thus, at fixed $\eta_{\rm c}$, higher
mass halos are probed at relatively smaller $D$ than are lower
mass halos.  \citet{cwc-massesI} showed (see their Figure~2)
that the covering fraction in fixed impact parameter bins, $f_c(D)$,
was higher for higher mass halos than for lower mass halos,
particularly for $D > 50$~kpc and $W_{\rm cut}=0.1$ and $0.3$~{\AA}.
For $D \leq 50$~kpc, $f_c(D)$ is effectively independent of virial mass
for $W_{\rm cut}=0.1$, $0.3$, and $0.6$~{\AA}, but is higher for
higher mass halos for $W_{\rm cut}=1.0$~{\AA}.  This behavior 
resembles the behavior of $f_c(\eta_{\rm c})$.  Given these
considerations, the virial mass dependence of $f_c(\eta_{\rm c})$ in
the range $0.5 < \eta_{\rm c} \leq 1.0$ is likely reflecting the
virial mass dependence of $f_c(D)$ on impact parameter.

As illustrated in Figure~\ref{fig:fDbar}, the average covering
fraction inside the theoretical cooling radius ($\eta_{\rm c} \leq
1$), exhibits little to no dependence on virial mass.  As
the $W_r(2796)$ absorption threshold is increased, the covering
fraction decreases from $\simeq 0.8$ down to $\simeq 0.2$.  The
average covering fraction outside the theoretical cooling radius does
not vanish, as would be expected if the absorbing gas originated from
cloud fragmentation and condensation from the hot coronal halo gas.
At projected distances where the density of the hot coronal gas is too
low to cool, $f_c(\eta_{\rm c})$ ranges from $\simeq 0.5$ down to
$\simeq 0.2$, decreasing as $W_{\rm cut}$ is increased from $0.1$ to
$1.0$~{\AA} and showing little evidence for a virial mass dependence,
especially for $W_{\rm cut}=0.6$ and $1.0$~{\AA}.

Comparing $f_c(\eta_{\rm c})$ inside and outside the theoretical
cooling radius, we can infer that the spatial properties of the
cool/warm CGM gas are {\it not\/} fundamentally connected to where the
cool/warm gas resides relative to the cooling radius of the hot
coronal halo gas\footnote{We also explored $Z_{\rm gas} \sim 0.03$ and
  $Z_{\rm gas} \sim 0.3$.  Note that the locus of points on
  Figure~\ref{fig:etacool} are virtually unchanged over the range
  $Z_{\rm gas} = 0.03$ to $Z_{\rm gas} = 0.3$; the higher (lower)
  $Z_{\rm gas}$ points have larger (smaller) $R_{\rm c}$, and thus
  there is a small upward (downward) shift in the points in
  Figures~\ref{fig:etacool}a and \ref{fig:etacool}b and a small
  leftward (rightward) shift in the points in
  Figures~\ref{fig:etacool}c and \ref{fig:etacool}d. The shifts are
  barely discernible and the results are qualitatively identical.  The
  covering fractions shown in Figure~\ref{fig:fDbar} are reduced
  (increased) by $\simeq 0.1$ for higher (lower) $Z_{\rm gas}$ for
  $\eta_{\rm c} \leq 1$.  For $\eta_{\rm c} > 1$, the covering
  fraction is unchanged with $Z_{\rm gas}$.  This behavior applies for
  all $W_{\rm cut}$.}.  The 20--50\% covering fraction {\it outside
  the cooling radius\/} indicates that all of the {\MgII} absorbing
CGM {\it inside the virial radius\/} may not originate in
fragmentation and condensation of the hot coronal gas phase inside the
cooling radius of galaxy halos.  It is of interest that as the
$W_r(2796)$ absorption threshold is increased, the difference between
the covering fractions inside and outside the cooling radius decrease,
such that for $W_{\rm cut} = 1.0$~{\AA}, they are virtually identical
with $f_c(\eta_{\rm c}) \simeq 0.2$.  This indicates that the more
optically thick and or kinematically complex the material is, the more
uniformly distributed it is with respect to the theoretical cooling
radius.  We further note that this trend is not sensitive to virial
mass.


\section{Discussion}
\label{sec:discuss}

In this work we have shown a picture of the cool/warm CGM where trends
become clearer once virial mass is taken into account. According to
the results presented in \S~\ref{sec:results}, virial mass determines
the extent and strength of the {\MgII} absorbing gas such that
equivalent width increases with increasing virial mass at fixed
distance and decreases with increasing distance from the galaxy at a
fixed virial mass.  In any given narrow range of impact parameter, the
equivalent widths are systematically smaller in the CGM of smaller
virial mass halos and systematically larger in the CGM of higher
virial mass halos.



The data reveal that trends and correlations are present between
several various quantities.  To examine this further, we explored the
multivariate behavior of the absorption and galaxy properties.  We
present this exercise and discuss the results in
Appendix~\ref{app:multivariate}.  As will be further discussed below,
these directly examined trends and correlations are the underlying
physical relationships that yielded the results presented in our
initial study \citep{cwc-massesI}, which clearly indicated a virial
mass segregation on the $W_r(2796)$--$D$ that is responsible for a
substantial component of scatter in the $W_r(2796)$ versus $D$
anti-correlation due to higher virial mass galaxies exhibiting
stronger absorption and larger impact parameter. Furthermore, we have
reported here that the mean $W_r(2796)$ is constant with $\eta_{\rm v}
= D/R_{\rm vir}$ (especially for $\eta_{\rm v} \leq 0.3$, see
Figure~\ref{fig:EWvsDR}b).  These results, and the vanishing of the
virial mass segregation on the $W_r(2796)$--$\eta_{\rm v}$ plane
\citep[see Figure~1c of][]{cwc-massesI}, which was quantified with a
high statistical significance, indicate that once distances are
scaled to the virial radius of each galaxy, the {\MgII} absorbing CGM
is self-similar with virial mass.

\subsection{The Absorption Radius}

In \S~\ref{sec:halorad} we investigated, $R(M_{\rm\, h})$, the
absorption radius, and the degree to which it shows dependence on
virial mass for various $W_r(2796)$ absorption thresholds (see
Figure~\ref{fig:DMhenv}).  The results indicate that when the
$W_r(2796)$ absorption threshold is small, $W_{r}(2796) \ge
0.1$~{\AA}, the absorption radius is highly proportional to virial
mass ($\gamma \simeq 0.45$) and the covering fraction is quite large
$f_c \simeq 0.8$.  As the absorption radius is examined for
progressively stronger absorption, the proportionality to virial mass
progressively decreases, such that for $W_r(2796) \ge 1.0$~{\AA},
$\gamma \simeq 0.2$.  In addition, the covering fraction decreases to
$f_c \simeq 0.35$.

For $W_r(2796) \ge 0.1$~{\AA}, the fitted absorption radius increases
by an order of magnitude (from $\sim 10$ kpc to $\sim 200$~kpc) over
four decades of virial mass ($10 \leq \log M_{\rm\, h}/M_{\odot} \leq
14$).  As such the fit predicts very extended absorption ($D >
200$~kpc) for $\log M_{\rm\, h}/M_{\odot} > 13$ when the weakest
absorption is included.  Since we do not probe impact parameters
greater than $D=200$~kpc, we do not have the data to verify this.
Interestingly, in a statistical study of 50,000 {\MgII} absorbers
($0.4 \leq z \leq 2.5$) compared with images of the quasar fields
using SDSS data, \citet{zhu-MgII} show that the mean $W_r(2796)$
follows a decreasing power law with impact parameter out to 10~Mpc.
At $D=100$~kpc, the mean equivalent width of their sample is
$W_r(2796) \simeq 0.2$~{\AA} (comparable to our mean equivalent width
at this impact parameter), and is $W_r(2796) \simeq 0.003$~{\AA} at
$D=10$~Mpc.  Assuming NFW density profiles, they find that the surface
density profile of {\MgII} absorbing gas, $\Sigma({\MgII})$
[M$_{\odot}$~pc$^{-2}$] is dominated by the single halo term out to
1~Mpc, outside of which the two-halo term dominates the gas surface
density profile.
The results of \citet{zhu-MgII} corroborate the idea that {\MgII}
absorption can be highly extended for weaker absorption and may have
implications for understanding the redshift path density of the
population of weak {\MgII} absorbers \citep{weakI, weakII, narayanan07,
  evans13}.

For $W_r(2796) \ge 1.0$~{\AA}, our fitted relation predicts that for
the most optically thick and/or kinematically complex absorbing gas,
the sensitivity of the absorption radius to virial mass is not as
pronounced, such that the radius increases by no more than a factor of
five (from $\sim 30$ kpc to $\sim 150$~kpc) over the range $10 \leq
\log M_{\rm\, h}/M_{\odot} \leq 14$.

The decrease in both the slope, $\gamma$, and the normalization,
$R_{\ast}$, with increasing $W_r(2796)$ absorption threshold is
strongly governed by the fact that the average covering fraction
decreases for stronger absorption and for increasing $D$
\citep[see][and references therein]{nikki-cat2}.  Most notably, it may
be the differential behavior with both $M_{\rm\, h}$ and $W_r(2796)$
absorption threshold in the ``rate'' at which the covering fraction,
$f_{\rm c}(D)$, decreases with impact parameter [i.e., the slope of
  $f_{\rm c}(D)$] that governs the behavior of $\gamma$ and $R_{\ast}$
with $W_r(2796)$ absorption threshold [see Figure~2 of
  \citet{cwc-massesI} for an illustration of this differential
  behavior in $f_{\rm c}(D)$].

For the lowest (highest) $W_r(2796)$ absorption threshold, the
relatively steeper (shallower) virial mass dependence of $R(M_{\rm\,
  h})$ reflects the steeper (shallower) decline in $f_{\rm c}(D)$.
Note that this possible effect is most pronounced in the regime $\log
M_{\rm\, h}/M_{\odot} < 12$, because the change in the slope of
$f_{\rm c}(D)$ with $W_r(2796)$ absorption threshold is most
pronounced for lower virial mass galaxies, being steepest for the
lowest $W_r(2796)$ absorption threshold.  This latter fact, in
particular, results in the relatively large value of $\gamma$ for the
absorption threshold $W_{\rm cut} = 0.1$~{\AA}.  Note that the
differential behavior in $f_{\rm c}(D)$ is naturally explained by the
self-similarity of the {\MgII} absorbing CGM with virial mass, as
discussed in \citet{cwc-massesI}.

The substantial uncertainties in $R_{\ast}$ reflect the degree of
fuzziness (both radially and spatially) in the mean absorption radius
for a given absorption threshold.  As such, the parameterizations of
the absorption radius with virial mass reflect a correlation between
impact parameter and virial mass with the interpretation that, on
average, higher mass halos have a more extended CGM with lower
geometric covering fractions.

We caution that the formalism of an absorption radius does imply a
well-defined boundary to the extent of the absorbing gas.  The
parameterization itself, as applied in this work, incorporates the
assumptions of a spherical radius (circular in projection) and that
the sky covering of the absorbing material is random.  In hydrodynamic
cosmological simulations, asymmetric filamentary structure in the
cool/warm phase of the CGM is a common feature of simulations
\citep{keres05, keres09, dekel06, ocvirk08, dekel09, ceverino10,
  vandevoort11, vandevoort+schaye11, goerdt13}.  \citet{ggk-sims}
showed that much of the {\MgII} absorbing gas in the outer regions of
the CGM is in the form of filaments.  Using 123 galaxies from the
{\magiicat} sample, \citet{ggk-orientations} reported that the
covering fraction is a maximum $f_c = 0.80$ along the projected minor
axis, is $f_c = 0.65$ along the projected major axis, and minimizes at
$f_c = 0.50$ at projections intermediate to these two galactic axes.
Thus, for {\MgII} absorbing gas, the assumptions of a spherical
geometry and random covering fraction are not supported by simulations
nor observations so that the above parameterization provides the mean
behavior of the {\MgII}-absorbing CGM with virial mass averaged over
all galaxy orientations.

By scaling the absorption radius by virial radius, we obtain the
remarkable result that the mass-normalized absorption envelope,
$\eta(M_{\rm\, h})$, is very weakly dependent on virial mass ($\gamma'
\leq \pm 0.1$) and has a value of $\eta_{\rm v}^{\ast} =
R_{\ast}/R^{\ast}_{\rm vir} \simeq 0.3$ independent of $W_r(2796)$
absorption threshold.  Inspection of Figure~\ref{fig:DMRhenv} reveals
that the number of galaxies with absorption above all $W_r(2796)$
absorption thresholds drops dramatically outside the inner 30\% of the
virial radius.  This suggests that both optically thin and/or
kinematically quiescent and optically thick and/or kinematically
complex absorbing gas is strongly concentrated within the inner 30\%
of the virial radius regardless of the virial mass of the galaxy.

The relatively weak dependence of $\eta(M_{\rm\, h})$ on $M_{\rm\, h}$
and the decreasing covering fraction with increasing $W_r(2796)$
absorption threshold are both consistent with the fact that the slope
of $f_{\rm c}(D/R_{\rm vir})$ is virtually identical for lower and
higher virial mass galaxies, but becomes shallower with increasing
$W_r(2796)$ absorption threshold [see Figure~2 of \citet{cwc-massesI}
  for an illustration of this behavior in $f_{\rm c}(D/R_{\rm vir})$].

The behavior of both the absorption radius and the mass-normalized
absorption envelope are consistent with the results presented in
Figures~\ref{fig:EWvsD} and \ref{fig:EWvsDR}, which show that the mean
$W_r(2796)$ increases with virial mass within a finite impact
parameter range, but is constant with $\eta_{\rm v} = D/R_{\rm vir}$,
especially for $\eta_{\rm v} \leq 0.3$.  There is a remarkable
self-similarity in the mean absorption relative to the virial radius
over the full range of virial masses represented in our sample.  The
{\MgII} column densities (governed by metallicity, density, and cloud
size), and/or kinematics (number of clouds) are, on average, highly
similar across virial mass within the inner 30\% of the virial radius,
and possibly out to the virial radius.  However, since the virial
radius is proportional to virial mass, the physical extent of the
absorbing gas is greater for galaxies with higher virial mass.  For
the CGM to have the similar average $W_r(2796)$ for all virial masses
as a function of $\eta_{\rm v} = D/R_{\rm vir}$, it implies that the
mean $W_r(2796)$ increases with virial mass in finite impact parameter
ranges (which is confirmed with our sample).

The observed increase in the mean $W_r(2796)$ with virial mass in
finite range of impact parameter is also apparent in the virial mass
dependence of the upper envelope of absorption, as shown in
Figure~\ref{fig:EW2D}a.  This implies a virial mass ``gradient'' in
the $W_r(2796)$--$D$ plane in the direction of increasing $W_r(2796)$.
The steepening in the relationship between $W_r(2796)$ and virial mass
as impact parameter is decreased, as shown in Figure~\ref{fig:EW2D}b,
reflects the fact that this mass gradient is steeper at smaller impact
parameters\footnote{This statement may seem to contradict the data
  presented in Figure~\ref{fig:EWvsD}, but we remind the reader that
  Figure~\ref{fig:EWvsD} presents $\log W_r(2796)$ versus $D$.}.  To a
large degree, this mass gradient provides insight into the systematic
scatter of $W_r(2796)$ on the $W_r(2796)$--$D$ plane, and the
significant reduction of scatter of $W_r(2796)$ on the
$W_r(2796)$--$\eta_{\rm v}$ plane shown in \citet{cwc-massesI}.
Since, on average, virial mass is correlated with galaxy luminosity
(per the formalism of halo abundance matching), this explains the
significant systematic luminosity segregation on the $W_r(2796)$--$D$
plane reported in Paper~II \citep{nikki-cat2}.

Thus, examination of the data using several methods, as presented in
Figures~\ref{fig:DMhenv}--\ref{fig:EWvsDR}, all corroborate a
picture in which the {\MgII}-absorbing CGM is self-similar with 
relative location with respect to the virial radius.

\subsection{The Cooling Radius}

We have investigated the behavior of $W_r(2796)$ with $\eta_{\rm c} =
D/R_{\rm c} $, the projected location where the absorption arises with
respect to the theoretical cooling radius.  Within the theoretical
formalism of the cooling radius, i.e., to the degree that it can be
viewed as a truly physical phenomenon associated with galaxy halos, we
find that {\MgII} absorption is found both inside and outside the
cooling radius.  For the optically thin and/or kinematically quiescent
gas, the covering fraction is roughly a factor of two higher inside
the cooling radius as compared to outside.  However, for stronger
absorption, the covering fraction is independent of whether the gas is
outside or inside the cooling radius.

The model of \citet{maller04} does not predict cool/warm clouds in the
CGM outside the cooling radius.  If the {\MgII} absorbing clouds have
a single origin of fragmentation and condensation out of the hot
coronal gas, then our findings might suggest the underlying physical
principles from which a theoretical cooling radius is derived should
be questioned.  However, the model of \citet{maller04}, by design,
does not include stellar feedback mechanisms nor accretion from the
intergalactic medium or mergers.

If there is reality to the theoretical cooling radius, we would then
infer that absorbing structures residing outside the cooling radius
are either recycled/processed clouds, winds, and/or infalling material
and that the properties of cool/warm CGM gas are not fundamentally
governed by where the gas resides relative to the cooling radius.
That is, the non-negligible covering fraction for {\MgII} absorption
outside the theoretical cooling radius corroborates a multiple
origins scenario for the cool/warm CGM provided by direct observation
of winds \citep{tremonti07, martin09, weiner09, rubin10, martin12},
infall \citep{rubin11, ggk-sims, ggk-q1317}, rotation kinematics
\citep{steidel02, kcems}, superbubble kinematics
\citep{cwc-garching05, bond01, ellison03}, and orientation effects
\citep{bordoloi11, bouche11, ggk-orientations}.

\begin{figure*}[thb]
\epsscale{1.1} 
\plotone{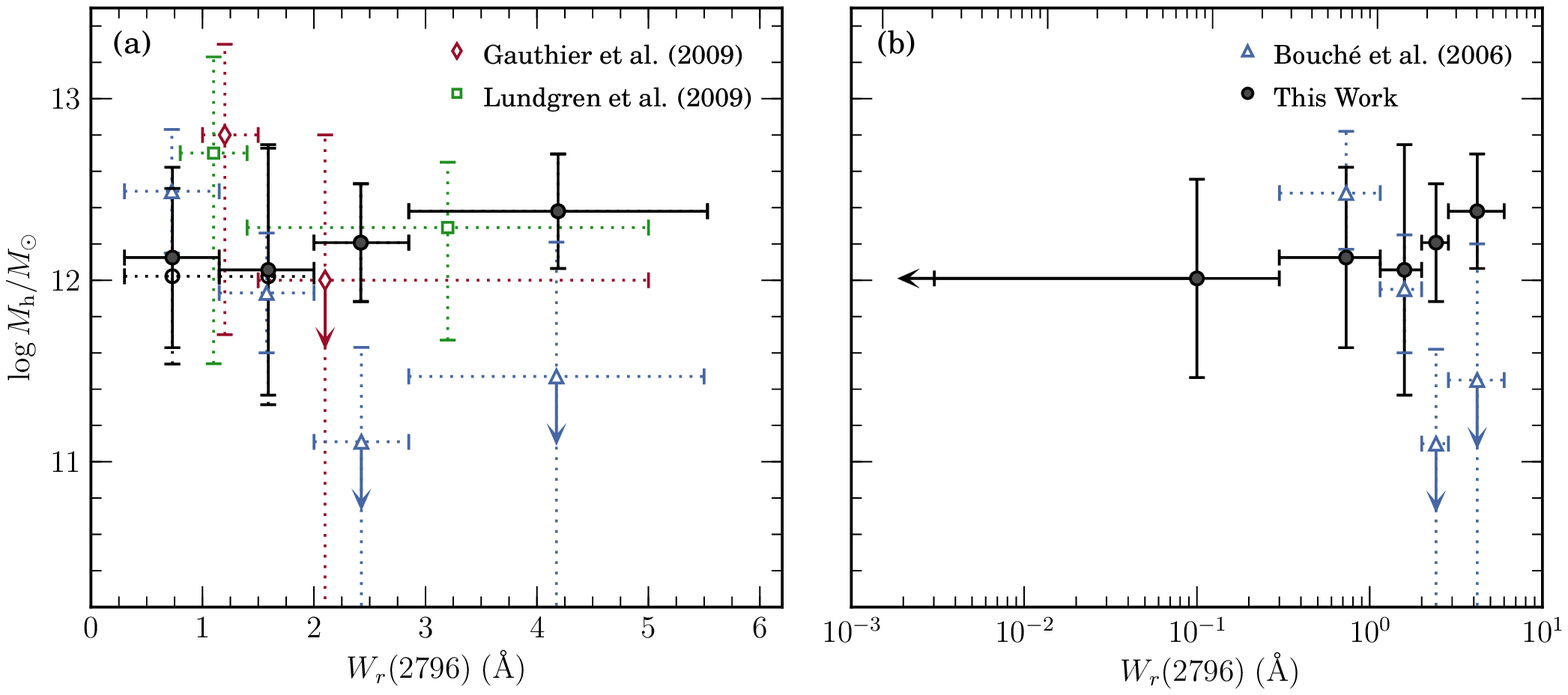}
\caption{(a) Mean virial mass, $\left< \log M_{\rm\, h}/M_{\odot}
  \right>$, versus $W_r(2796)$ for $W_r(2796) \geq 0.3$~{\AA}. Shown
  are the data from \citet{gauthier09} [red], \citet{lundgren09}
  [green], \citet{bouche06} [blue], and this work [black].  Downward
  arrows indicate upper limits on virial mass.  For comparison, the
  data from this work have been presented using the equivalent width
  bins defined by \citet{bouche06}.  Solid black data points include
  all impact parameters, whereas the open black data points include
  only those galaxies probed at $D < 140$ kpc for direct comparison
  with the sample of \citet{bouche06}.  Error bars for our data are
  the standard deviation in $\left< \log M_{\rm\, h}/M_{\odot}
  \right>$.  Our data do not reproduce the $M_{\rm\, h}$--$W_r(2796)$
  anti-correlation claimed by authors using absorber-galaxy
  cross-correlation techniques. (b) Same as for (a), but including the
  mean virial mass in the range $W_r(2796) < 0.3$~{\AA}.  Since many
  of the points with $W_r(2796)\leq 0.3$~{\AA} are upper limits, we
  annotate the binned point with an arrow.  No trend between $\left<
  \log M_{\rm\, h}/M_{\odot} \right>$ and $W_r(2796)$ is found even
  with the inclusion of the weakest absorbers and ``non-absorbers''.}
\label{fig:EWvsMh-compare}
\end{figure*}

\subsection{Comparison to Previous Works}
\label{sec:previousworks}

\citet{bouche06} used a large ($\simeq 1800$) sample of $z\simeq 0.5$
      {\MgII} absorbers with $W_r(2796) > 0.3$~{\AA} and some 250,000
      luminous red galaxies (LRGs) to obtain a statistical relation
      between equivalent width and virial mass for their flux-limited
      LRG sample. They estimated the virial masses of the absorbers by
      measuring the bias in the absorber-LRG cross-correlation
      relative to the LRG auto-correlation function.  They reported a
      $3 \sigma$ anti-correlation between virial mass and equivalent
      width, which they interpret as showing that {\MgII} absorbers are
      not virialized within their host halos but instead originate
      from galactic winds in star forming galaxies.

In a study using a similar sample size and covering a similar redshift
range, \citet{gauthier09} applied an essentially identical method and
report a $\sim 1~\sigma$ anti-correlation between equivalent width and
virial mass for their volume-limited LRG sample. Defining stronger
absorbers to have $W_r(2796) > 1.5$~{\AA} and weaker absorbers to have
$1.0 < W_r(2796) \leq 1.5$~{\AA}, they conclude that their weaker
{\MgII} absorbers (associated with $\log M_{\rm\, h}/M_{\odot} <
13.4$) are clustered more than their stronger absorbers (associated
with $\log M_{\rm\, h}/M_{\odot} < 12.7$).  Such a result would be
consistent with the halo occupation model of \citet{tinker08} in which
the strongest {\MgII} absorbers are suppressed in the most massive
halos, which would reduce the clustering of strong {\MgII} absorbers.

Motivated by the model of \citet{tinker08} and the
$W_r(2796)$--$M_{\rm\, h}$ anti-correlation of \citet{bouche06},
\citet{lundgren09} also undertook a similar analysis of $0.36 \leq z
\leq 0.8$ {\MgII} absorbers with $W_r(2796) > 0.8$~{\AA}, which they
cross correlated with some 1.5 million LRGs (volume-limited sample).
They report a ``marginal'' anti-correlation (significance level not
stated) between $W_r(2796)$ and virial mass and conclude that their
weaker {\MgII} absorbers occupy halos some 25 times more massive than
their stronger absorbers.  With a substantially larger and more
controlled sample than that of \citet{bouche06}, they were
unsuccessful at obtaining a higher significance in the
$W_r(2796)$--$M_{\rm\, h}$ anti-correlation; they actually found a
weaker signal than \citet{bouche06}.  The \citet{lundgren09} work
actually calls into question the veracity of a $W_r(2796)$--$M_{\rm\,
h}$ anti-correlation.

In direct conflict with the $3~\sigma$ result of \citet{bouche06} and
the less statistically significant follow-up results of
\citet{gauthier09} and \citet{lundgren09}, we have shown that our data
are highly consistent with no correlation between $W_r(2796)$ and
$M_{\rm\, h}$ when all impact parameters are considered; the rank
correlation tests are highly consistent with the null hypothesis of no
correlation ($0.1~\sigma$).

In Figure~\ref{fig:EWvsMh-compare}a, we directly compare our sample to
those of \citet{bouche06}, \citet{gauthier09}, and \citet{lundgren09}.
Since \citet{bouche06} reports the most significant
$W_r(2796)$--$M_{\rm\, h}$ anti-correlation, we binned our data to
match theirs, however, we plot the \citet{gauthier09} and
\citet{lundgren09} data as presented in those works.  For absorbers
with $W_r(2796) > 0.3$~{\AA}, our results are consistent within
uncertainties with all three absorber-LRG cross-correlation studies.
Interestingly, there is a slight discrepancy at $W_r(2796) \sim
2.4$~{\AA} where \citet{bouche06} obtain an upper limit on the mean
virial mass that is $\sim 2\sigma$ lower than our value.  Since the
\citet{bouche06} sample is limited to $D \leq 140$~kpc, we recomputed
$\left< \log M_{\rm\, h}/M_{\odot} \right>$ for the subsample of our
data for which $D \leq 140$~kpc.  These points are plotted as open
points.  There is no change in the mean virial mass for $W_r(2796) >
2$~{\AA} and only an insignificant reduction in the mean virial mass for
$W_r(2796) < 2$~{\AA}.

Since we probe well below $W_r(2796) = 0.3$~{\AA}, where the majority
of our equivalent width measurements are upper limits ($3~\sigma$), we
computed $\left< \log M_{\rm\, h}/M_{\odot} \right>$ for the
non-absorbers and the weakest absorbers.  In
Figure~\ref{fig:EWvsMh-compare}b, we compare our data directly to
those of \citet{bouche06} and add the data point for $W_r(2796) <
0.3$~{\AA} absorbers and non-absorbers.  Our data are consistent with
no $W_r(2796)$--$M_{\rm\, h}$ anti-correlation over this much broader
equivalent width range.  
We caution that $\simeq 20$\% of the $W_r(2796) < 0.3$~{\AA} absorbers
and non-absorbers reside at $D > 140$~kpc, whereas the
\citet{bouche06} systems all reside at $D < 140$~kpc.  However, as we
showed above, the inclusion of larger impact parameters has a
negligible effect on the mean virial mass.

In summary, we find no anti-correlation between $W_r(2796)$ and
$M_{\rm\, h}$, whether we binned our data to obtain the means (per
Figure~\ref{fig:EWvsMh-compare}) or performed statistical tests on
the unbinned data.  In fact, we find trends that suggest $W_r(2796)$
and $M_{\rm\, h}$ are correlated in finite ranges of impact parameter.
Our data, as presented in Figure~\ref{fig:EWvsMh-compare}, are
statistically consistent with the cross correlation clustering
analyses of \citet{bouche06}, \citet{gauthier09}, and
\citet{lundgren09}; however, our results do not imply any suggestion
of a $W_r(2796)$--$M_{\rm\, h}$ anti-correlation.

\subsection{Implications for Clustering}

If, per the halo occupation model of \citet{tinker08}, higher mass
halos have a CGM environment that suppresses {\MgII} absorbers, then
the covering fraction of the CGM would be observed to decline for
higher virial mass galaxies.  This would reduce the clustering of
stronger absorbers.  However, the model of \citet{tinker08} is tuned
to match the $W_r(2796)$--$M_{\rm\, h}$ anti-correlation reported by
\citet{bouche06}.

In this work, we have shown that the data do not support a
$W_r(2796)$--$M_{\rm\, h}$ anti-correlation\footnote{We add that {\sc
    bhk}-$\tau$ non-paramteric rank correlation tests on the unbinned
  data also indicate no correlation between $L_K/L_K^{\ast}$ and
  $W_r(2796)$ [$P(\tau_k)=0..52$, $N(\sigma) = 0.6$], nor $M_K$ and
  $W_r(2796)$ [$P(\tau_k)=0.65$, $N(\sigma) = 0.5$].}, but do support
a strong positive trend in finite impact parameter ranges.
\citet{cwc-massesI} showed that the covering fraction is effectively
invariant with virial mass (see their Figures 3 and 4), even for
different $W_r(2796)$ absorption thresholds.  This fact places tension
on the halo occupation model, and it remains to be worked out how this
would change the argument for weaker clustering of stronger {\MgII}
absorbers.

We expect that the lack of a $W_r(2796)$--$M_{\rm\, h}$
anti-correlation and the invariance in the {\MgII} covering fraction
with virial mass would nullify a weaker clustering of stronger {\MgII}
absorbers. Interestingly, \citet{rogerson12}, using paired quasar
sightlines of {\MgII} absorbers, were able to show that the
\citet{tinker08} model was ruled out because the model failed to 
reproduce the observed variation in {\MgII} absorption between
sightlines.

\subsection{Interpretations}

In addition to quantifying the extent of {\MgII} absorbing gas,
previous works have also investigated physical interpretations to
understand the CGM, both within the context of other halo gas
statistics and the theoretical framework of gas accretion and galactic
outflows.  Along these lines, \citet{bouche06} claim that an
anti-correlation between $W_r(2796)$ and $M_{\rm\, h}$ is a natural
consequence of the behavior of absorber statistics and that since
there are no strong {\MgII} absorbers at large distances from
galaxies, equivalent width must be inversely proportional to virial
mass. They argue that this physical relation follows from the
combined results that the absorption radius is proportional to galaxy
luminosity, $R(L) = R_{\ast} (L/L^{\ast})^\beta$ with $\beta>0$, and
that the equivalent width is inversely proportional to absorption
radius, $W_r(2796) = \alpha_1 \log R(L) + \alpha_2$ (with $\alpha_1 <
0$), the latter reflecting the $W_r(2796)$--$D$ anti-correlation.  

The \citet{bouche06} argument is based upon the assumption that there
is no virial mass dependence on the slope and normalization
($\alpha_1$ and $\alpha_2$) of the upper absorption envelope
(absorption radius), which forces the inference that there is a
horizontal virial mass gradient on the $W_r(2796)$--$D$ plane such
that higher mass halos are preferentially found at larger impact
parameter.  As shown in Figure~\ref{fig:EW2D}a, a galaxy with strong
{\MgII} absorption usually has absorption at small impact
parameter. However, the fact that absorption occurs closer to a galaxy
does not imply that the halo is low mass and the fact that absorption
occurs far from a galaxy does not imply that it has high virial
mass. What we find is that stronger (weaker) absorption in a finite
impact parameter range implies the host galaxy has a higher (lower)
virial mass, and this applies regardless of whether the impact
parameter is small or large (also see Figure~\ref{fig:EW2D}b).

Galaxies that inhabit more massive dark matter halos simply have
stronger absorption at a given distance, even though $W_r(2796)$
decreases with distance at a given mass.  Our data directly show that
the virial mass gradient on the $W_r(2796)$--$D$ plane is vertical,
such that the slope of the upper absorption envelope steepens with
increasing virial mass.  This behavior results in a flat relationship
between the mean $W_r(2796)$ as a function of virial mass when all
impact parameters are included in the mean.


\subsection{Abundance Matching Considerations}

Since the halo abundance matching method yields the average virial mass
for a galaxy with a measured $M_r$, the behavior of the {\MgII}
absorbing CGM reported in this work reflects an averaged behavior with
virial mass, and is not based upon a 1:1 correspondence between
$W_r(2796)$ and a dynamically measured virial mass, $M_{\rm\, h}$.
However, we note several reasons that the averaged behavior is an
accurate representation.  

First, the analysis does provide a 1:1 correspondence between
$W_r(2796)$ and $M_r$.  Each galaxy associated with a measurement of
{\MgII} absorption has a measured $r$-band luminosity.  Second, for
halo abundance matching, the relationship between the average virial
mass and $M_r$ at a given redshift is monotonic and smooth (see
Figure~\ref{fig:MhvsMr}); thus, the main difference between results
obtained using the average virial mass as opposed to the measured
$M_r$ is contained in the slope of the $M_{\rm\, h}$--$M_r$ relation
as a function of $M_r$.  This slope steepens at the bright end, which
effectively provides a stretching in the dynamic range of the galaxy
property being compared to the {\MgII} absorption, essentially
increasing the leverage and/or moment arm over which the CGM-galaxy
connection can be explored.  Third, the general behavior of
$W_r(2796)$ with virial mass, including mass dependence of the
covering fraction, $f_c(D)$, mass segregation on the $W_r(2796)$--$D$
plane, etc., is also seen directly with $L_B/L_B^{\ast}$,
$L_K/L_K^{\ast}$, $M_B$, and $M_K$ \citep{nikki-cat2}.  For the
luminosities, the trends with $M_K$ (a proxy for stellar mass) are
invariably the most statistically significant.

The higher statistical significance for these trends and correlations
with virial mass is a consequence of the rapid increase in virial mass
at the bright end of the luminosity function, which, as stated above,
provides added leverage for exploring the galaxy-CGM connection.  The
added benefit of employing the average virial mass is that the average
virial radius of an $M_r$ galaxy can be incorporated into the
analysis.  Knowing the physical extent of the dark matter halo has
provided enhanced insight.  For example, the significantly reduced
scatter and vanishing of virial mass segregation on the
$W_r(2796)$-$\eta_{\rm v}$ plane as compared to the $W_r(2796)$-$D$
plane, the tight scaling of $W_r(2796)$ with $(\eta_{\rm v})^{-2}$,
the self-similarity of the covering fraction, $f_c(\eta_{\rm v})$, and
the invariance of the mean $W_r(2796)$ with virial mass in finite
$\eta_{\rm v}$ ranges.  Virial mass also provides a formalism for
estimating the theoretical cooling radius, which has provided some
insight into the fact that {\MgII} absorption strength and covering
fraction shows no dependence on this theoretically based quantity.


\section{Summary and Conclusions}
\label{sec:conclude}

Using 182 {\MgII} absorbing galaxies from {\magiicat}, we have
examined the behavior of the {\MgII} absorbing CGM in relation to the
virial mass of the host galaxy.  Details of the sample are described
fully in Paper~I \citep{nikki-cat1} and Paper~II \citep{nikki-cat2}.

In this work, we have presented additional details of the halo
abundance matching technique previous applied by \citet{cwc-massesI}.
Calculation of the virial radii was described in that study. In this
work, we incorporated the theoretical cooling radius, which we
computed using the multi-phase halo model of \citet{maller04} adopting
a cool/warm gas component ($10^4 \leq T \leq 10^5$~K) and a hot gas
coronal component ($T \geq 10^{5.5}$~K).  The cool/warm component
corresponds to the {\MgII} absorbing gas probed in this study.  Since
the median virial mass of our galaxy sample is $\log M_{\rm\,
  h}/M_{\odot}=12$, we will refer to $\log M_{\rm\, h}/M_{\odot} < 12$
as lower mass halos and $\log M_{\rm\, h}/M_{\odot} >12$ as higher
mass halos.

In an effort to understand the relationships between the presence and
strength of {\MgII} absorption in the CGM of these galaxies, we
examined the behavior of the {\MgII} $\lambda 2796$ rest-frame
equivalent width, $W_r(2796)$, with virial mass, $M_{\rm\, h}$, impact
parameter, $D$, projected location relative to the virial radius,
$\eta_{\rm v} = D/R_{\rm vir}$, and projected location relative to the
theoretical cooling radius, $\eta_{\rm c} = D/R_{\rm c}$.  Highlights
of our findings include:

(1) Assuming a Holmberg-like virial mass dependence to the {\MgII}
``absorbing radius'', $R(M_{\rm\, h}) = R_{\ast}(M_{\rm\,
  h}/M^{\ast}_{\rm\, h})^{\gamma}$, where $M^{\ast}_{\rm\, h} =
10^{12}$~M$_{\odot}$, we found a factor of two steepening in the power
law index, from $\gamma \simeq 0.2$ to $\gamma \simeq 0.4$, as the
$W_r(2796)$ absorption threshold is decreased from
$W_r(2796)=1.0$~{\AA} to $0.1$~{\AA} (see Figure~\ref{fig:DMhenv}).
We also found that the normalization at $M^{\ast}_{\rm\, h}$, slightly
increases with decreasing $W_r(2796)$ absorption threshold.  These
behaviors indicate cool/warm gas is more extended around higher virial
mass galaxies than around lower mass galaxies and the CGM is patchier
(more highly structured) as $W_r(2796)$ absorption threshold
increases.

(2) The absorption radius parameterizations were applied to determine
the ``mass-normalized absorption envelope'', $\eta_{\rm v}(M_{\rm\,
  h}) = \eta^{\ast}_{\rm v}(M_{\rm\, h}/M^{\ast}_{\rm\,
  h})^{\gamma'}$, where $\gamma' = \gamma - 1/3$ and $\eta^{\ast}_{\rm
  v}$ is the ratio $R_{\ast}$ to $R_{\rm vir}$ for an $M^{\ast}_{\rm\,
  h}$ galaxy (see Figure~\ref{fig:DMRhenv}).  We found that the mass
dependence of the mass-normalized absorption envelope is very weak,
ranging from $\gamma' \simeq 0.1$ to $\gamma' \simeq -0.14$, as the
$W_r(2796)$ absorption threshold is increased from
$W_r(2796)=0.1$~{\AA} to $1.0$~{\AA}.  The mean extent for all
$W_r(2796)$ absorption thresholds is $\eta^{\ast}_{\rm v} = 0.3$.
Given the weak virial mass dependence, this implies that the majority
of {\MgII} absorption, regardless of virial mass or absorption
strength, resides within the inner 30\% of the virial radius (in
projection).

(3) In finite impact parameter ranges, we found that the mean
$W_r(2796)$ shows a strong trend (greater than $2.5~\sigma$
significance) to increase with increasing virial mass in a power-law
fashion (see Figure~\ref{fig:EWvsD}).  The slope of the
maximum-likelihood fit increases with increasing impact parameter,
whereas the zero point decreases \citep[reflecting the
  anti--correlation on the $W_r(2796)$--$D$ plane; see][]{nikki-cat1}.
On average, at a given impact parameter, optically thicker, higher
column density and/or more kinematically complex cool/warm gas is
associated with higher mass halos, whereas weaker absorption and
optically thinner gas is associated with lower mass halos.  However,
in finite $\eta_{\rm v} = D/R_{\rm vir}$ ranges, the mean $W_r(2796)$
is constant with virial mass (see Figure~\ref{fig:EWvsDR}).  These
findings imply a self-similarity in the behavior of the {\MgII}
absorbing CGM properties with virial mass, consistent with the
covering fraction behavior reported by \citet{cwc-massesI}.  The mean
absorption strength fundamentally depends upon where the gas resides
relative to the virial radius.

(4) To the degree that the theoretical cooling radius, $R_{\rm c}$, is
a physically real location, the projected distance where the CGM is
probed with respect to the cooling radius, $\eta_{\rm c} = D/R_{\rm
  c}$, is a poor indicator of {\MgII} absorption strength (see
Figure~\ref{fig:etacool}).  On the $W_r(2796)$--$\eta_{\rm c}$ plane,
cool/warm absorbing gas is commonly found outside the theoretical
cooling radius and the range of $\eta_{\rm c}$ over which absorption
is found increases with increasing $W_r(2796)$.  Taking into account
the scaling between virial mass and the theoretical cooling radius, we
found that the covering fraction inside the cooling radius mirrors the
behavior of the covering fraction as a function of impact parameter
\citep[see][]{cwc-massesI}.  If the cooling radius is a real entity,
the presence of {\MgII} absorbing clouds outside the cooling radius
implies that the cool/warm CGM gas likely does not originate {\it
  only\/} from fragmentation and condensation out of the hot coronal
gas halo.

(5) Though we report a strong trend for increasing $W_r(2796)$ with
increasing virial mass in finite impact parameter ranges, the mean
$W_r(2796)$ is independent of virial mass when averaged over all
impact parameters (see Figure~\ref{fig:EWvsMh-compare}).  A {\sc
  bhk}-$\tau$ rank-correlation test on the unbinned equivalent widths
yields a less than $0.1~\sigma$ significance for ruling out no
correlation with virial mass (see \S~\ref{sec:discuss} for additional
details).  The lack of correlation between mean $W_r(2796)$ and mean
virial mass is contrary to the $W_r(2796)$--$M_{\rm\, h}$
anti-correlations reported by \citet{bouche06}, \citet{gauthier09},
and \citet{lundgren09} using virial mass bias galaxy-absorber cross
correlation techniques.  We note that, statistically, our data are not
inconsistent with their data, but our data clearly suggest no
anti-correlation between $W_r(2796)$ and virial mass.  This places
tension on halo occupation models of {\MgII} absorbing gas
\citep[cf.,][]{tinker08} and would suggest that stronger {\MgII}
absorbers are not necessarily less clustered than weaker absorbers.


\subsection{What Drives the Self-Similarity of the CGM?}

A main result of this work is that the properties of the cool/warm
component of the CGM are self-similar with virial mass and
fundamentally connected to the parameter $\eta_{\rm v}=D/R_{\rm vir}$,
the projected galactocentric distance of the gas relative to the
virial radius.  Regardless of viral mass, the mean {\MgII}
absorption strength is first and foremost governed by where it resides
with respect to the virial radius of the halo.  We found that the
majority of the {\MgII} absorbing cool/warm CGM is located within the
inner 30\% of the virial radius.

Though the mean $W_r(2796)$ strongly trends toward a positive
correlation with virial mass in finite impact parameter ranges, the
overall lack of a correlation between the mean $W_r(2796)$ and virial
mass when all impact parameters are included is due to the highly
significant anti-correlation between $W_r(2796)$ and impact parameter
at fixed virial mass.  Most remarkable is that the mean $W_r(2796)$ is
constant as a function of $\eta_{\rm v} = D/R_{\rm vir}$ for
$\eta_{\rm v} \leq 0.3$, and may be constant all the way out to the
virial radius.

\citet{nikki-cat2} showed that the {\MgII} covering fraction decreases
with increasing $W_r(2796)$ absorption threshold at all impact
parameters and decreases with increasing distance from the central
galaxy.  \citet{cwc-massesI} showed that the {\MgII} absorption
covering fraction is effectively invariant as a function of virial
mass for all $W_r(2796)$ absorption thresholds, though it decreases as
the $W_r(2796)$ absorption threshold is increased.  The latter result
places tension on the notion that ``cold-mode'' accretion is
suppressed in higher mass halos ($\log M_{\rm\, h}/M_{\odot} \geq 12$)
as purported by \citet{birnboim03}, \citet{keres05}, \citet{dekel06},
and \citet{stewart11}.  One solution is that much of the {\MgII}
absorbing gas in higher mass halos arises in outflowing winds and/or
infalling metal enriched sub-halos (low-mass satellite galaxies, some
of which may be embedded in filaments).

Whatever the reasons that explain invariance of the covering fraction
with galaxy virial mass, the data indicate that the mean $W_r(2796)$
scales as an inverse-square power law, $(D/R_{\rm vir})^{-2}$, with
remarkably low scatter over several decades of virial mass
\citep{cwc-massesI}.  Combined, the virial mass invariance of the
covering fraction and the inverse-square profile of the mean
equivalent width place strong constraints on the nature of the
low-ionization CGM and are highly suggestive that the CGM is
self-similar with the virial mass of the host galaxy.

We note that our results are remarkably consistent with the findings
of \citet{stocke13}, who examined the CGM in multiple low- and
high-ionization transitions for $\simeq 70$ galaxies at $z \leq 0.2$.
They find that the majority of the metal-line absorbing gas in the CGM
resides within the inner 50\% of the virial radius.  They also report
that, {\it once virial radius scaling is applied}, there is little
distinction between CGM clouds as a function of galaxy luminosity,
radial location, or relative velocity.  Furthermore, \citet{stocke13}
find that the absorbing cloud diameters decrease with $(D/R_{\rm
  vir})^{-1.7\pm0.2}$, which is very close to an inverse square
relationship (however, the derived diameters exhibit considerable
scatter about the relation).

Stating that the cool/warm CGM is self-similar across a wide range of
virial mass is not equivalent to stating that the CGM is identical for
all galaxies vis-\`a-vis a simple scaling with virial radius.
Evidence for multiple origins of {\MgII} absorbing clouds, such as
winds \citep{tremonti07, martin09, weiner09, rubin10, martin12,
  bradshaw13, bordoloi13}, superbubbles \citep{cwc-garching05, bond01,
  ellison03}, infall \citep{rubin11, ggk-sims, ggk-q1317}, and
evidence for rotation kinematics \citep{steidel02, kcems} and
orientation dependencies \citep{bordoloi11, bouche11,
  ggk-orientations} precludes such a notion.  Furthermore, galaxies of
different dark matter halo masses live in different overdensities and
therefore local environments.

We are faced with the central question: how is it that the various
physical processes governing the cool/warm baryons in the CGM, which
respond to their host dark matter density profile and local over-dense
environment, yield mass invariant covering fractions and self-similar
radial profiles of the mean {\MgII} absorption strengths with location
relative to the virial radius, $R/R_{\rm vir}$, over a large range of
galaxy virial mass?

It is well established that higher mass halos live in higher
overdensity regions and are surrounded by greater numbers of sub-halos
\citep[e.g.,][]{mo-white96, klypin11}.  The more massive sub-halos can
form stars and chemically enrich their immediate surrounding, such
that absorbing gas in sub-halos likely comprises some component of
what we call the cool/warm CGM of higher mass halos \citep{kepner99,
  gnat04, vandevoort+schaye11}.  Lower mass halos, in contrast, live
in lower overdensity regimes, where the contribution of enriched gas
to their CGM from sub-halos is presumably lower.  On the other hand,
the influence of stellar feedback may have a relatively more important
influence on the CGM of lower mass galaxies
\citep[e.g.,][]{dallavecchia08, weinmann12, ceverino13,
  trujillo-gomez13}.  A great deal of theoretical work is pushing the
frontiers of our understanding of the different dominating physical
processes governing the ISM-CGM-IGM cycle as a function of galaxy
stellar and virial mass, and we will highlight some of these in the
below discussion.

As speculated by \citet{werk13}, the decrease in the average
absorption strength in low-ionization metals with increasing impact
parameter could imply a decreasing surface density in these ions, a
decreasing metallicity, and/or an increasing ionization state with
increasing galactoccentric distance, $R$, from the central galaxy.
Interestingly, \citet{stocke13}, having performed ionization modeling
of CGM absorbing clouds for their sample, find no clear trends in
cloud ionization parameter, density, metallicity, or temperature with
distance from the central galaxy or with galaxy luminosity (though
they do find trends for decreasing cloud sizes and masses with
distance from the central galaxy, which reflects the decreasing {\HI}
column density).  This would suggest that neither a metallicity
gradient nor an ionization gradient in the CGM is the driving
mechanism governing the $W_r(2796)$--$D$ anti-correlation out to the
virial radius.

\citet{werk13} also conclude that it is unlikely that the star
formation rate in the central galaxy is a dominant factor governing
the strength of low ion absorption.  Instead, they find that the
column densities of the low-ionization species correlate with stellar
mass, indicating that there is more circumgalactic gas in more massive
galaxy halos.  Similarly, \citet{zhu13} find that the amount of
{\CaII} in halos is larger for galaxies with higher stellar mass.
Since stellar mass and virial mass are correlated, these results are
consistent with our findings of a more extended upper envelope
to the absorption with increasing virial mass (see
Figure~\ref{fig:EW2D}a) and a proportionality between $W_r(2796)$ and
virial mass in finite impact parameter ranges (see
Figure~\ref{fig:EWvsD}).

Examining the average properties $z=0.25$ galaxies in low-resolution
cosmological simulations that include momentum-driven winds,
\citet{ford13a} employed ionization modeling of the CGM gas and found
that {\MgII} absorbing gas column density is centrally concentrated
and decreases with increasing $R$ from the central galaxy with very
little qualitative morphological difference in the radial profiles
with halo mass (filaments, satellite galaxies, and sub-halos are
azimuthally smoothed out).  They find virtually no temperature
gradient with impact parameter for the low-ionization species, which
arises in $T = 10^{4-4.5}$ K gas.  Though more massive halos have
larger hot gas fractions \citep{keres05, vandevoort+schaye11},
\citet{ford13a} find the overdensity of gas where {\HI} absorption
arises increases with halo mass from $\Delta \rho/\rho = 10^2$ for
$\log M_{\rm\, h}/M_{\odot} = 11$ to $\Delta \rho/\rho = 10^3$ for
$\log M_{\rm\, h}/M_{\odot} = 13$ ($D\sim 100$~kpc); most of the {\HI}
arises in cool/warm $T <10^{5}$ K gas, even in the highest mass halos.

The simulations and modeling of \citet{ford13a} clearly show that the
extent of the cool/warm CGM increases with increasing galaxy virial
mass; galaxies in more massive halos have a more extended CGM than
galaxies in lower mass halos.  Inspection of their {\MgII} column
density profiles show a predicted mean {\MgII} column density of
$\left< N({\MgII}) \right> = 10^{13}$~cm$^{-2}$ at $D \simeq 20$ kpc
for $\log M_{\rm\, h}/M_{\odot} = 11$ and $D \simeq 100$ kpc for $\log
M_{\rm\, h}/M_{\odot} = 13$.  Assuming thermal broadening for
$T=10^{4.5}$ K, this corresponds to a mean absorption strength of
$\left< W_r(2796) \right> =0.15$~{\AA}, a value consistent with our
measurements at these impact parameters.  Thus, the simulations and
ionization modeling predict that a given $W_r(2796)$ value, on
average, will be measured at a larger impact parameter in higher mass
halos in a manner that is consistent with our findings.  We also note
that these example points probe the same $\eta_{\rm v}= D/R_{\rm
  vir}$, since the ratio of the virial radii of the two simulated
galaxies, $(10^{11}/10^{13})^{1/3} \simeq 1/5$, equals the ratio of
the impact parameters of the respective galaxies.  A constant mean
$W_r(2796)$ with $\eta_{\rm v}$ is precisely the behavior we showed in
Figure~\ref{fig:EWvsDR}.


As such, the results of \citet{ford13a} provide a natural explanation
for the self-similarity of the cool/warm CGM with virial mass.  Their
simulations imply that winds contribute to the presence of {\MgII}
absorption in the CGM such that higher mass halos have larger mean
$W_r(2796)$ at a fixed impact parameter.  Sub-halos (satellites) and
filamentary infall play a role as well, since these contributions were
simply smoothed (averaged out) in the radial gas profiles presented by
\citet{ford13a}.  That is, predominantly photoionized outflowing winds
and accreting sub-halo clumps could go a long way toward explaining
the average self-similar properties of the low ionization CGM across
several decades of virial mass.

It would be of interest to examine the column density profiles from
the \citet{ford13a} work as a function of $D/R_{\rm vir}$ and virial
mass to quantify the degree that they are self-similar with virial
mass.  The foremost observational constraints that any model would be
required to satisfy are the inverse-square profile, $W_r(2796) \propto
(D/R_{\rm vir})^{-2}$, and the virial mass invariant covering fraction
that decreases with increasing $W_r(2796)$ absorption threshold, both
of which should be measured using ``mock'' absorption line analysis
through the simulated halos.

Though our data show that the average {\MgII} absorption strength in
the CGM obeys clear trends with virial mass, impact parameter, and
virial radius, there is a great deal of spread in the distribution of
$W_r(2796)$.  The spread is significant enough that the frequency of
non-detections increases with increasing impact parameter.  This could
possibly be due to substantial variations in the metallicity of the
CGM from sightline to sightline, even in the absence of a clear
metallicity gradient out to the virial radius.  We now consider the
possibility of a correlation between stellar/virial mass with the
metallicity of the cool/warm CGM in gas and plausible explanations for
strong variations in metallicity in the CGM from sightline to
sightline.

\citet{tremonti04} reported a tight correlation between stellar mass,
$M_{\ast}$, and gas-phase metallicity, $Z_{\hbox{\tiny ISM}}$, of the
interstellar medium (ISM) spanning three decades of stellar mass and a
factor of ten in metallicity.  \citet{mannucci10} examined the more
general relation between stellar mass, gas-phase metallicity, and star
formation rate (SFR) and found a tight surface in this 3D space,
dubbed the fundamental metallicity relation (FMR).  At low stellar
mass, metallicity decreases with increasing SFR, while at high stellar
mass, metallicity is independent of SFR. \citet{bothwell13} showed
that SFR may not be the fundamental third parameter of the FMR; they
find that {\HI} mass drives the stellar-mass metallicity relation such
that metallicity continues to correlate with stellar mass as {\HI}
mass increases.  Stellar mass and {\HI} mass correlations with gas
phase metallicity in the ISM may suggest a similar relation, on
average, in the CGM. However, the CGM being much more extended, and
being an interface with the IGM, is undoubtedly more complex such that
sightline to sightline variations could mask a galaxy/halo mass
metallicity correlation.

Indeed, in order to understand the stellar-mass metallicity
correlations, the flow and recycling of ISM and IGM gas through the
CGM must be invoked and tuned.  \citet{dave11a} determined that gas
content is regulated by a competition between inflow and gas
consumption within the interstellar medium, which is governed by the
star formation law.  That is, star-forming galaxies develop via a
slowly evolving equilibrium balanced by inflows (driven by
gravity/mass), wind recycling, star formation rates, and outflows, the
latter regulating the fraction of inflow that gets converted into
stars \citep{dave11b}.  \citet{dayal13} found that for more massive
galaxies, ISM metal enrichment due to star formation is diluted by
inflow of metal-poor IGM gas that yields a constant value of the ISM
gas metallicity with SFR (thereby reproducing the FMR at high mass).
In these massive galaxies, the effects of outflows are severely
mitigated due to the deep gravity wells. Conversely, lower mass
galaxies, which have smaller SFR, produce lower metallicity outflows,
but they are more efficiently distributed throughout the CGM due to
the shallower potential wells.  A similar model by \citet{lilly13}
indicates that the $M_{\ast}$--$M_{\rm\, h}$ relation, established by
baryonic processes within galaxies, suggests a significant fraction
(40\%) of baryons coming into the halos are being processed through
the galaxies.

Thus, we see that the mass-metallicity relationships of galaxies
\citep[e.g.,][]{tremonti04, mannucci10, bothwell13} theoretically
suggest a regulatory physical cycle between the ISM and the CGM that
involves lower metal enrichment of the CGM in lower mass galaxies and
higher metal enrichment of the CGM in higher mass galaxies; however,
the wind/outflowing material is more efficiently distributed into the
CGM in lower mass galaxies and less efficiently distributed in higher
mass galaxies.  This general behavior of the wind/recycled gas,
coupled with the rates at which clumpy and filamentary accretion is
mixed in the CGM, likely provides an excellent first-order physical
understanding of how the CGM of galaxies living in different mass
halos can be self-similar in their mean {\MgII} absorption properties.
We remind the reader that self-similar means the projected
profile of {\MgII} absorption strength with respect to the virial
radius is universal, i.e., $W_r(2796) \propto (D/R_{\rm vir})^{-2}$,
and the covering fraction of the CGM is independent of virial mass.

\citet{turnshek05} reported that gas-phase metallicity strongly
correlates with the velocity spread of {\MgII} [$W_r(2796)$ expressed
  in velocity units] for large $N({\HI})$ absorbers.  Using zCOSMOS
galaxies in the redshift interval $1.0 \leq z \leq 1.5$,
\citet{bordoloi13} report that the {\MgII} equivalent width of the
outflowing component increases with both galaxy stellar mass and star
formation rate. At similar stellar masses, the blue galaxies exhibit a
significantly higher outflow equivalent width as compared to red
galaxies.  In the UKIDSS Ultra-Deep Survey, \citet{bradshaw13} found
that the highest velocity outflows are found in galaxies with the
highest stellar masses and the youngest stellar populations.  They
conclude that high-velocity galactic outflows are mostly driven by
star-forming processes consistent with a mass-metallicity relation.

On the other hand, \citet{lehner13} reported a bimodality in the
metallicity of the CGM of luminous galaxies and conclude that the more
metal-rich absorbers likely originate from the nearby large galaxy in
the form of outflowing or recycling gas while the lower metallicity
gas is infall from the IGM.  Interestingly, of the galaxies for which
\citet{stocke13} could constrain the cool/warm CGM metallicities, nine
absorbers have $Z_{\hbox{\tiny CGM}} \simeq Z_{\hbox{\tiny ISM}}$ and
velocity offsets, $\Delta v$, from the galaxies that are $\sim 10$\%
of the halo escape velocity, $v_{\rm esc}$; these are identified as
bound clouds, possibly recycling material.  Five absorbers have
$Z_{\hbox{\tiny CGM}} \simeq Z_{\hbox{\tiny ISM}}$ and velocities
indicating $\Delta v> v_{\rm esc}$; these are identified as unbound
outflows.  Three absorbers have $Z_{\hbox{\tiny CGM}} \leq 0.2 \,
Z_{\hbox{\tiny ISM}}$ and are identified with infall.  In several
cases geometrical constraints confirm the flow direction of the
studied clouds.  \citet{stocke13} find no discernible differences in
the densities, ionization parameters, cloud sizes or masses between
the inflowing and outflowing absorbers.

In summary, the stellar mass gas-phase metallicity correlations places
strong constraints on the outflow and recycling of metal-enriched gas,
the inflow of metal-poor gas, and the incomplete mixing of the gas
through the CGM, while absorption line observations show there are
variations in gas metallicity that are consistent with these physical
processes.  The theory and observations indicate a connection between
the mean $W_r(2796)$, galaxy stellar mass, and gas phase metallicity
and kinematics.  Higher metallicity gas in chemically processed wind
material gives rise to larger equivalent widths in absorption.
Conversely, unmixed inflow material gives rise to smaller absorption
equivalent widths due to the lower metallicity.  

One would then expect the scatter in the $W_r(2796)$ distribution to
be primarily due to metallicity variations from sightline to sightline
(recalling little variation in cloud ionization, temperature, and
density from ionization models).  The decrease in {\MgII} covering
fraction with increasing distance from the central galaxy (and
absorption threshold) would then quantify the scatter in the
metallicity of the cool/warm CGM and imply that the metallicity is
more uniformly distributed from cloud to cloud at smaller
galactocentric distances but highly variable from absorbing cloud to
absorbing cloud at larger galactocentric distances.  If this scenario
is correct, it could partially explain why the frequency of sightlines
with upper limits on $W_r(2796)$ is higher at larger impact
parameters.

Further insight is gleaned from the simulations of
\citet{vandevoort+schaye11}, who conducted a thorough study of the CGM
parameter space with virial mass, stellar feedback, and distance from
the central galaxy\footnote{We quote results for their model
  REF\_L050N512.}.  For galaxies in $ \log M_{\rm\, h}/M_{\odot}
\simeq 12$ halos, their ``cold'' CGM gas exhibits four orders of
magnitude spread in metallicity at $R \simeq R_{\rm vir}$, with
density-weighted mean $Z_{\rm gas} \simeq 0.01$, while the ``hot'' gas
has only a single order of magnitude spread with $Z_{\rm gas} \simeq
0.1$.  Deeper inside the virial radius at $R \simeq 0.1 R_{\rm vir}$,
the metallicity of the inflowing gas has the narrow range $0.1 \leq
Z_{\rm gas} \leq 1$; this spread broadens to lower metallicities,
$10^{-2} \leq Z_{\rm gas} \leq 0.3$, by $R \simeq 0.3 R_{\rm vir}$ and
to $10^{-4} \leq Z_{\rm gas} \leq 0.5$ by $R \simeq R_{\rm vir}$.  On
the other hand, the outflow metallicity spread remains at a constant
$0.1 \leq Z_{\rm gas} \leq 1$ with radius out to $R \simeq R_{\rm
  vir}$.  The outflow fraction is $\simeq 0.4$, holding constant out
to $R_{\rm vir}=1.0$ (mass weighted).  Most of the gas mass is in the
inflow.  Across virial mass over the range $10 \leq \log M_{\rm\,
  h}/M_{\odot} \leq 13$, just inside the virial radius, the outflow
metallicity remains constant with galaxy virial mass within the range
$0.03 \leq Z_{\rm gas} \leq 0.5$.  However, the lower envelope on this
large range in the inflow metallicity rises to a higher minimum
metallicity as mass increases (the spread narrows toward a higher mean
metallicity).

To the degree that {\MgII} absorption probes the CGM, the general
increasing spread in the metallicity with increasing $R$ found by
\citet{vandevoort+schaye11} would be consistent with a growing
frequency of sightlines with upper limits on $W_r(2796)$ as the CGM is
probed closer to the virial radius.  This is consistent with a
covering fraction that decreases with increasing distance from the
central galaxy.

The behavior of the mass-normalized absorption envelope,
$\eta(M_{\rm\, h})$, remains to be understood.  This envelope has a
mean value of $\eta_{\rm v}^{\ast} \simeq 0.3$ (see
Figure~\ref{fig:DMRhenv}), is weakly dependent upon virial mass, and is
independent of absorption threshold.  This could be explained by a
narrower range of metallicity within $R \simeq 0.3 R_{\rm vir}$, as
suggested by the simulations of \citet{vandevoort+schaye11}.  Beyond
this scaled radius, the spread in the metallicity increases, and the
mean column density of {\MgII} absorbing gas has declined
\citep{ford13a}, which could be due to decreasing cloud sizes as
impact parameter increases \citep{stocke13}.  A second possibility is
that wind material, whether bound or unbound, remains in an ionization
state that is detectable in {\MgII} absorption primarily within $R =
0.3R_{\rm vir}$.  However, the lack of an ionization gradient in the
cool/warm CGM gas studied by \citet{stocke13}, and the simulations of
\citet{ford13a}, do not support this idea.  A third possibility is
that the majority of the {\MgII} absorbing gas is bound and recycles
such that the gas is confined within a ``turnaround'' radius of
$R/R_{\rm vir} \simeq 0.3$.  This would imply that wind material, on
average, would be required to reach $R/R_{\rm vir} \simeq 0.3$
regardless of galaxy virial mass.  Though this is not entirely
consistent with the findings of \citet{stocke13}, who find higher
metallicity unbound CGM clouds in several instances, it is
commensurate with the results of \citet{ford13b}, who find that the
majority of gas associated with {\MgII} absorption is ``recycled
accretion'', meaning that, regardless of the origin of the gas, it
will accrete onto the galaxy within the time span of $\sim 1$ Gyr.

The low frequency of sightlines with large $W_r(2796)$ found outside
$R/R_{\rm vir} = 0.3$, may, on average, be due to enriched sub-halos
(satellites) surrounding the more massive galaxies.  For $R/R_{\rm
  vir} > 0.3$, the growing frequency of very weak {\MgII} absorption
clouds and sightlines along which upper limits on $W_r(2796)$ are
measured, may be due to lower metallicity infalling material that has
not fully mixed with the recycling material inside the putative
``turnaround'' radius.  Accounting for satellites (around the more
massive galaxies) that enrich their local medium as they infall and
accounting for ``pristine'' infalling filaments, a wide range of
$W_r(2796)$ could arise from infalling gas.  Though the observational
evidence is quite compelling that the majority of strong {\MgII}
absorbers arise from metal-enriched wind driven material (described
above), some large $W_r(2796)$ values at large impact parameters could
be due to enriched infalling satellites in the more massive galaxy
halos.

For the mean $W_r(2796)$ to be constant with virial mass inside this
putative ``turnaround'' radius, we would need to invoke physics that
conspires to yield a degeneracy between gas metallicity, velocity
spread, and ionization conditions as a function of galaxy virial mass.
That is, in the final analysis, we should view the self-similarity of
the cool/warm CGM with virial mass as a reflection of a global
quasi-equilibrium regulation in which cool/warm cloud creation,
destruction and/or recycling timescales, hydrodynamical physics, and
external reservoirs of CGM gas balance so as to yield the simple
result that, on average, the equivalent width of {\MgII} absorbing CGM
gas is strongly connected to galactocentric distance with respect to
the virial radius, especially within the inner 30\%.



Though speculative in nature and only a qualitative picture, the
scenario we outline illustrates the possibility that the
self-similarity of the {\MgII} absorbing CGM with virial mass could
result from multiple processes that are consistent with what is
currently known about galaxies, the ISM, the CGM, and the local IGM
and environment.  Furthermore, this view of the CGM is one that is
fully consistent with simulations and models \citep[cf.,][]{dave11a,
  dave11b, dayal13, lilly13} that go far to explain a holistic
interconnectedness between star formation, stellar feedback, galaxy
stellar and virial mass, and the gas cycles of the ISM, CGM, and IGM
as constrained by observations.

The results we found for {\MgII} absorption should equally apply for
cool/warm gas absorption from other low-ionization potential metallic
species such as {\SiII}, {\CII}, and {\FeII}.  In fact, as we mention
previously, \citet{zhu13} have presented evidence that {\CaII} CGM
absorption is stronger in galaxies with higher stellar masses.  It
would be of interest to study the metallicity and abundance ratios as
a function of galactocentric distance relative and with respect to the
virial radius in order to discern whether the abundance gradient is
flat with $\eta_{\rm v}=D/R_{\rm vir}$, declines smoothly, falls off
precipitously at some point, or exhibits increasing scatter with
increasing distance from the central galaxy.  However, considering the
findings of \citet{stocke13} for a sample of $\simeq 70$ galaxies, such
an analysis will likely require an order of magnitude increase in the
number of quasar sightlines through the CGM environment of galaxies.



\acknowledgments

We thank Anatoly Klypin for insightful comments and discussion.  We
also thank the anonymous referee for comments that led to an improved
manuscript.  This work makes use of the {\magiicat} data
\citep{nikki-cat1} downloadable from {\it
  http://astronomy.nmsu.edu/cwc/Group/magiicat/}, which is hosted
by The Department of Astronomy at New Mexico State University.  CWC
and NMN were partially supported through grant HST-GO-12252 provided
by NASA's Space Telescope Science Institute, which is operated by AURA
under NASA contract NAS 5-26555.  CWC was also partially supported by
a NASA New Mexico Space Grant Consortium (NMSGC) Research Enhancement
Grant.  NMN was also partially supported by a NMSGC Graduate
Fellowship and by a three-year Graduate Research Enhancement Grant
(GREG) sponsored by the Office of the Vice President for Research at
New Mexico State University.


\begin{appendix}
\section{Halo Abundance Matching with Bolshoi}
\label{app:ham}

In this appendix, we present our findings from our explorations to
quantitatively understand the statistical and systematic uncertainties
inherent in our application of halo abundance matching as described in
\S~\ref{sec:ham}.  Since the $r$-band luminosity function is published
as $M_r - 5\log h$ in the Vega system, we performed the abundance
matching using this quantity (we presented $M_r$ in the {\sc ab}
system in Table~\ref{tab:obsprops}).  Thus, in this appendix, all
references to the $r$-band absolute luminosity refer to $M_r - 5\log
h$ in the Vega system, for which the range is $-23.6 \leq M_r - 5\log
h \leq -16.0$.  The conversion is $M_r(\hbox{{\sc ab}}) = [ M_r -5\log
  h \, ]_{\hbox{\tiny Vega}} + 0.1429$.

\subsection{Systematics and Scatter due to Luminosity Bin Size}

As mentioned in \S~\ref{sec:ham}, there is scatter in the $V_c^{\rm
max}$--$M_{\rm\, h}$ relation in the Bolshoi halo catalogs due to the
different formation times of halos of a given mass.  We treat this
scatter by calculating the mean virial mass, $M_{\rm\, h}$, within a
fixed luminosity bin, $\Delta M_r$, and assign the standard deviation
as the statistical uncertainty in the average virial mass. We compute
one-sided standard deviations to obtain insight into the asymmetry of
the virial mass distribution within the $\Delta M_r$ bin; as such, we
are not presenting formally proper statistical uncertainty
measurements but are quantifying the degree of scatter and skew in the
underlying distribution of $M_{\rm\, h}$ employed in obtaining the
mean value.

Since the LF has variable slope with $M_r - 5\log h$, the mean
$M_{\rm\, h}$ will have a systematic dependence on the width of
$\Delta M_r$.  The expectation is that the broader the bin size, the
more $M_{\rm\, h}$ will be skewed toward smaller values due to the
increased abundance of fainter galaxies in the LF.  Since the LF
slopes are different at different redshifts, this systematic skew will
be different at each redshift.  In addition, increasing the width of
$\Delta M_r$ results in the inclusion of more halos being averaged,
which affects the adopted uncertainties in $M_{\rm\, h}$.

\begin{figure*}[thb]
\epsscale{1.15} 
\plotone{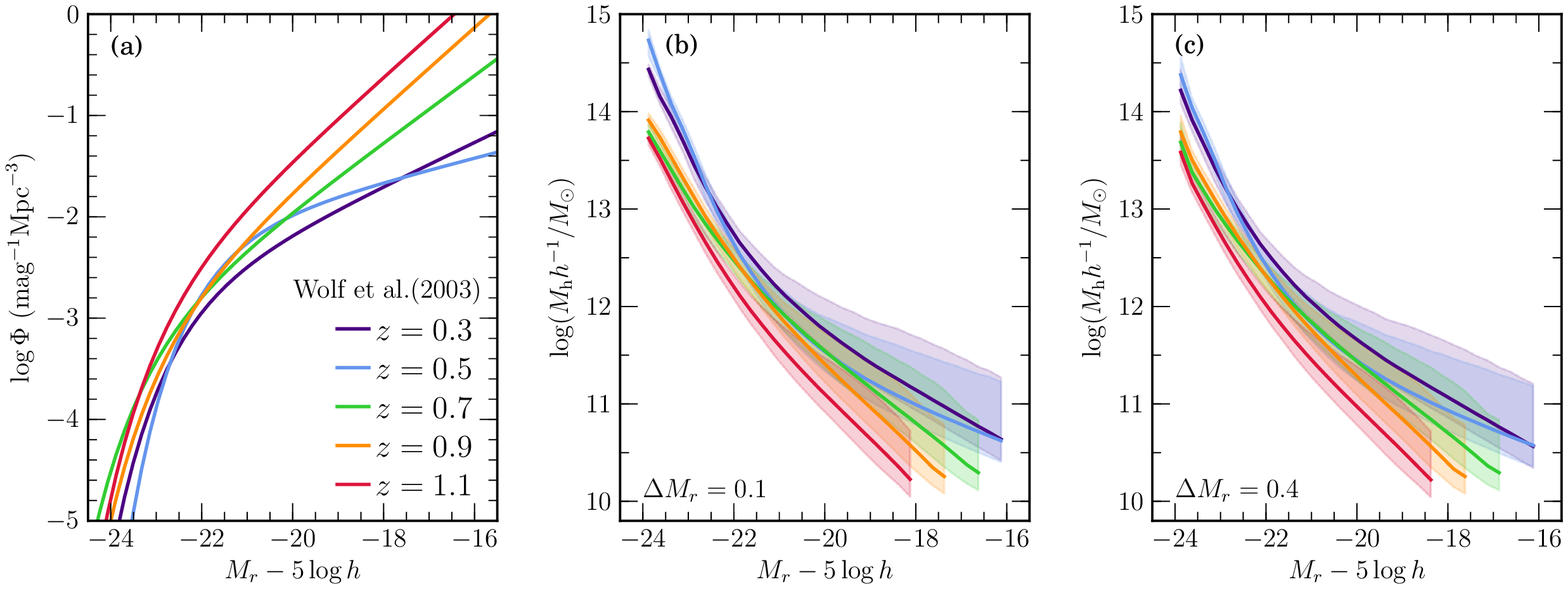}
\caption{(a) The fitted curves to the COMBO-17 $r$-band LFs of
  \citet{wolf03}, presented for the five redshifts, $z=0.3$, 0.5, 0.7,
  0.9, and 1.1.  (b,c) The mean virial mass and its standard deviation
  in units $h^{-1}$~M$_{\odot}$, determined by halo abundance matching
  the halos in the Bolshoi \citep{klypin11} simulations, versus
  $r$-band luminosity, $M_r - 5\log h$ (Vega).  The mean virial mass
  as determined using (b) a luminosity binning of $\Delta M_{\rm
    r}=0.1$ , and (c) a binning of $\Delta M_{\rm r}=0.4$. }
\label{fig:MhvsMr}
\end{figure*}


In Figure~\ref{fig:MhvsMr}a, we present the COMBO-17 $r$-band LFs
based upon the Schechter function parameter fits of \citet{wolf03}.
To examine how the width of $\Delta M_r$, affects the systematics of
and scatter in $M_{\rm\, h}$, we varied $\Delta M_r$ over the range
$0.1 \leq \Delta M_r \leq 0.4$ and performed the halo abundance
matching over the range $-24 \leq M_r - 5\log h \leq -16$.  In
Figures~\ref{fig:MhvsMr}b and \ref{fig:MhvsMr}c, we present $\log
M_{\rm\, h}h^{-1}/M_{\odot}$ versus $M_r-5\log h$ for $\Delta M_r =
0.1$ and $\Delta M_r = 0.4$, respectively.  The solid curves are the
mean $M_{\rm\,h}$ and the shaded regions are the one-side standard
deviations of the distribution of virial masses in the bin $\Delta
M_r$.  The redshift dependence is shown by the individual curves.

Consider Figure~\ref{fig:MhvsMr}b.  Note that the minimum and maximum
$M_{\rm\, h}$ are different for each redshift bin.  The maximum virial
mass, which increases with decreasing redshift, is dictated by the
distribution of virial masses in the halo catalog.  The increase in the
maximum virial mass with decreasing redshift reflects virial mass growth
evolution.  The minimum mass is dictated by the completeness of the
velocity function, $n(V_c^{\rm max})$, at small $M_{\rm\, h}$, as
discussed in both \citet{trujillo-gomez11} and \citet{klypin11}.  The
truncation of $n(V_c^{\rm max})$ is at brighter luminosity at higher
redshift because of the steeper LF at high redshift, i.e., the minimum
virial mass in the catalog gets assigned to a brighter galaxy.

Comparing Figures~\ref{fig:MhvsMr}b and \ref{fig:MhvsMr}c, we find
that the adopted bin size of $\Delta M_r$ has virtually no effect on
the scatter of each mass estimate and no more than a 0.3 dex
systematic lowering of $M_{\rm\, h}$ for $\Delta M_r = 0.4$ as
compared to $\Delta M_r = 0.1$ in the regime of $M_r - 5\log h < -23$.
This systematic is due to the steepness of the LF at the very bright
end.  For our methods, we find that $M_{\rm\, h}$ is sensitive to the
width of the luminosity bin to no more than $\simeq 0.35$ dex for the
highest masses when we also account for statistical uncertainties.  We
adopted $\Delta M_r \leq 0.1$ for this work in order to minimize the
effect of the variable slopes of the LF.

\subsection{Systematics due to Observational Uncertainty in the LF}

Since the abundance of dark matter halos is known to high precision in
the $\Lambda$CDM cosmology, a substantial source of potential
systematic uncertainty in the derived $M_{\rm\, h}$ could arise from
systematic errors in the evolution of the measured LF.  To examine the
range of possible systematics in our adopted virial mass calculations,
we explored variations in $M_{\rm\, h}$ under the presumption that
evolution in the observed LF from $z=1$ to $z=0$ is dominated by
systematic measurement errors.

To emulate systematics in the LF, we abundance matched to the observed
LF over the range $-24 \leq M_r - 5\log h \leq -16$ in each of the
five redshift bins using only the $z=0.1$ Bolshoi halo catalog.  We
thus evolve the LF while holding the halo population constant.  We
then performed the identical exercise using only the $z=1.0$ Bolshoi
halo catalog, thereby holding the halo population constant but with
abundance matching to a different virial mass distribution (separated
by $\sim 7$~Gyr of cosmic time).  The exercise emulates plausible
systematics in the evolution of the LF.  We also varied the width of
the luminosity bin used for obtaining the mean $M_{\rm\, h}$,
illustrating the $\Delta M_{\rm r}=0.1$ and 0.4 cases.

\begin{figure*}[thb]
\epsscale{0.9} 
\plotone{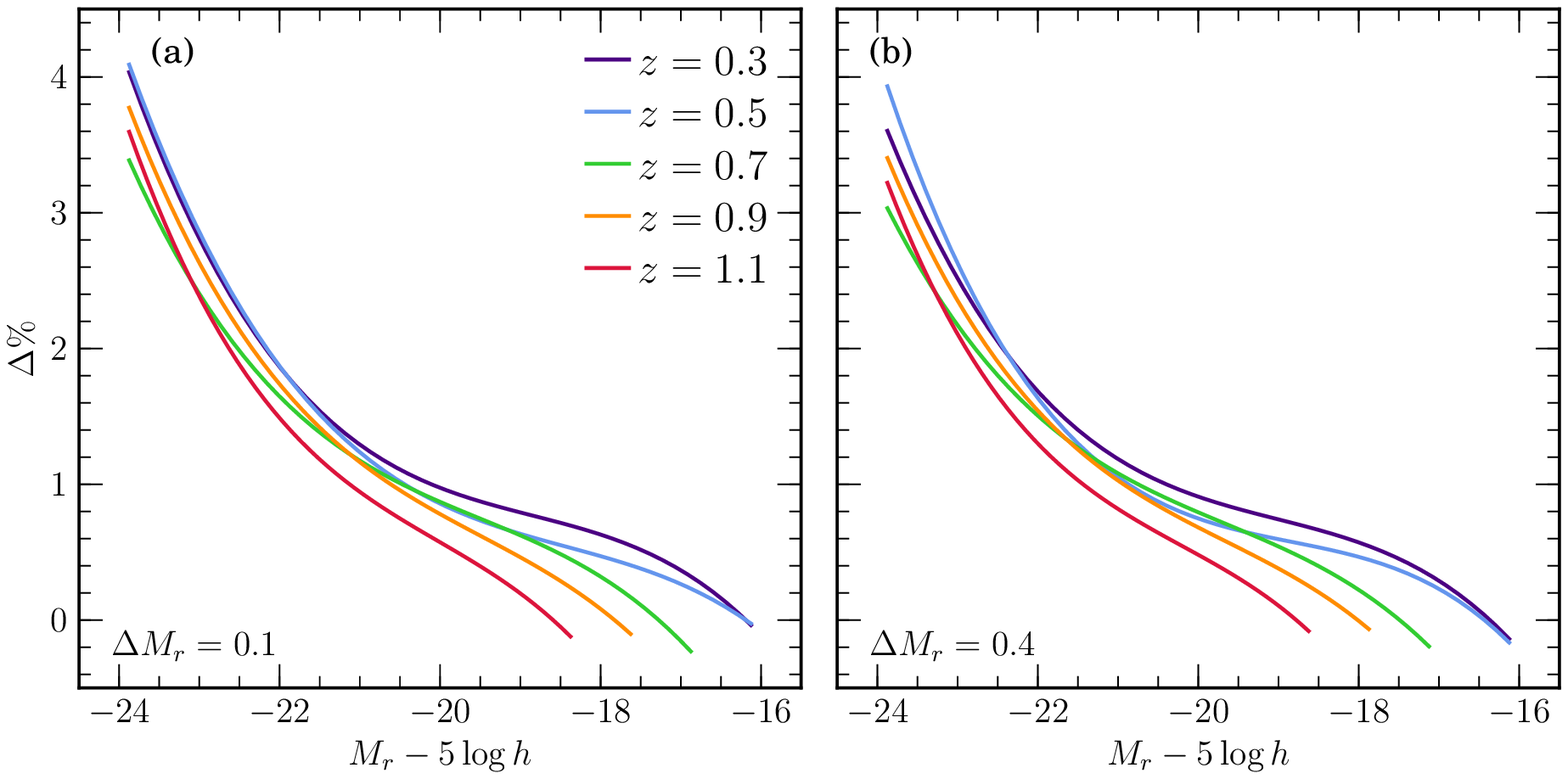}
\caption{The percent difference, $\Delta\%$, between the $M_{\rm\, h}$
  as a function of $M_r - 5 \log h$ (Vega) obtained for the
  exploration of systematic uncertainties in the evolution of the
  $r$-band LF (see text).  (a) The results for luminosity bin $\Delta
  M_r = 0.1$ used for averaging the scatter.  (b) The results for
  luminosity bin $\Delta M_r = 0.4$ used for averaging the scatter.
  The exercise indicates that no more than a 4\% systematic difference
  in $M_{\rm\, h}$ is likely to be present in our adopted values over
  the redshift range of our study. }
\label{fig:Test-PhiLvsz}
\end{figure*}

In Figure~\ref{fig:Test-PhiLvsz}, we plot the percent difference,
\begin{equation}
\Delta\% = 100 \cdot \left[ \frac{M_{\rm\, h}(z\!=\!0.1) - M_{\rm\,
h}(z\!=\!1.0)}{M_{\rm\, h}(z\!=\!0.1)} \right] \, ,
\end{equation}
between the $M_{\rm\, h}$ obtained with the $z=0.1$ and $z=1.0$
Bolshoi halo catalogs.  The results are shown for each redshift bin of
the LF (colored as in Figure~\ref{fig:MhvsMr}).
Figure~\ref{fig:Test-PhiLvsz}a illustrates the $\Delta M_r = 0.1$
exercise, and Figure~\ref{fig:Test-PhiLvsz}b, illustrates the $\Delta
M_r = 0.4$ exercise.

The exploration indicates that no more than a 4\% systematic
difference in $M_{\rm\, h}$ is likely to be present in our adopted
values over the redshift range of our study.  The effect monotonically
increases toward the bright end of the LF and is somewhat independent
of the shape of the LF in this luminosity regime.  There is also some
dependence on the faint end slope of the LF in the range $M_r -5 \log
h > -20$ at the level of 1\% percent difference.  We should have no
more than a $\Delta\% =4\%$ systematic error at the highest mass end
based upon the reasonable assumptions we have incorporated to model
systematic uncertainty in LF evolution.

\section{The Cooling Radius}
\label{app:rcool}

As described in \S~\ref{sec:rcooldiscuss}, the cooling radius is the
radial distance, $r=R_{\rm c}$, at which the initial hot gas density,
$\rho_{\rm gas}(r)$, equals the cooling density, $\rho_{\rm c}$, the
characteristic density at which gas has time to cool since the halo
formed.  That is, the cooling radius is defined when $\rho_{\rm
  gas}(r) = \rho_{\rm c}$ is satisfied.
We adopt the two-phase halo model of \citet[][hereafter
  MB04]{maller04}.  Equating their Eqs.~9 and 12, we solve for the radius
$r=R_{\rm c}$ that zeros the relation
\begin{equation}
\frac{f_b \, M_{\rm\, h}}{4\pi R_s^3 \, g [C_{\rm v}( M_{\rm\, h},z_{\rm gal})]}
\frac{R_s^3}{\left\{ r+(3/4)R_s \right\} \left( r + R_s \right)^2}
-
\frac{3\mu_e^2 m_p kT }{2\mu_{\hbox{\tiny N}} \tau_{\rm f} \Lambda (T,Z_{\rm gas})} 
= 0 \, ,
\label{eq:rootsolve}
\end{equation}
where the first term on the left hand side is the initial gas density
profile having a thermal core of $3R_s/4$ and the second term on the
right hand side is the cooling density.  Though Eq.~\ref{eq:rootsolve}
can be rearranged into a cubic equation that can be root solved, we
obtained the solution using Brent's method \citep[see][]{recipes} to a
fractional accuracy $\Delta R_{\rm c}/R_{\rm c} \leq 10^{-7}$.

The central density, which provides the amplitude of the radial gas
density profile, depends upon $f_b = \Omega_b/\Omega_m = 0.17$, the
cosmic mean baryon mass fraction, and the scale radius, $R_s = R_{\rm
  vir}/C_{\rm v}$.  The concentration parameter depends upon both
virial mass and redshift due to an evolving dark matter density
profile in response to mass growth.  \citet{bullock01a} show that the
median value is well approximated by the relation
\begin{equation}
\log C_{\rm v}( M_{\rm\, h},z_{\rm gal}) \simeq 0.9823 - 0.13 \log (
M_{\rm\, h}/10^{13}\, \hbox{M}_{\odot} ) - \log (1+z_{\rm gal}) \, .
\label{eq:concentration}
\end{equation}
The amplitude of the radial gas density profile also depends upon the
concentration according to the function 
\begin{equation}
g[x] = 9\ln (1 + 4x/3) - 8 \ln (1+x) - 4x (1+x)^{-1} \, .
\end{equation}

\begin{figure*}[thb]
\epsscale{1.15} 
\plotone{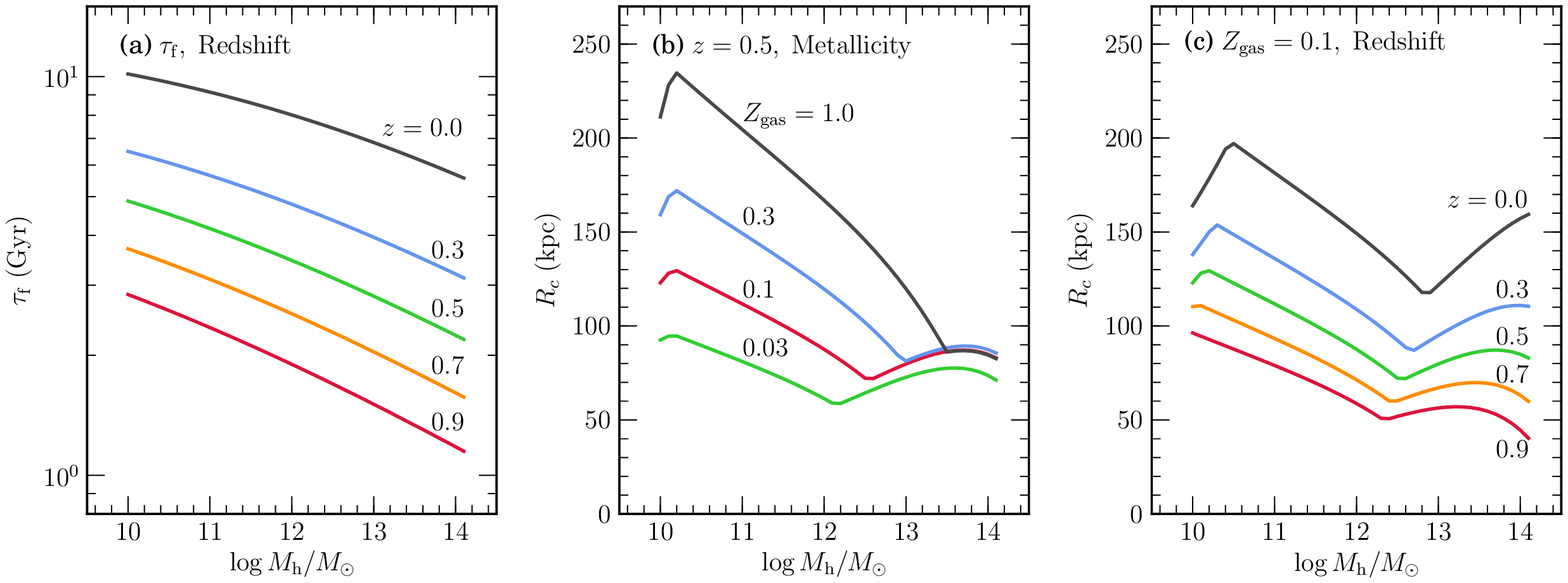}
\caption{(a) The formation time, $\tau_{\rm f}$, computed from
  Eq.~\ref{eq:tform}, as a function of virial mass, $M_{\rm\, h}$, at
  a given redshift over the range $0 \leq z \leq 1$ for $z=0.0$, 0.3,
  0.5, 0.7, and 0.9.  (b) The dependence of the cooling radius,
  $R_{\rm c}$, on the metallicity of the hot halo gas for $z=0.5$. (c)
  The dependence of $R_{\rm c}$ on redshift for a hot halo gas
  metallicity $Z_{\rm gas} = 0.1$.  The turnover points in the curves
  are due to the different cooling regimes in the cooling function
  $\Lambda (T,Z_{\rm gas})$.}
\label{fig:rcool}
\end{figure*}

The cooling density depends upon $\mu_e$, the mean mass per electron,
$\mu_{\hbox{\tiny N}}$, the mean mass per nuclear particles,
$\tau_{\rm f}$, the mean formation time for a halo of mass $M_{\rm\,
  h}$ for a galaxy at redshift $z_{\rm gal}$, and $\Lambda (T,Z_{\rm
  gas})$, the volume averaged cooling rate [cm$^3$ erg s$^{-1}$].  For
the mean masses per particle, we follow MB04 and adopt $\mu_e = 1.18$
and $\mu_{\hbox{\tiny N}} = 0.62$ for a fully ionized gas with a
helium mass fraction of $Y=0.3$.  The halo formation time is computed
from $\tau_{\rm f} = t_{\hbox{\tiny LB}}(z_{\rm f}) - t_{\hbox{\tiny
    LB}}(z_{\rm gal})$, where (see MB04, Eq.~8)
\begin{equation}
z_{\rm f} = z_{\rm gal} + 0.122 \, C_{\rm v}( M_{\rm\, h},z_{\rm gal}) 
\ln \left[  {M_{\rm\, h}(z_{\rm gal})} /  {M_{\rm\, h}(z_{\rm f})} \right] \, .
\label{eq:tform}
\end{equation}
The look-back time is computed from $ t_{\hbox{\tiny LB}}(z) =
\int _{0}^{z} dz/[(1+z)E(z)]$, where $E^2(z) = \Omega_m (1+z)^3 +
\Omega _{\Lambda}$.  Following MB04, we adopt $M_{\rm\, h}(z_{\rm
  gal}) / M_{\rm\, h}(z_{\rm f}) = 2$.  In Figure~\ref{fig:rcool}a, we
plot $\tau_{\rm f}$ as a function of virial mass at various redshifts.
Since $\rho_{\rm c} \propto \tau_{\rm f}^{-1}$.  it is clear that
overestimating the formation time would have the effect of
systematically underestimating the cooling radius of our sample
galaxies.  If for example, we directly applied Eq.~8 from MB04
assuming that our measured virial mass is applied at $z=0$, we would
underestimate $\rho_{\rm c}$ by as much as a factor of five for the
highest redshift galaxies in our sample.

The temperature of the hot gas is $T = \mu_{\hbox{\tiny N}} (V_c^{\rm
  max})^2 /2\gamma k$, where $V_c^{\rm max} = [GM_{\rm\, h}(R_{\rm
    max})/R_{\rm max}]^{1/2}$ is the peak (maximum) circular velocity,
which is computed for the mass inside $R_{\rm max} = 2.15 R_s$.  We
compute $V_c^{\rm max}$ assuming an NFW dark matter density profile
\citep{nfw95, klypin01} normalized to the measured $M_{\rm\, h}(R_{\rm
  vir})$ and adopting the concentration parameter given by
Eq.~\ref{eq:concentration}.  Following MB04, we adopt an adiabatic
index of $\gamma = 1$ for isothermal gas.  For the cooling function,
$\Lambda (T,Z_{\rm gas})$, we employ the approximate piece-wise
power law function of MB04 as outline in their Equation~A2\footnote{We
  found two consequential typographical errors in Appendix A of MB04.
  In their Equation A2, the power-law index $\alpha$ for the
  temperature regime $T_r < T \leq T_m$ has a sign error.  It should
  be expressed $\alpha = 1 + (1/3) \ln Z_{\rm gas}$. For the
  temperature regime $T_m < T \leq T_b$, the term $(T/T_b)^{-1}$
  should be $(T/T_m)^{-1}$, which is required of the piece-wise
  approximation function if the amplitudes are to match across
  temperature regimes at $T=T_m$.}.  The behavior of the cooling
function has a non-trivial dependence upon the gas metallicity,
$Z_{\rm gas}$, which remains an important unconstrained quantity.

In Figure~\ref{fig:rcool}b, we plot the metallicity dependence of
$R_{\rm c}$ as a function of virial mass for halos at $z=0.5$.  Shown
are $Z_{\rm gas} = 0.03$, 0.1, 0.3, and 1.0 in solar units.  The
changes in slopes in $R_{\rm c}$ as a function of virial mass are due
to the different cooling regimes, which are defined by the temperature
ranges $T_r < T \leq T_m$ (recombination cooling by hydrogen), $T_m <
T \leq T_b$ (metal-line cooling), and $T>T_b$ (bremsstrahlung
cooling), where $T_r = 1.5 \times 10^4$~K, $T_m = 1.5 \times 10^5$~K,
and $T_b = 10^6 + (1.5\times 10^7) \cdot Z_{\rm gas}^{2/3}$~K (see MB04,
Appendix A).  At fixed redshift and virial mass, the magnitude of
$R_{\rm c}$ is due only to the metallicity dependence of the cooling
function, which has a power-law slope that steepens with $Z_{\rm gas}$
for $T_r < T \leq T_m$. Higher metallicity results in a higher
electron density, and therefore an increased recombination rate; thus
increases the cooling rate, which lowers the cooling density and
therefore increases $R_{\rm c}$.  At $\log M_{\rm\, h}/M_{\odot} >
12$, the upward turn in $R_{\rm c}$ occurs at different virial masses
because metal-line cooling dominates to higher temperature
in higher metallicity gas, thereby elevating the temperature at which
bremsstrahlung cooling begins to dominate ($T>T_b$).  As virial mass
increases, the formation time is shorter, and this manifests as a
turn down in $R_{\rm c}$ at the very highest masses.
Figure~\ref{fig:rcool}b clearly illustrates the sensitivity of $R_{\rm
  c}$ to the gas metallicity of the hot phase and that this
sensitivity is most pronounced at the lowest virial masses, roughly a
factor of 1.5 increase in $R_{\rm c}$ for a 1~dex increase in $Z_{\rm
  gas}$ at $\log M_{\rm\, h}/M_{\odot} = 11$.

\subsection{Metallicity and the Fiducial Model}

Given the metallicity dependence of $R_{\rm c}$, there can be uncertainty
in the size of the cooling radius for fixed virial mass.
However, as we discuss below, observations and theory corroborate that
the average metallicity of the hot gas of galaxy halos can be well
approximated as $Z_{\rm gas} \simeq 0.1$ for a large range of virial
mass and galaxy morphological type.

Observations of hot halos of spirals and ellipticals indicate they
both obey the same $L_{\hbox{\tiny X}}$--$L_K$ (0.5--2.0 keV X-ray and
$K$-band luminosity) relations and the same $L_{\hbox{\tiny
    X}}$--$T_{\hbox{\tiny X}}$ relations, from which a common origin
of hot coronal gas in both early and late type galaxies is inferred
\citep{crain10}.  The similar correlations arise because
$L_{\hbox{\tiny X}}$, $L_K$, and $T_{\hbox{\tiny X}}$ are all
proportional to virial mass.

Comparing observations to their {\sc gimic} simulations
\citep{crain09}, drawn from the Millennium Simulation
\citep{springel05}, \citet{crain13} infer that the hot CGM observed
via X-ray emission has its origins in both hierarchical accretion and
stellar recycling in that the majority of $L_{\ast}$ galaxies develop
quasi-hydrostatic coronae through shock heating and adiabatic
compression of gas accreted from the intergalactic medium (IGM),
supplemented by relatively small amounts of gas recycled through the
galaxy by stellar feedback.  They conclude that the hot corona is
primarily primordial gas and is forged via accretion during galaxy
assembly.

Though the range of hot coronal metallicities determined from X-ray
luminosities might suggest near solar enrichment, \citet{crain13} find
that luminosity-weighting of X-ray measurements bias the perceived
metallicity of hot coronal gas.  In their simulations, $L_X$ weighted
metallicities are $Z_{\rm gas} \sim 1$, but gas mass-weighted
metallicities are $Z_{\rm gas} \sim 0.1$ (with a very shallow trend
for metallicity to decrease with increasing virial mass).

\citet{hodges13} reported $Z_{\rm gas} \sim 0.1$ for NGC~891 (a late
type galaxy), and argue that the primary source of the gas is IGM
accretion.  Even at higher redshifts than covered by our sample of
galaxies, simulations suggest that the hot coronal metallicity is
consistent with $Z_{\rm gas} \sim 0.1$.  \citet{shen12}, using their
ErisMC simulations, find total gas metallicities of $Z_{\rm gas}
\simeq 0.08$ at $\simeq 100$ kpc at $z=3$, consistent with recent
observations of circumgalactic metals around Lyman Break Galaxies.  In
the {\sc owls} simulations, \citet{vandevoort+schaye11} find $Z_{\rm
  gas} \sim 0.1$ in the ``hot mode'' phase of the CGM ($T_{\rm
  max} > 10^{5.5}$~K) for the virial mass range $\log M_{\rm\,
  h}/M_{\odot} = 10$--13 at $z=2$.

Based upon the above considerations, we adopt $Z_{\rm gas}=0.1$ for
our fiducial model for computing the cooling radius for our galaxy
sample.  In Figure~\ref{fig:rcool}c, we plot the redshift evolution of
$R_{\rm c}$ for $Z_{\rm gas} = 0.1$.  Shown are $z = 0$, 0.3,
0.5, 0.7, and 0.9.  At fixed mass and metallicity, evolution in the
cooling radius is dominated by the formation time of the halo and the
concentration parameter, which sets the hot gas density scale via
$g[C_{\rm v}(M_{\rm\, h},z_{\rm gal})]$, the scale radius via
$R_s=R_{\rm vir}/C_{\rm v}$, and the gas temperature via $V_c^{\rm
  max}$, since $R_{\rm max}$ is proportional to $R_s$.

\section{Multivariate Behavior}
\label{app:multivariate}

To further elucidate the multivariate relationships between
$W_r(2796)$, virial mass, impact parameter, virial radius, and
theoretical cooling radius, we performed bivariate Kendall-$\tau$ and
{\sc bhk}-$\tau$ non-parametric rank correlation tests between these
quantities.  We remind the reader that the {\sc bhk}-$\tau$ test
applies when upper limits must be taken into account.  In
Table~\ref{tab:rankcorr}, we present our results, where $N_{\rm sys}$
is the number of galaxies in the test, $\tau_k$ is the Kendall-$\tau$
(which ranges between $-1$ for a 1:1 anti-correlation and $+1$ for a
1:1 correlation), $P(\tau_k)$ is the probability of that value of
$\tau_k$ under the null-hypothesis assumption, and $N(\sigma)$ is the
significance level for the normal distribution of non-parametric
rankings of the $N_{\rm sys}$ data points.  For additional insight, we
separated the full sample into ``absorbers'' and ``non-absorbers''
[those with upper limits on $W_r(2796)$].


The correlation between impact parameter and virial mass provides much
insight into the cool/warm CGM.  All previous {\MgII} surveys
\citep[see][and references therein]{nikki-cat1, nikki-cat2} have
indirectly reported a correlation between galaxy luminosity and impact
parameter, which has been interpreted as a fundamental relationship
between the absorption radius and luminosity, i.e., the Holmberg
relationship $R(L) = R_{\ast}(L/L^{\ast})^{\beta}$, where $\beta > 0$.
It is thus no surprise this correlation is present with virial mass,
as we quantified and showed in Figure~\ref{fig:DMhenv}.  Note that
there is no statistically significant correlation between $D$ and
$M_{\rm\, h}$ for the non-absorber subsample; however, there is a
positive trend which is consistent with higher mass halos having
higher covering fraction at fixed $D$ \citep{cwc-massesI}.

Since $R_{\rm vir}$ is proportional to $M_{\rm \, h}$, the lack of a
significant correlation between $\eta_{\rm v}=D/R_{\rm vir}$ and
$M_{\rm \, h}$ is also a consequence of the correlation between $D$
and $M_{\rm \, h}$. Since $R_{\rm c}$ is inversely proportional to
$M_{\rm \, h}$, the highly significant correlation between $\eta_{\rm
  c}=D/R_{\rm c}$ and $M_{\rm \, h}$ is also a consequence of the
correlation between $D$ and $M_{\rm \, h}$.  For non-absorbers, the
$\eta_{\rm c}$--$M_{\rm \, h}$ correlation induced by the positive
trend between $D$ and $M_{\rm \, h}$ is a consequence of the
virial mass dependence of covering fraction at fixed $D$
\citep[see][Figure~2]{cwc-massesI}.  In essence, these statistics
reflect a cool/warm CGM {\MgII}-absorption radius that scales in
proportion to virial mass and a virial mass dependent covering
fraction at fixed $D$.

\begin{deluxetable}{ccccrccccrccccrcc}
\tablecolumns{17}
\tablewidth{0pt}
\setlength{\tabcolsep}{0.06in}
\tablecaption{Non-Parametric Correlation Tests\label{tab:rankcorr}}
\tablehead{
  \colhead{}                   &
  \colhead{}                   &
  \colhead{}                   &
  \multicolumn{4}{c}{Full Sample} &
  \colhead{}                   &
  \multicolumn{4}{c}{Absorbers Only} &
  \colhead{}                   &
  \multicolumn{4}{c}{Non-Absorbers Only\tablenotemark{b}} \\
  \colhead{(1)}                &
  \colhead{(2)}                &
  \colhead{\phantom{x}}                   &
  \colhead{(3)}                &
  \colhead{(4)\tablenotemark{a}}     &
  \colhead{(5)}                &
  \colhead{(6)}                &
 \colhead{\phantom{x}}                   &
  \colhead{(7)}                &
  \colhead{(8)\tablenotemark{a}}     &
  \colhead{(9)}                &
  \colhead{(10)}               &
 \colhead{\phantom{x}}                   &
  \colhead{(11)}               &
  \colhead{(12)\tablenotemark{a}}     &
  \colhead{(13)}               &
  \colhead{(14)}               \\                
  \colhead{Prop 1}             &
  \colhead{Prop 2}             &
  \colhead{}                   &
  \colhead{$N_{\rm sys}$}        &
  \colhead{$\tau_k$}           &
  \colhead{$P(\tau_k)$}        &
  \colhead{$N(\sigma)$}        &
  \colhead{}                   &
  \colhead{$N_{\rm sys}$}        &
  \colhead{$\tau_k$}           &
  \colhead{$P(\tau_k)$}        &
  \colhead{$N(\sigma)$}        &
  \colhead{}                   &
  \colhead{$N_{\rm sys}$}        &
  \colhead{$\tau_k$}           &
  \colhead{$P(\tau_k)$}        &
  \colhead{$N(\sigma)$}       }
\startdata
 $D$           &  $M_{\rm\, h}$  && 182 & $+0.24$     & $1.2\times 10^{-6}$  & {\bf 4.9} && 123 & $+0.29$ & $2.1\times 10^{-6}$ &  {\bf 4.7} && 59 & $+0.21$ & $1.7\times 10^{-2}$ &       2.4  \\  
 $\eta_{\rm v}$  & $M_{\rm\, h}$  && 182 & $-0.11$     & $2.1\times 10^{-2}$  &        2.3 && 123 & $-0.09$ & $1.5\times 10^{-1}$  &       1.4 && 59 & $-0.34$ & $1.1\times 10^{-4}$ &  {\bf 3.9} \\  
 $\eta_{\rm c}$  & $M_{\rm\, h}$  && 182 & $+0.37$     & $< 10^{-11}$         & {\bf 7.5} && 123 & $+0.39$ & $1.8\times 10^{-10}$ &  {\bf 6.4} && 59 & $+0.41$ & $4.0\times 10^{-6}$ &  {\bf 4.6} \\  
 $\eta_{\rm c}$  & $\eta_{\rm v}$ && 182 & $+0.43$     & $< 10^{-11}$         &  {\bf 8.7} && 123 & $+0.43$ &   $< 10^{-11}$       &  {\bf 6.9} && 59 & $+0.19$ & $3.4\times 10^{-2}$ &      2.1 \\ 
 $M_{\rm\, h}$   & $W_r(2796)$   && 182 & $\cdots$    & $9.6\times 10^{-1}$  &        0.1 && 123 & $+0.04$ & $5.1\times 10^{-1}$  &       0.7  && $\cdots$  & $\cdots$  & $\cdots$  & $\cdots$ \\
 $D$            & $W_r(2796)$  && 182 & $(-)\cdots$ & $1.2\times 10^{-10}$  & {\bf 7.9} && 123 & $-0.19$ & $1.8\times 10^{-3}$ &  {\bf 3.1} && $\cdots$  & $\cdots$  & $\cdots$  & $\cdots$   \\
 $R_{\rm vir}$   & $W_r(2796)$   && 182 & $\cdots$    & $6.7\times 10^{-1}$  &       0.4 && 123 & $+0.05$ & $3.8\times 10^{-1}$  &       0.9  && $\cdots$  & $\cdots$  & $\cdots$  & $\cdots$   \\
$ R_{\rm c}$     & $W_r(2796)$   && 182 & $\cdots$    & $3.1\times 10^{-2}$  &      2.2 && 123 & $-0.07$ & $2.4\times 10^{-1}$  &       1.2   && $\cdots$  & $\cdots$  & $\cdots$  & $\cdots$   \\
 $\eta_{\rm v}$  & $W_r(2796)$   && 182 & $(-)\cdots$ & $1.1\times 10^{-10}$ & {\bf 8.8} && 123 & $-0.27$ & $9.9\times 10^{-6}$  &  {\bf 4.4}  && $\cdots$  & $\cdots$  & $\cdots$  & $\cdots$  \\
 $\eta_{\rm c}$  & $W_r(2796)$   && 182 & $(-)\cdots$ & $3.8\times 10^{-8}$  & {\bf 5.5} && 123 & $-0.11$ & $6.4\times 10^{-2}$  &      1.9    && $\cdots$  & $\cdots$  & $\cdots$  & $\cdots$ \\[-7pt]
\enddata
\tablenotetext{a}{The {\sc bhk}-$\tau$ test does not provide a value for
  the Kendall-$\tau$.  When $N(\sigma) \geq 3$, we include a
  ``$+$'' for a correlation and ``$-$'' for an anti-correlation.}
\tablenotetext{b}{Since all $W_r(2796)$ measurements are upper limits,
  tests with absorption strength could not be performed.}
\end{deluxetable}

The correlation on the $\eta_{\rm v}$--$\eta_{\rm c}$ plane is
primarily an impact parameter sequence.  As seen in
Figure~\ref{fig:etacool}c, the data trace out increasing impact
parameter from the lower left to the upper right [small $(\eta_{\rm
    c},\eta_{\rm v})$ pairs to larger $(\eta_{\rm c},\eta_{\rm v})$
  pairs].  The effect of virial mass is to scatter an $(\eta_{\rm
  c},\eta_{\rm v})$ pair up and to the left relative to the sequence
for smaller virial mass, or to scatter the point downward and to the
right for higher virial mass.  The reason the slope and locus of
points is slightly shallower and below the 1:1 correlation line is due
to the $D$--$M_{\rm\, h}$ correlation.  Note that the $\eta_{\rm
  v}$--$\eta_{\rm c}$ correlation weakens below $3~\sigma$
significance for non-absorbers, which all have $W_r(2796) \leq
0.3$~{\AA}, primarily reside at $\eta_{\rm v} > 0.3$, and are found
both inside and outside the theoretical cooling radius.  This behavior
indicates that regions devoid of strong absorption do not commonly
persist within the inner 30\% of the virial radius, and at the same
time are not physically governed by their location with respect to the
cooling radius.

Of interest is the total lack of correlation between $W_r(2796)$ and
virial mass.  We have shown that the mean $W_r(2796)$ in fixed impact
parameter bins correlates with virial mass (see
Figure~\ref{fig:EWvsD}).  However, there is a strong anti-correlation
between $W_r(2796)$ and impact parameter \citep[($7.9~\sigma$ for the
  full {\magiicat} sample,][]{nikki-cat2}, so that when all impact
parameters are included, the correlations between $W_r(2796)$ and
$M_{\rm\, h}$ at fixed impact parameter are averaged out.  This point
is central to our discussion in \S~\ref{sec:previousworks}, where we
compare our findings to those of previous works.

Finally, we see that $W_r(2796)$ and $\eta_{\rm v}=D/R_{\rm vir}$ are
anti-correlated at high significance \cite[originally presented
  in][]{cwc-massesI}.  As compared to the $W_r(2796)$--$D$
anti-correlation, the increased significance of this anti-correlation
cannot be induced by the correlation between $D$ and $M_{\rm\, h}$
(i.e., the proportionality between the absorption radius and virial
mass), since this would have the effect of {\it reducing\/} its
significance relative to the $W_r(2796)$--$D$ anti-correlation.  On
the other hand, the anti-correlation between $W_r(2796)$ and
$\eta_{\rm c}=D/R_{\rm c}$ is anticipated because $R_{\rm c}$ is
inversely proportional to $M_{\rm h}$.

The upshot is that, giving full consideration to cross-correlation
effects, especially the proportionality between the absorption radius
and virial mass and the proportionality between covering fraction and
virial mass at fixed impact parameter, the location of the cool/warm
gas in relation to the virial radius is the strongest indicator of the
{\MgII} absorption equivalent width.  Furthermore, we know that this
fact generally applies across the full range of virial masses due to
the fact that [1] the mass-normalized absorption envelope, $\eta_{\rm
  v}(M_{\rm\, h})$, has a low sensitivity to virial mass and has a
value $\eta^{\ast}_{\rm v} = 0.3$ (see Figure~\ref{fig:DMRhenv}), [2]
the mean $W_r(2796)$ is independent of virial mass as a function of
$\eta_{\rm v}$, especially in the regime $\eta_{\rm v} \leq 0.3$ (see
Figure~\ref{fig:EWvsDR}), and [3] the significant virial mass
segregation on the $W_r(2796)$--$D$ plane vanishes on the
$W_r(2796)$--$\eta_{\rm v}$ plane \citep[see Figure~1c
  of][]{cwc-massesI}.


\end{appendix}


\end{document}